\documentclass[11pt]{article}

\usepackage[final]{acl}

\usepackage{times}
\usepackage{latexsym}

\usepackage[T1]{fontenc}

\usepackage[utf8]{inputenc}

\usepackage{microtype}

\usepackage{inconsolata}

\usepackage{graphicx}

\usepackage{amsmath,amsfonts,amssymb}
\usepackage{bm}
\usepackage{soul}
\usepackage{array}
\usepackage{float}
\usepackage{dsfont}
\usepackage{pifont}
\usepackage{amsmath}
\usepackage{amssymb}
\usepackage{amsfonts}
\usepackage{graphicx}
\usepackage{pdfpages}
\usepackage{enumitem}
\usepackage{booktabs}
\usepackage{multirow}
\usepackage{multicol}
\usepackage{fontawesome5}
\usepackage[most]{tcolorbox}
\usepackage{colortbl}
\tcbuselibrary{breakable}

\definecolor{yellow}{HTML}{F6BD60}
\definecolor{white}{HTML}{FFE0C1}
\definecolor{pink}{HTML}{F5CAC3}
\definecolor{tale}{HTML}{84A59D}
\definecolor{red}{HTML}{F28080}
\definecolor{orange}{HTML}{FF7F00}
\definecolor{green1}{HTML}{72C3A3}
\definecolor{green2}{HTML}{70B48F}
\definecolor{orange}{HTML}{FE8019}
\definecolor{grey}{HTML}{EBDBB2}
\definecolor{brain}{HTML}{FFABBE}
\definecolor{blue}{HTML}{076678}
\definecolor{purple}{HTML}{5861AC}
\definecolor{narrative}{HTML}{458588}
\definecolor{white2}{HTML}{F8F5E9}
\definecolor{white3}{HTML}{F5F2F2}
\definecolor{white4}{HTML}{FAF9EE}
\definecolor{tablewhite}{HTML}{EAEFEF}
\definecolor{purewhite}{HTML}{FFFFFF}
\definecolor{bargreen}{HTML}{85C79A}
\definecolor{barred}{HTML}{FDACAC}
\definecolor{textgreen}{HTML}{84B179}

\newcommand{\prism}{\ensuremath{%
    \mathchoice{\raisebox{-0.15em}{\includegraphics[height=0.8em]{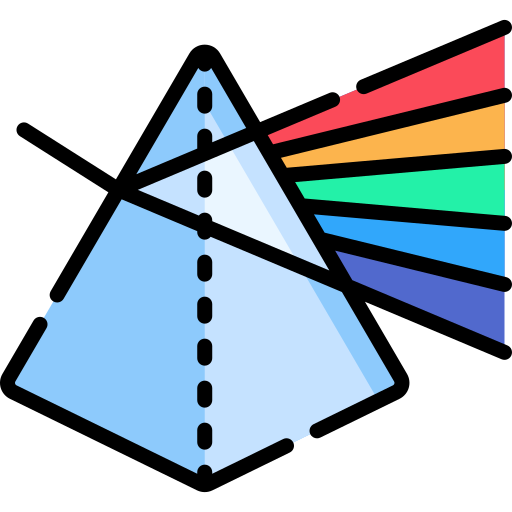}}}
    {\raisebox{-0.1em}{\includegraphics[height=0.8em]{figure/prism.png}}}
    {\raisebox{-0.1em}{\includegraphics[height=0.8em]{figure/prism.png}}}
    {\raisebox{-0.05em}{\includegraphics[height=0.8em]{figure/prism.png}}}
}}

\newcommand{\ourdata}{{\fontfamily{qag}\selectfont{\small{AppraiSal}}}}
\newcommand{\ourmethod}{{\fontfamily{qag}\selectfont{\small{PRISM}}}}
\usepackage{lipsum}

\title{
\textit{Why It Hurts:} Identifying the Drivers of Negative Thoughts in \\ Emotional Support Conversations
}

\author{
    Hainiu Xu$^{\clubsuit}$ \quad
    Zhaoyue Sun$^{\clubsuit}$ \\
    \textbf{Hanqi Yan}$^{\clubsuit}$ \quad
    \textbf{Jinhua Du}$^{\heartsuit}$ \quad
    \textbf{Caroline Catmur}$^{\clubsuit}$ \quad
    \textbf{Yulan He}$^{\clubsuit,\diamondsuit}$ \\[0.5em]
    $^\clubsuit$King's College London \\
    $^\heartsuit$ JinFlow Intelligence Ltd. \\
    $^\diamondsuit$The Alan Turing Institute \\
    {\tt \{hainiu.xu, yulan.he\}@kcl.ac.uk} \\
}

\begin{document}
\maketitle
\begin{abstract}
  Large Language Models (LLMs) are increasingly used for emotional support tasks, such as negative thought reframing. This task relies on modifying cognitive appraisals, the subjective interpretation of events that elicit negative emotions, which is typically conceptualized along multiple discrete dimensions. Current LLM-based frameworks model cognitive appraisal by exhaustively evaluating all possible dimensions, but they fail to account for the varying saliency of these dimensions across different contexts. In this work, we investigate a vital yet overlooked question: "\emph{Can LLMs infer the salient appraisal dimensions from emotional support conversations?}" To address this question, we introduce the \ourdata~benchmark, containing 996 emotional support conversations with human-annotated mental states, including salient cognitive appraisal dimensions. Furthermore, we propose \prism~\ourmethod, a multi-agent probabilistic framework grounded in Bayesian Inverse Planning, designed to improve LLMs' ability to identify context-specific appraisal dimensions. Experimental results show that \ourmethod~brings improvements to LLMs across various sizes, particularly in identifying the most salient appraisal dimensions\footnote{Data and code are available at \url{https://github.com/seacowx/AppraiSal-and-PRISM}.}.
\end{abstract}

\begin{figure} [t]
    \centering
    \includegraphics[width=0.9\linewidth]{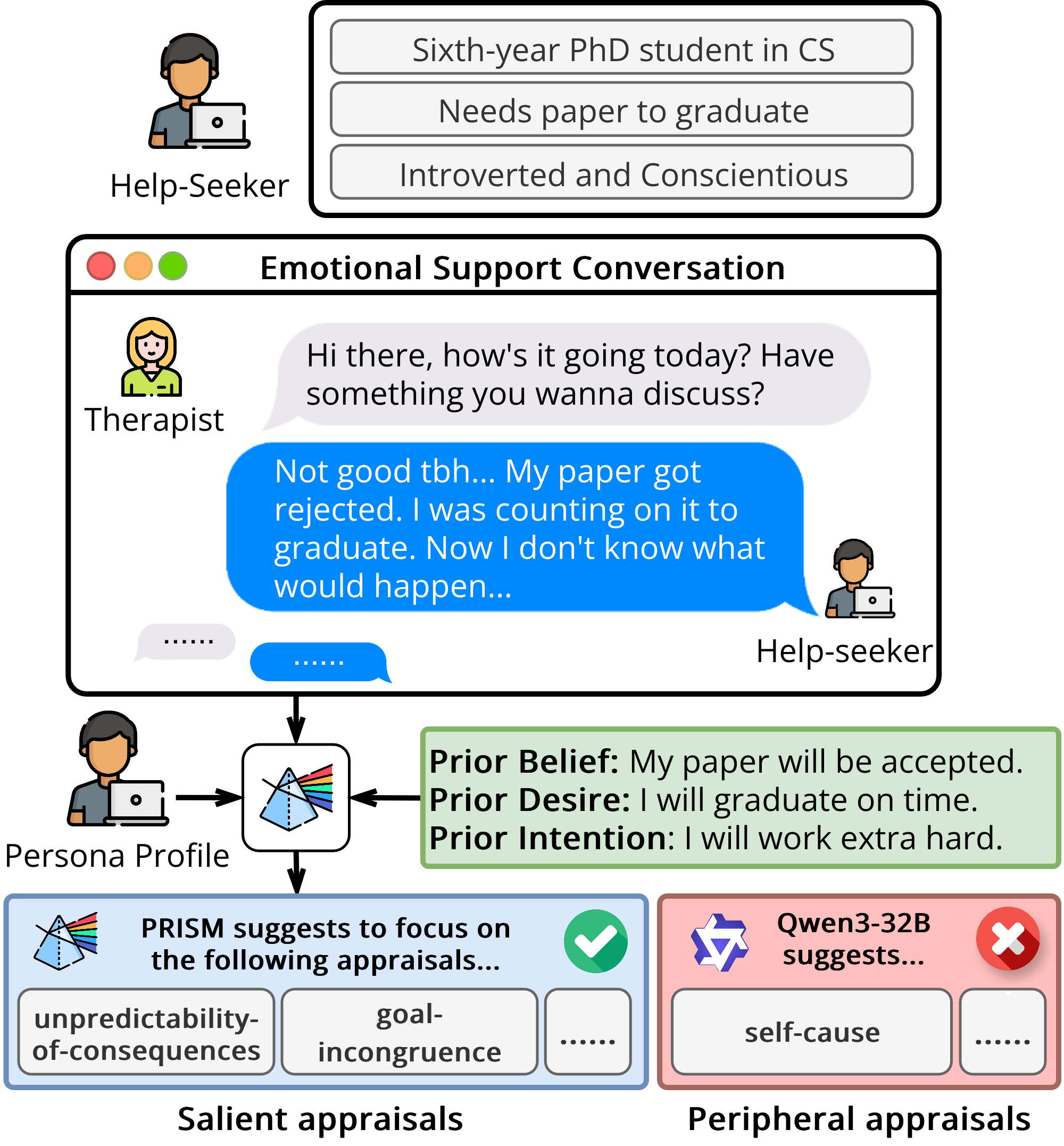}
    \caption{The salient appraisal reasoning task and the \prism~framework. Given a help-seeker’s persona profile, ESC trajectory, and pre-event BDI, \ourmethod~enables models to identify the salient appraisals more accurately (bottom left), whereas models without it often fail to recover the most relevant appraisals (bottom right).}
    \label{fig:figure1}
    \vspace{-1em}
\end{figure}

\section{Introduction}
Consider the event of a research paper rejection. A final-year PhD student may experience distress because the rejection blocks a critical goal (\emph{they need this paper to graduate}) or the consequences are uncertain (\emph{they do not know what will happen next}). In contrast, a researcher who attributes the rejection to reviewer misunderstanding or unfair evaluation may instead feel frustrated because someone else controlled the outcome. Each of these interpretations point to a different \textit{cognitive appraisal} \cite{arnold1960emotion, roseman2001appraisal, ortony2022cognitive, yeo2024associations}, as exemplified in Figure~\ref{fig:figure1}. Cognitive appraisals are typically conceptualized along discrete dimensions such as \textit{goal-incongruence}, \textit{unpredictability-of-consequences}, and \textit{unacceptable-consequences}\footnote{We use "appraisal dimensions" and "appraisals" interchangeably thereafter.}.

Effective emotional support requires identifying which appraisals are actually \emph{salient}. That is, what are the specific dimensions of subjective interpretation that are most responsible for a person's negative emotional response \cite{gross2015emotion}. A growing body of work incorporates cognitive appraisal theory into LLM-based frameworks to carry out empathic communications, such as cognitive reframing \cite{zhan2023evaluating, zhanlarge, zhou2025modeling}.
However, these approaches typically treat all appraisal dimensions as equally relevant, exhaustively evaluating every dimension for every situation, which may diminish the efficacy of the reframing process by failing to prioritize and focus on the most impactful appraisal dimensions.

The difficulty of the problem is that the salient appraisals are rarely stated explicitly in emotional support conversations (ESCs) and must therefore be inferred from an individual's utterances and contextual information \cite{troiano2023dimensional}. Doing so requires reasoning about an individual's latent mental states \cite{ortony2022cognitive}, such as their beliefs about the situation, their desired outcomes, and their intended actions, a capability commonly referred to as Theory of Mind (ToM). While much work has investigated LLMs' ToM capabilities \cite{ma2023towards, chen2025theory}, largely focused on classic false belief tasks, relatively little attention has been paid to ToM reasoning in the context of ESCs.
Consequently, a fundamental yet largely unexplored question remains: "\textit{Can LLMs reliably identify salient appraisal dimensions from emotional support conversations?}"

To address this question, we introduce the \textbf{Apprai}sal \textbf{Sal}iency (\ourdata) dataset. To the best of our knowledge, \ourdata~is the first benchmark designed to assess LLMs' ability to infer salient cognitive appraisals from ESCs. As an initial step toward improving LLMs' capability in this task, we propose \textbf{PR}obalistic \textbf{I}nference of \textbf{S}alient \textbf{M}ental States (\prism~\ourmethod), a multi-agent, probabilistic framework grounded in Bayesian Inverse Planning (BIP). Evaluation results from \ourdata~show that humans can consistently identify salient appraisal dimensions from ESCs while both open-source and proprietary models fall behind. While our \ourmethod~framework brings significant improvements, particularly in the case of identifying the most salient appraisal, LLMs' ability to infer salient cognitive appraisal dimensions still falls behind that of humans. Furthermore, our proof-of-concept study (\S\ref{sec:case_study}) shows that people generally prefer emotional support responses that are grounded in salient cognitive appraisal dimensions.

We summarize our contributions as follows:
\begin{itemize}[leftmargin=6mm, noitemsep]
\item[1.] We introduce \ourdata, a novel benchmark for assessing LLMs' ability to infer salient cognitive appraisals from ESCs.
\item[2.] We propose \prism~\ourmethod, a multi-agent, probabilistic framework grounded in BIP for identifying salient appraisals from ESCs.
\item[3.] We provide extensive experiments and a proof-of-concept study to demonstrate the effectiveness of \ourmethod~in inferring salient appraisals and the importance of salient appraisals in ESCs.
\end{itemize}

\section{The \ourdata~Dataset}
\paragraph{Task Formulation}
\label{sec:task_formulation}
We formulate the central task supported by \ourdata~as \emph{salient appraisal reasoning}. Consider an emotional support conversation between a help-seeker ($h$) and a therapist ($t$), represented as a dialogue trajectory $\mathcal{D}=\{(u_{h,1}, u_{t,1}), \ldots\}$, where $u_{h,1}$ denotes the help-seeker’s utterance at the first turn. Given the dialogue trajectory, the help-seeker’s persona information $\mathcal{P}$, and a predefined set of $n$ cognitive appraisal dimensions $\mathcal{A}=\{a_1,\ldots,a_n\}$, the goal is to identify the appraisal dimensions most salient for explaining the help-seeker’s negative emotional response. Formally, the model estimates the saliency of each appraisal dimension as
$
\mathbb{P}(a_i \mid \mathcal{D}, \mathcal{P}), \quad a_i \in \mathcal{A}.
$
It then returns the top-$k$ (in \ourdata, $k \leq 5$) appraisal dimensions with the highest estimated saliency:
\begin{equation}
\{\hat{a}_1, \ldots, \hat{a}_k\} = \text{Top-}k_{a_i \in \mathcal{A}} \, \mathbb{P}(a_i \mid \mathcal{D}, \mathcal{P}).
\end{equation}

\begin{figure*} [th]
    \centering
    \includegraphics[width=\textwidth]{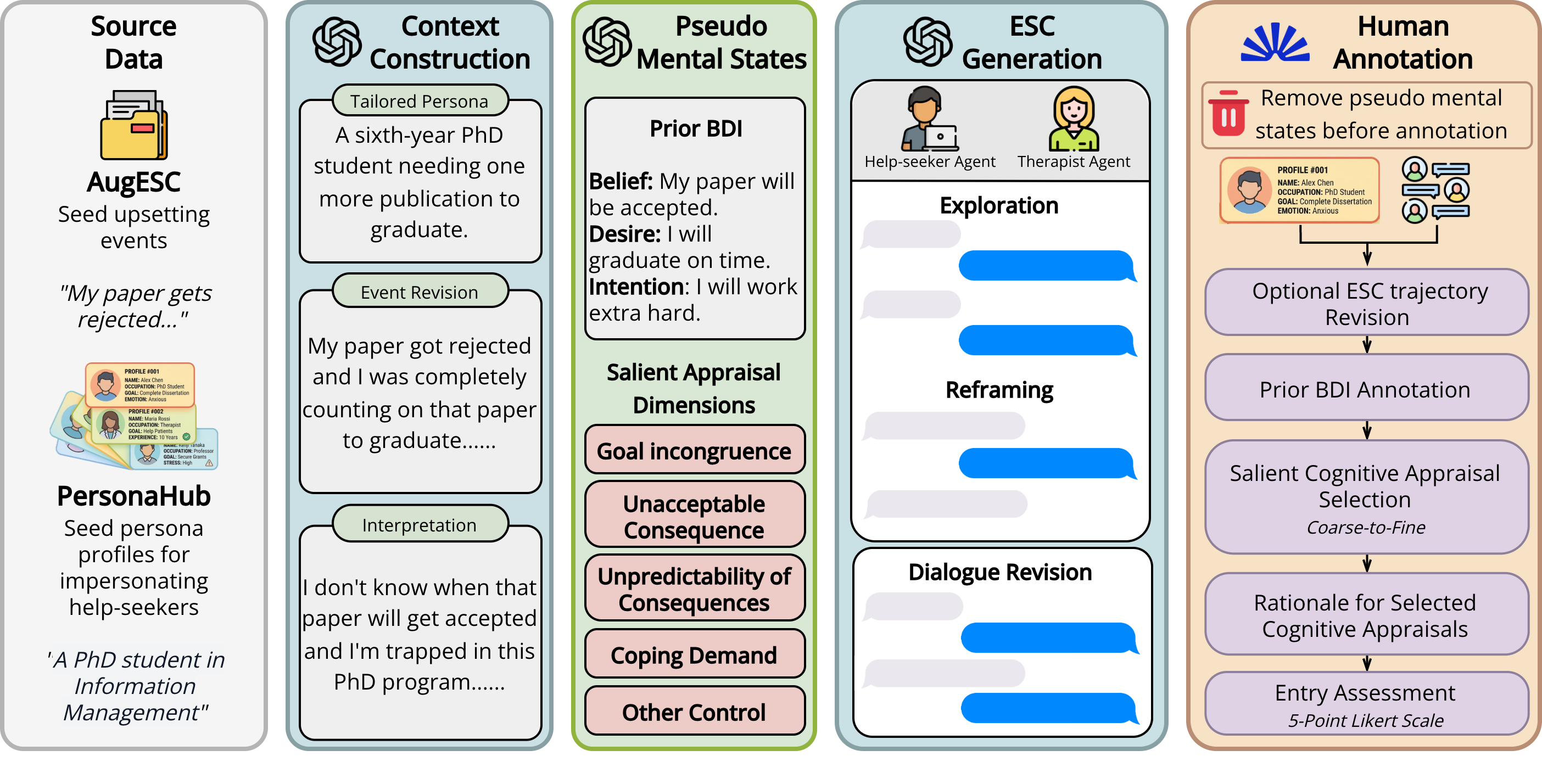}
    \vspace{-2em}
    \caption{An overview of the \ourdata~construction process. Upsetting events and persona profiles are combined to construct persona-grounded contexts, from which pseudo mental states are generated to support synthetic ESC generation. The generated conversations are then revised, stripped of pseudo appraisal labels, and annotated by humans for pre-event BDI, salient appraisals, and quality, followed by filtering to obtain the final dataset.}
    \label{fig:appraisal_construction}
    \vspace{-1em}
\end{figure*}

\paragraph{Cognitive Appraisals}
\label{sec:cogapps}
We define a set of appraisals based on the 21 appraisal dimensions given by \citet{hofmann2020appraisal}. Specifically, we removed dimensions that are redundant under ESC\footnote{Dimensions such as \textit{pleasantness} and \textit{unpleasantness} are redundant in ESC as all events in ESC are upsetting.} and dichotomized dimensions to reduce ambiguity, which resulted in a set of 14 appraisal dimensions. See Appendix~\ref{app:cogapps} for detailed construction of our appraisal dimensions.

\subsection{Data Construction}
\label{sec:data_construction}
Figure~\ref{fig:appraisal_construction} provides an overview of the \ourdata~construction pipeline.
Following prior works \cite{zheng2023augesc, zheng2023building, lee2024cactus, xiao2024healme, zhou2025crisp}, we first generate ESC trajectories using a dual-agent framework and subsequently recruit human annotators to evaluate dialogue quality and provide annotations.

\paragraph{Source Data \& Context Construction}
Each conversation of \ourdata~centers around an upsetting event, which we seed from AugESC \cite{zheng2023augesc}. To increase the fidelity of the original events, we construct a persona tailored for each event using profiles from the PersonaHub\footnote{We use the \href{https://huggingface.co/datasets/proj-persona/PersonaHub/viewer/persona}{persona subset of PersonaHub}.} \cite{ge2024scaling}. Each constructed persona includes basic demographic attributes such as name, gender, education level, and occupation, along with the Big Five personality traits.
Based on the tailored persona profiles, we rewrite\footnote{See Appendix~\ref{app:rewritten_situations} for examples of the rewritten entries.} the event from the help-seeker’s perspective and generate the help-seeker’s subjective negative interpretation using \texttt{GPT-5-mini}. See Appendix~\ref{app:persona_construction} for the detailed persona construction process.

\vspace{-0.5em}
\paragraph{Pseudo Mental States}
To ensure that the generated conversations are realistic and grounded, we construct pseudo mental states for each data instance. Building on the Ortony–Clore–Collins (OCC) model of emotion \cite{ortony2022cognitive}, we first generate the help-seeker’s \textit{beliefs}, \textit{desires}, and \textit{intentions}\footnote{The OCC model does not consider \textit{intentions}; we introduce them to make the underlying mental state more explicit.} prior to the occurrence of the upsetting event (pre-event BDI), representing the help-seeker’s mental state immediately before the upsetting event. We then prompt \texttt{GPT-5-mini} to generate 5 salient appraisals conditioned on the event, persona profile, and pre-event BDI. These machine-generated pseudo mental states are used to ground the ESC generation in the next step.

\paragraph{ESC Generation}
To produce high-quality ESC trajectories, we employ a two-stage generation pipeline. First, we use a dual-agent framework consisting of a help-seeker agent and a therapist agent to generate dialogue trajectories, with \texttt{gpt-oss-120b} \cite{openai2025gptoss120bgptoss20bmodel} as the backbone model. The help-seeker agent is conditioned on the tailored persona, their subjective interpretation of the upsetting event, their pre-event BDI, and 5 salient appraisals. The therapist agent is guided by the standard Cognitive Behavioral Therapy (CBT) procedure \cite{wenzel2017basic}. Specifically, the therapist begins with an \textit{Exploration} stage to understand and evaluate the help-seeker's thoughts and beliefs. This is followed by a \textit{Reframing} stage, in which the therapist attempts to reframe the help-seeker's negative thoughts using the \textit{reAppraisal} framework\footnote{See Appendix~\ref{app:reAppraisal} for the reAppraisal framework.}  \cite{uusberg2023reappraising}. The duration of each dialogue is determined autonomously by the help-seeker agent, which decides after each turn whether the conversation should continue.

After generating ESC trajectories, we perform trajectory revision using \texttt{GPT-5-Mini} with three objectives. First, each conversation must follow the Exploration--Reframing procedure typical of CBT. Second, the help-seeker must not voluntarily reveal their internal mental states unless explicitly asked. Third, the conversational style must correspond to the help-seeker's persona profile. For example, an \textit{antagonistic} person is expected to be less cooperative than an \textit{agreeable} person.

\paragraph{Human Annotation}
Once the ESC trajectories are revised, we remove the pseudo labels associated with the help-seeker's cognitive appraisals before annotation. Annotators are then provided with the help-seeker's persona profile and the generated ESC trajectory as context. Based on this information, each annotator performs four sub-tasks\footnote{The annotation task received ethical clearance in 2026 and is granted for a period of 5 years.}

First, the annotator reads the dialogue and optionally revises it to improve coherence and realism. Second, the annotator infers the help-seeker’s pre-event BDI, with the pseudo BDI provided as a reference. Third, the annotator identifies the salient cognitive appraisal dimensions using a coarse-to-fine protocol: the annotator first selects high-level appraisal groups and then refines the selection by choosing specific fine-grained dimensions.
To discourage careless annotation, we require annotators to select exactly five appraisal dimensions they consider salient in the given context and to provide a rationale for each choice.
Finally, the annotator rates the overall quality of the data entry on a 5-point Likert scale. See Appendix~\ref{app:esc_tom_annotation_interface} for the demongraphic information of the annotators, quality assurance questions, and the annotation interface.

\paragraph{Data Statistics}
 In total, we recruited 145 annotators from Prolific\footnote{https://www.prolific.com} who contributed 1,120 annotated entries. On average, each annotator was compensated with \$1.52 per data entry. We further filter the data by removing entries with an average rating falling below $3.0$, yielding the final \ourdata~with 996 entries. We present detailed data statistics in Appendix~\ref{app:detailed_data_stats}. To measure inter-annotator agreement, we randomly sampled 50 entries, which were annotated by three annotators from Prolific. As the annotations come in as a set of appraisal dimensions, we assess their agreement using average Jaccard Similarity as well as Krippendorff's $\alpha$. Results from Table~\ref{tab:iaa} show that the annotators are able to achieve a relatively strong agreement in their judgment of salient cognitive appraisal dimensions.

 \begin{table}[tp]
     \centering
     \small
    \resizebox{\columnwidth}{!}{
     \begin{tabular}{c c c c c}
     \toprule
          \multicolumn{4}{c}{\textbf{Jaccard Similarities}} & \multirow{2}{*}{\textbf{Krippendorff's $\boldsymbol{\alpha}$}} \\
          \cmidrule{1-4} A1|A2 & A1|A3 & A2|A3 & Mean & \\
     \midrule
     0.749 & 0.787 & 0.791 & 0.776 & 0.776 \\
     \bottomrule
     \end{tabular}
     }
     \caption{Inter-annotator agreement results on \ourdata. A1, A2, A3 correspond to the three annotators.}
     \label{tab:iaa}
 \end{table}

\begin{figure*} [t]
    \centering
    \includegraphics[width=\textwidth]{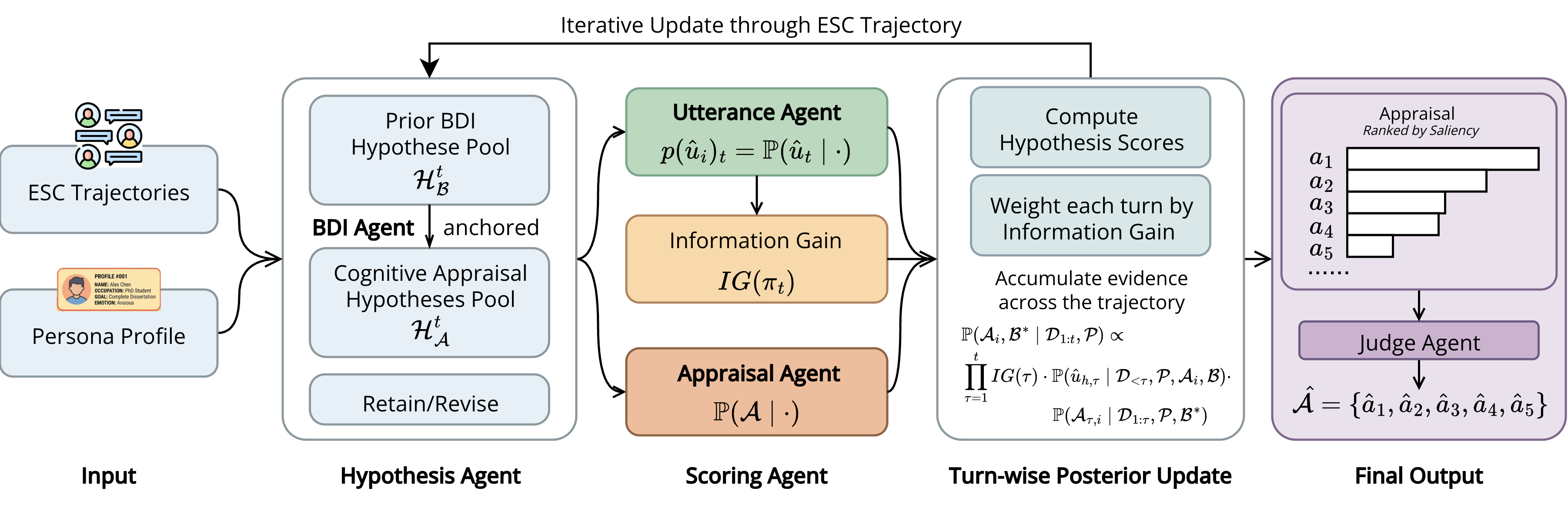}
    \caption{Overview of the \prism~\ourmethod~framework. Given an ESC trajectory and the help-seeker’s persona, PRISM performs hypothesis-driven Bayesian inverse planning over latent pre-event BDI and appraisal states. A hypothesis agent maintains candidate BDI and appraisal sets, three specialized agents score them turn by turn, and information gain weights the contribution of each dialogue turn to produce final saliency scores and the top-5 salient appraisals.}
    \label{fig:prism_framework}
    \vspace{-0.5em}
\end{figure*}

\section{The \ourmethod~Framework}
\label{sec:method}
As an initial step toward improving LLMs' ability to infer salient cognitive appraisals from ESC trajectories, we draw inspiration from recent ToM reasoning approaches \cite{jin2024mmtom, ying2024grounding, kimhypothesis, zhangautotom} and propose \ourmethod, a multi-agent framework grounded in Bayesian Inverse Planning, as shown in Figure~\ref{fig:prism_framework}.

\subsection{Preliminaries}

\paragraph{Forward Generative Model}
We model the help-seeker's action (utterance) in an ESC as a generative process. Following the OCC model \cite{ortony2022cognitive}, we assume that the help-seeker's utterance at each turn $t$ reflects their understanding of the dialogue context $\mathcal{D}_{<t}$, their persona profile $\mathcal{P}$, and their latent mental states, including pre-event BDI $\mathcal{B}$ and salient cognitive appraisals $\mathcal{A}$. Formally, we model the help-seeker's utterance as:
\begin{equation}
\hat{u}_{t} \sim \mathbb{P}(u_{t} \mid \mathcal{D}_{<t}, \mathcal{P}, \mathcal{B}, \mathcal{A}).
\label{eq:forward}
\end{equation}

\paragraph{Backward BIP Inference} Based on the forward generative process, we derive the backward inference under the BIP framework as follows:
\begin{align}
\begin{split}
   \mathbb{P}(\mathcal{A}, \mathcal{B}\mid&\mathcal{D}_{1:t}, \mathcal{P}) \propto \mathbb{P}(\hat{u}_{t} \mid \mathcal{D}_{<t}, \mathcal{P}, \mathcal{B}, \mathcal{A}) \cdot \\
   &\mathbb{P}(\mathcal{B}\mid\mathcal{D}_{<t}, \mathcal{P}) \cdot \mathbb{P}(\mathcal{A}\mid\mathcal{D}_{<t}, \mathcal{P}, \mathcal{B}),
\end{split}
\label{eq:bip_main}
\end{align}
where $\mathbb{P}(\hat{u}_{t} \mid \mathcal{D}_{<t}, \mathcal{P}, \mathcal{B}, \mathcal{A})$ denotes the likelihood of the help-seeker producing the current utterance given the dialogue history, persona profile, pre-event BDI, and cognitive appraisals. $\mathbb{P}(\mathcal{B}\mid\mathcal{D}_{<t}, \mathcal{P})$ denotes the posterior over pre-event BDI states given the dialogue history and persona profile up to turn $t-1$. $\mathbb{P}(\mathcal{A}\mid\mathcal{D}_{<t}, \mathcal{P}, \mathcal{B})$ denotes the posterior over appraisal sets given the dialogue history, persona profile, and pre-event BDI. A detailed derivation of Equation~\ref{eq:bip_main} is provided in Appendix~\ref{app:bip_derivation}.

\subsection{The \ourmethod~Workflow}

\paragraph{Hypothesis-Driven BIP Inference} Most BIP-based ToM reasoning methods assume a confined and small label space. 
In \ourdata, however, the model infers five salient appraisal dimensions from a set of 14 candidates, which leads to a combinatorial explosion of the hypothesis space. To constrain this search space, we first deploy a \texttt{hypothesis agent} to generate and manage candidate hypotheses, similar to \citet{kimhypothesis}. 

At the first turn of the ESC dialogue, the hypothesis agent is instructed to produce a batch of hypotheses. In the subsequent turns, instead of prompting the hypothesis agent to produce new sets of hypotheses, the agent is instructed to revise the existing hypotheses. Specifically, we define three actions that the hypothesis agent can make on each of the hypothesis: \texttt{Retain}, where the hypothesis is kept as-is; \texttt{Replace}, where the hypothesis is overwritten by a new one; and \texttt{Rerank}, where the rankings of the hypotheses are adjusted. Further, \ourmethod~maintains two hypothesis pools, one for the pre-event BDI state $\mathcal{B}$ and one for the cognitive appraisal set $\mathcal{A}$, each of size $h=8$. We denote these pools at turn $t$ by $\mathcal{H}_{\mathcal{B}}^{t}$ and $\mathcal{H}_{\mathcal{A}}^{t}$.

\paragraph{Posterior Modeling}
Conditional distributions in Equation~\ref{eq:bip_main}, namely $\mathbb{P}(\hat{u}_{t}\mid\cdot)$, $\mathbb{P}(\mathcal{B}\mid\cdot)$, and $\mathbb{P}(\mathcal{A}\mid\cdot)$, are instantiated by three LLM agents: the \texttt{utterance}, \texttt{bdi}, and \texttt{appraisal agent}.

For each agent, we instruct it to generate the target content associated with the conditional distribution it models. For instance, the \texttt{utterance agent} is tasked with generating the help-seeker’s next utterance. The conditioning variables, namely $\mathcal{D}_{<t}$, $\mathcal{P}$, $\mathcal{B}_t$, and $\mathcal{A}_{t, i}$, are incorporated into the instruction prompt $\mathcal{I}$. Suppose the observed help-seeker’s utterance at turn $t$ consists of $m$ tokens, denoted by $\hat{u}_{t} = \{w_1, w_2, \ldots, w_m\}$. We treat the observed utterance $\hat{u}_{t}$ as the target response and compute the log-likelihood score\footnote{See Appendix~\ref{app:prompt_logprobs} for details.}:
\begin{align}
\begin{split}
&p(\hat{u}_{t,i})
= \log \mathbb{P}(\hat{u}_{t} \mid \mathcal{D}_{<t}, \mathcal{P}, \mathcal{B}_t, \mathcal{A}_{t,i}) \\
&= \sum_{j=1}^{m}
\log \mathbb{P}\Big(
w_j | w_{<j}, \mathcal{I}(\mathcal{D}_{<t}, \mathcal{P}, \mathcal{B}_t, \mathcal{A}_{t,i})
\Big),
\label{eq:utterance_posterior}
\end{split}
\end{align}
where $\mathbb{P}(w_j \mid w_{<j}, \mathcal{I})$ denotes the token-level probability obtained from the LLM. We compute $p(\hat{u}_{t,i})$ for each $\mathcal{A}_{t,i} \in \mathcal{H}_{\mathcal{A}}^{t}, i = 1, \ldots, h$, yielding a set of likelihood scores, each corresponding to an appraisal hypothesis. We apply the same procedure to the \texttt{bdi agent} and the \texttt{appraisal agent}.

\paragraph{Information Gain}
In ESC, the amount of information conveyed about the help-seeker's appraisals can vary vastly across dialogue turns. To account for this variation, we use Shannon entropy to quantify the turn-level informativeness. Specifically, given the set of utterance log-likelihood scores $\{p(\hat{u}_{t,i})\}_{i=1}^h$, we first convert them into a probability distribution over hypotheses:
\begin{equation}
\pi_{t,i}
=
\frac{\exp\big(p({\hat{u}_{t,i}})\big)}{\sum_{k=1}^{h}\exp\big({p(\hat{u}_{t,k}})\big)}.
\label{eq:hypothesis_prob}
\end{equation}
We then define the information gain at turn \(t\) as
\begin{equation}
IG(\pi_{t})=1 - \frac{H(\pi_t)}{\ln h},
\label{eq:ig}
\end{equation}
where \(H(\pi_t) = -\sum_{i=1}^{h}\pi_{t,i}\ln \pi_{t,i}\) is the Shannon entropy of the hypothesis distribution.

$IG(\pi_t)$ measures how much more informative the utterance likelihood distribution at turn $t$ is than a uniform distribution over $h$ hypotheses. Intuitively, a more informative utterance $\hat{u}_t$ is more sensitive to differences among appraisal hypotheses $\mathcal{A}_{t, i}$, thereby inducing a sharper distribution over $p(\hat{u}_t)$; less informative utterances, in contrast, yield distributions closer to uniform.

\begin{table*}[t]
    \centering
    \resizebox{0.92\textwidth}{!}{
    \begin{tabular}{l c c c c c c}
    \toprule
        & \textbf{Precision@1} & \textbf{Precision@2} & \textbf{Precision@3} & \textbf{Precision@4} & \textbf{Precision@5}
        & \textbf{Jaccard Index} \\
        \toprule
        \rowcolor{white2}
        \textbf{Human} & $1.000$ & $0.990$ & $0.920$ & $0.775$ & $0.748$ &
        $0.612$   \\
        \midrule
        \textbf{Random} & $0.361$ & $0.342$ & $0.350$ & $0.358$  &
        $0.357$ & 
        $0.234$ \\
        \textbf{Majority} & $0.755$ & $0.655$ & $0.561$ & $0.530$ & $0.522$ &
        $0.379$ \\
        \midrule
        \multicolumn{7}{c}{Instruct Models} \\
        \midrule
        \textbf{\texttt{Qwen3-4B}} &	$0.722$ & $0.669$ &	$0.645$ & $0.593$ &	$0.570$ &
        $0.422$ \\
        \textbf{\texttt{Qwen3-32B}} & $0.758$ & $0.706$ & $0.686$ &	$0.648$ & $0.611$ & 
        $0.465$ \\
        \textbf{\texttt{Qwen3-30B-A3B}} &	$0.763$ & $0.694$ &	$0.655$ & $0.617$ &	$0.595$ & 
        $0.448$ \\
        \textbf{\texttt{Gemma3-4B}} & $0.438$ & $0.574$ & $0.602$ & $0.582$ & $0.566$ & 
        $0.416$ \\
        \textbf{\texttt{Gemma3-27B}} & $0.746$ & $0.684$ & $0.652$ & $0.593$ & $0.554$ &
        $0.406$  \\
        \midrule
        \multicolumn{7}{c}{Reasoning Models} \\
        \midrule
        \textbf{\texttt{Qwen3-4B}} & $0.759$ & $0.723$ &	$0.677$ & $0.632$ &	$0.598$ &
        $0.452$ \\
        \textbf{\texttt{Qwen3-32B}} & $0.739$ &	$0.709$ &	$0.676$ &	$0.636$ &	$0.608$ & 
        $0.461$ \\
        \textbf{\texttt{Qwen3-30B-A3B}} & $0.713$ &	$0.680$ &	$0.667$ &	$0.624$ &	$0.599$ & 
        $0.452$ \\
        \textbf{\texttt{gpt-oss-20B}} &	$0.563$ &	$0.596$ &	$0.609$ &	$0.595$ &	$0.565$ & 
        $0.414$ \\
        \textbf{\texttt{gpt-oss-120B}} &	$0.799$ &	$0.739$ &	$0.694$ &	$0.652$ &	$0.617$ &	
        $0.477$ \\
        \textbf{\texttt{DeepSeek-V3.2}}
        & $0.731$ & $0.727$ & $0.694$ & $0.638$ & $0.603$ &  
        $0.455$ \\
        \textbf{\texttt{GPT-5-Nano}} & $0.700$ &	$0.685$ &	$0.653$ &	$0.621$ &	$0.594$ &	
        $0.445$ \\
        \textbf{\texttt{GPT-5-Mini}} & $0.664$ &	$0.672$ &	$0.650$ &	$0.622$ &	$0.591$ &	
        $0.444$  \\
        \midrule
        \rowcolor{tablewhite}
        \multicolumn{7}{c}{with the \ourmethod~Framework} \\
        \midrule
        \rowcolor{tablewhite}
         \textbf{\texttt{Qwen3-4B}} & $0.851_{\text{ }0.129\uparrow}$ & $0.760_{\text{ }0.091\uparrow}$ & $0.688_{\text{ }0.043\uparrow}$ & $0.641_{\text{ }0.048\uparrow}$ & $0.616_{\text{ }0.046\uparrow}$ &	
         $0.469_{\text{ }0.047\uparrow}$ \\
        \rowcolor{tablewhite}
        \textbf{\texttt{Qwen3-32B}} & $0.820_{\text{ }0.062\uparrow}$ & $\underline{0.782}_{\text{ }0.076\uparrow}$ & $\boldsymbol{0.724}_{\text{ }0.038\uparrow}$ & $\boldsymbol{0.671}_{\text{ }0.023\uparrow}$ & $\underline{0.629}_{\text{ }0.018\uparrow}$ & 
        $\underline{0.488}_{\text{ }0.023\uparrow}$ \\
        \rowcolor{tablewhite}
        \textbf{\texttt{Qwen3-30B-A3B}} & $\boldsymbol{0.882}_{\text{ }0.119\uparrow}$ & $0.767_{\text{ }0.073\uparrow}$ & $0.684_{\text{ }0.029\uparrow}$ & $0.641_{\text{ }0.024\uparrow}$ & $0.616_{\text{ }0.021\uparrow}$ & 
        $0.471_{\text{ }0.023\uparrow}$ \\
        \rowcolor{tablewhite}
        \textbf{\texttt{Gemma3-4B}} & $0.734_{\text{ }0.296\uparrow}$ & $0.705_{\text{ }0.131\uparrow}$ & $0.660_{\text{ }0.058\uparrow}$ & $0.621_{\text{ }0.039\uparrow}$ & $0.584_{\text{ }0.018\uparrow}$ & 
        $0.441_{\text{ }0.025\uparrow}$ \\
        \rowcolor{tablewhite}
        \textbf{\texttt{Gemma3-27B}} & $\underline{0.878}_{\text{ }0.132\uparrow}$ & $\boldsymbol{0.797}_{\text{ }0.113\uparrow}$	& $\underline{0.723}_{\text{ }0.071\uparrow}$ & $\underline{0.664}_{\text{ }0.071\uparrow}$ & $0.623_{\text{ }0.069\uparrow}$ & 
        $0.481_{\text{ }0.075\uparrow}$ \\
        \rowcolor{tablewhite}
        \textbf{\texttt{gpt-oss-20B}} & $0.762_{\text{ }0.199\uparrow}$ & $0.729_{\text{ }0.133\uparrow}$ & $0.688_{\text{ }0.079\uparrow}$ & $0.643_{\text{ }0.048\uparrow}$ & $0.613_{\text{ }0.048\uparrow}$ & 
        $0.467_{\text{ }0.053\uparrow}$ \\
        \rowcolor{tablewhite}
        \textbf{\texttt{gpt-oss-120B}} & $0.823_{\text{ }0.024\uparrow}$ & $0.757_{\text{ }0.018\uparrow}$ & $0.707_{\text{ }0.013\uparrow}$ & $0.660_{\text{ }0.008\uparrow}$ & $\boldsymbol{0.632}_{\text{ }0.015\uparrow}$ & 
        $\boldsymbol{0.492}_{\text{ }0.015\uparrow}$ \\
        \midrule
        \rowcolor{tablewhite}
        \textbf{Avg. Change} & $+0.137$ & $+0.091$ & $+0.047$ & $+0.037$ & $+0.034$ & 
        $+0.037$ \\
    \bottomrule
    \end{tabular}
    }
    \caption{Evaluation results of \ourdata~on various LLMs and the \ourmethod~framework. The \textbf{best} and \underline{second-best} results are marked in bold and underlined, respectively. The average changes brought by the \ourmethod~framework are shown in the last row. Performance gains from the \ourmethod~framework are shown in the subscript. All improvements are statistically significant under the two-sample t-test with $p < 0.05$. See Appendix~\ref{app:main_results_ttest} for detailed test statistics.}
    \label{tab:main_results}
    \vspace{-0.5em}
\end{table*}

\paragraph{Overall Workflow}
The \ourmethod~framework processes the ESC trajectory turn by turn, as illustrated in Figure~\ref{fig:prism_framework}. At each turn, the \texttt{hypothesis agent} maintains a pool of $h$ hypotheses, denoted by $\mathcal{H}_{\mathcal{B}}^{t}$, over the help-seeker’s pre-event BDI state $\mathcal{B}$. We first select the most likely pre-event BDI, $\mathcal{B}_t^{*}$, by taking the $\arg\max$ of $\mathbb{P}(\mathcal{B}\mid\mathcal{D}_{<t}, \mathcal{P})$, as estimated by the \texttt{bdi agent}. Conditioned on $\mathcal{B}_t^{*}$, we then generate hypotheses of salient appraisals as\footnote{Our experimental results show that many LLMs can reliably infer the help-seeker’s pre-event BDI from ESC trajectories. We present detailed results in Appendix~\ref{app:bdi_analysis}.}
\begin{align}
\mathcal{H}_{\mathcal{A}}^{t} = \texttt{LLM}(\mathcal{D}_{<t}, \mathcal{P}, \mathcal{B}_t^*).
\end{align}
Once the initial hypotheses are generated, the \texttt{hypothesis agent} performs one of two operations at each subsequent dialogue turn: \textit{Retain} or \textit{Revise}. The \textit{Retain} operation preserves the original hypothesis, whereas \textit{Revise} modifies it by replacing one or more appraisal dimensions.

We compute the unnormalized posterior score for each appraisal hypothesis set $\mathcal{A}_{i} \in \mathcal{H}_{\mathcal{A}}^{t}$ using Equation~\ref{eq:bip_main}, scaled by information gain\footnote{We omit the $\mathbb{P}(\mathcal{B}_t \mid \cdot)$ term because we directly use the point estimate $\mathcal{B}_t^*$.}:
\begin{align}
\begin{split}
    &\mathbb{P}(\mathcal{A}_i, \mathcal{B} | \mathcal{D}_{1:t}, \mathcal{P})
    \propto
    \prod_{\tau=1}^{t}
    IG(\tau)\cdot  \\
    &  \mathbb{P}(\hat{u}_{\tau} | \mathcal{D}_{<\tau}, \mathcal{P}, \mathcal{B}, \mathcal{A}) \cdot \mathbb{P}(\mathcal{A}_{\tau, i} | \mathcal{D}_{1:\tau}, \mathcal{P}, \mathcal{B}_\tau^*),
\end{split}
\end{align}
where $\mathcal{A}_{\tau,i}$ denotes the $i$-th hypothesis at turn $\tau$. The cumulative score can therefore be computed recursively across dialogue turns, as the \texttt{hypothesis agent} does not overwrite previously generated hypotheses and hence maintains alignment between the hypotheses at turns $t-1$ and $t$.

After each turn, we flatten the hypothesis sets and propagate each set-level score to its contents, yielding saliency logits over appraisal dimensions. We then normalize the resulting logits across dimensions using the softmax function.

At the end of the ESC trajectory, we provide the appraisal dimensions and their posterior densities to a \texttt{judge agent}, which then selects the final set of five salient appraisal dimensions.

\section{Experiments}
\label{sec:eval_setup}
\paragraph{Models} We evaluate the \ourdata~dataset on a range of open-sourced and proprietary LLMs, covering both instruction models and reasoning models. Specifically, we use the \texttt{7B} and \texttt{27B} variants of the \texttt{Gemma3} model \cite{gemmateam2025gemma3technicalreport}, the \texttt{4B}, \texttt{32B}, and \texttt{30B-A3B} variants of the \texttt{Qwen3} model \cite{yang2025qwen3technicalreport}, the \texttt{20B} and \texttt{120B} variants of the \texttt{gpt-oss} model \cite{openai2025gptoss120bgptoss20bmodel}, and DeepSeek V3.2-Thinking \cite{deepseekai2025deepseekv32pushingfrontieropen}. For proprietary models, we use the  \texttt{nano} and \texttt{mini} variants of \texttt{GPT-5} \cite{singh2025openaigpt5card}.
For the \prism~\ourmethod~framework, we disable \textit{thinking} for all the reasoning models for computational efficiency.

\paragraph{Metrics} As the salient appraisal reasoning task requires predicting a set of five appraisal dimensions, we evaluate performance using set-based metrics. Specifically, we report \textit{Precision}@$k$ and \textit{Jaccard Index}. Note that we do not report \textit{Recall} and \textit{F1} as the \textit{Recall} is equivalent to \textit{Precision@5} for our task. Further, we do not consider rank-based metrics as our analysis shows that while annotators are able to reach good agreement on the set of salient appraisals, they do not tend to agree on the ranking of the appraisals. We provide detailed metric definition, justification, and rank-based evaluation in Appendix~\ref{app:rank_explained} and \ref{app:eval_metrics}.

\paragraph{Main Results}
\label{sec:eval_results}
As shown in Table~\ref{tab:main_results}, humans perform strongly at identifying salient appraisals from ESC trajectories, particularly among the top three appraisals. In contrast, LLMs consistently underperform humans across all metrics. We observe two main failure modes. First, many LLMs perform reasonably well at identifying the single most salient appraisal dimension, as reflected by \textit{Precision@1}, but their performance declines as the number of target appraisal dimensions increases. Second, some models, such as \texttt{Gemma3-4B} and \texttt{gpt-oss-20B}, fail to place the most salient appraisals at the top of the ranking, resulting in poor \textit{Precision@1}. In these cases, the correct appraisals are often ranked second or third, causing performance to peak at \textit{Precision@3}.
Our error analysis shows that both failure modes are associated with LLMs' inherent bias in cognitive appraisal inference. We present the error analysis results in Appendix~\ref{app:error_analysis}.

Beyond zero-shot prompting, we also evaluate \emph{Self-Consistency} prompting \cite{wangself} by sampling  8 CoT trajectories from each LLM, matching the number of hypotheses used in \ourmethod. We also consider an \emph{Iterative Self-Consistency with Final Judge} variant, where the model performs self-consistency sampling at each turn of the dialogue and then makes the final prediction using the same \texttt{Judge} agent as in \ourmethod. Both variants produce only marginal improvements, and we therefore defer the complete results to Appendix~\ref{app:full_eval_results}.

Experimental results show that \ourmethod~brings significant improvement for inferring the most salient appraisal dimensions, with an improvement of up to $0.296$ for \textit{Precision@1} and up to $0.133$ for \textit{Precision@2}. In addition, \ourmethod~is particularly effective for small LLMs, such as \texttt{Qwen3-4B}, where \ourmethod~boosts their performance across all metrics to exceed that of \texttt{gpt-oss-120B} as well as that of proprietary models such as  \texttt{GPT-5-mini}.

\section{Analysis}

\begin{table} [t]
\resizebox{\columnwidth}{!}{
    \begin{tabular}{l c c c}
    \toprule
    Agent & A.A.$\uparrow$ & S.U.$\uparrow$ & C.R.$\uparrow$ \\
    \midrule
    Ordinary
    & $5.910_{\pm1.524}$ & $6.810_{\pm1.294}$ & $5.200_{\pm1.400}$ \\
        Exhaustive
    & $\boldsymbol{6.760_{\pm1.282}}$ & \underline{$7.120_{\pm1.042}$} & $\boldsymbol{6.420_{\pm1.097}}$ \\
       Saliency-Guided
    & $\boldsymbol{7.020_{\pm0.980}}$ & $\boldsymbol{7.360_{\pm1.063}}$ & $\boldsymbol{6.430_{\pm0.951}}$ \\
        Oracle-Guided
    & $\boldsymbol{7.070_{\pm1.194}}$ & $\boldsymbol{7.440_{\pm1.061}}$ & $\boldsymbol{6.450_{\pm1.014}}$ \\
    \bottomrule
    \end{tabular}
}

\vspace{2em}
\centering
{\small\textbf{(a) Saliency-Guided vs.\ Ordinary}}

\begin{tikzpicture}
    \newlength{\barwidth}
    \setlength{\barwidth}{\columnwidth}
    \def\barheight{0.42}
    \def\roundrad{3pt}

    \def\winA{50}
    \def\tieA{12}
    \def\lossA{38}
    \def\yA{0}

    \node[anchor=west, font=\small\bfseries] at (0,{0.65+\yA}) {Overall Human Preference};

    \begin{scope}
        \clip[rounded corners=\roundrad] (0,\yA) rectangle (\barwidth,{\yA+\barheight});
        \fill[bargreen] (0,\yA) rectangle ({\winA/100*\barwidth},{\yA+\barheight});
        \fill[gray!20] ({\winA/100*\barwidth},\yA) rectangle ({(\winA+\tieA)/100*\barwidth},{\yA+\barheight});
        \fill[barred] ({(\winA+\tieA)/100*\barwidth},\yA) rectangle (\barwidth,{\yA+\barheight});
    \end{scope}

    \draw[rounded corners=\roundrad, line width=0.5pt, black!70]
        (0,\yA) rectangle (\barwidth,{\yA+\barheight});

    \draw[line width=0.6pt, white]
        ({\winA/100*\barwidth},\yA) -- ({\winA/100*\barwidth},{\yA+\barheight});
    \draw[line width=0.6pt, white]
        ({(\winA+\tieA)/100*\barwidth},\yA) -- ({(\winA+\tieA)/100*\barwidth},{\yA+\barheight});

    \node[black, font=\scriptsize\bfseries] at ({\winA/200*\barwidth},{\yA+0.5*\barheight}) {\winA\%};
    \node[black, font=\scriptsize\bfseries] at ({(\winA+\tieA/2)/100*\barwidth},{\yA+0.5*\barheight}) {\tieA\%};
    \node[black, font=\scriptsize\bfseries] at ({(\winA+\tieA+\lossA/2)/100*\barwidth},{\yA+0.5*\barheight}) {\lossA\%};

    \def\winB{38}
    \def\tieB{28}
    \def\lossB{34}
    \def\yB{-1.05}

    \node[anchor=west, font=\small\bfseries] at (0,{0.65+\yB}) {Situation Analysis Preference};

    \begin{scope}
        \clip[rounded corners=\roundrad] (0,\yB) rectangle (\barwidth,{\yB+\barheight});
        \fill[bargreen] (0,\yB) rectangle ({\winB/100*\barwidth},{\yB+\barheight});
        \fill[gray!20] ({\winB/100*\barwidth},\yB) rectangle ({(\winB+\tieB)/100*\barwidth},{\yB+\barheight});
        \fill[barred] ({(\winB+\tieB)/100*\barwidth},\yB) rectangle (\barwidth,{\yB+\barheight});
    \end{scope}

    \draw[rounded corners=\roundrad, line width=0.5pt, black!70]
        (0,\yB) rectangle (\barwidth,{\yB+\barheight});

    \draw[line width=0.6pt, white]
        ({\winB/100*\barwidth},\yB) -- ({\winB/100*\barwidth},{\yB+\barheight});
    \draw[line width=0.6pt, white]
        ({(\winB+\tieB)/100*\barwidth},\yB) -- ({(\winB+\tieB)/100*\barwidth},{\yB+\barheight});

    \node[black, font=\scriptsize\bfseries] at ({\winB/200*\barwidth},{\yB+0.5*\barheight}) {\winB\%};
    \node[black, font=\scriptsize\bfseries] at ({(\winB+\tieB/2)/100*\barwidth},{\yB+0.5*\barheight}) {\tieB\%};
    \node[black, font=\scriptsize\bfseries] at ({(\winB+\tieB+\lossB/2)/100*\barwidth},{\yB+0.5*\barheight}) {\lossB\%};

    \def\winB{52}
    \def\tieB{22}
    \def\lossB{26}
    \def\yB{-2.10}

    \node[anchor=west, font=\small\bfseries] at (0,{0.65+\yB}) {Mental Analysis Preference};

    \begin{scope}
        \clip[rounded corners=\roundrad] (0,\yB) rectangle (\barwidth,{\yB+\barheight});
        \fill[bargreen] (0,\yB) rectangle ({\winB/100*\barwidth},{\yB+\barheight});
        \fill[gray!20] ({\winB/100*\barwidth},\yB) rectangle ({(\winB+\tieB)/100*\barwidth},{\yB+\barheight});
        \fill[barred] ({(\winB+\tieB)/100*\barwidth},\yB) rectangle (\barwidth,{\yB+\barheight});
    \end{scope}

    \draw[rounded corners=\roundrad, line width=0.5pt, black!70]
        (0,\yB) rectangle (\barwidth,{\yB+\barheight});

    \draw[line width=0.6pt, white]
        ({\winB/100*\barwidth},\yB) -- ({\winB/100*\barwidth},{\yB+\barheight});
    \draw[line width=0.6pt, white]
        ({(\winB+\tieB)/100*\barwidth},\yB) -- ({(\winB+\tieB)/100*\barwidth},{\yB+\barheight});

    \node[black, font=\scriptsize\bfseries] at ({\winB/200*\barwidth},{\yB+0.5*\barheight}) {\winB\%};
    \node[black, font=\scriptsize\bfseries] at ({(\winB+\tieB/2)/100*\barwidth},{\yB+0.5*\barheight}) {\tieB\%};
    \node[black, font=\scriptsize\bfseries] at ({(\winB+\tieB+\lossB/2)/100*\barwidth},{\yB+0.5*\barheight}) {\lossB\%};

    \node[anchor=west, font=\small] at (0,-2.45)
    {\textcolor{bargreen}{\rule{0.9em}{0.9em}}~\textbf{Win} \hspace{0.8em}
     \textcolor{gray!20}{\rule{0.9em}{0.9em}}~\textbf{Tie} \hspace{0.8em}
     \textcolor{barred}{\rule{0.9em}{0.9em}}~\textbf{Loss}};
\end{tikzpicture}

\vspace{0.8em}
{\small\textbf{(b) Saliency-Guided vs. \ Exhaustive}}

\begin{tikzpicture}
    \setlength{\barwidth}{\columnwidth}
    \def\barheight{0.42}
    \def\roundrad{3pt}

    \def\winA{50}
    \def\tieA{16}
    \def\lossA{34}
    \def\yA{0}

    \node[anchor=west, font=\small\bfseries] at (0,{0.65+\yA}) {Overall Human Preference};

    \begin{scope}
        \clip[rounded corners=\roundrad] (0,\yA) rectangle (\barwidth,{\yA+\barheight});
        \fill[bargreen] (0,\yA) rectangle ({\winA/100*\barwidth},{\yA+\barheight});
        \fill[gray!20] ({\winA/100*\barwidth},\yA) rectangle ({(\winA+\tieA)/100*\barwidth},{\yA+\barheight});
        \fill[barred] ({(\winA+\tieA)/100*\barwidth},\yA) rectangle (\barwidth,{\yA+\barheight});
    \end{scope}

    \draw[rounded corners=\roundrad, line width=0.5pt, black!70]
        (0,\yA) rectangle (\barwidth,{\yA+\barheight});

    \draw[line width=0.6pt, white]
        ({\winA/100*\barwidth},\yA) -- ({\winA/100*\barwidth},{\yA+\barheight});
    \draw[line width=0.6pt, white]
        ({(\winA+\tieA)/100*\barwidth},\yA) -- ({(\winA+\tieA)/100*\barwidth},{\yA+\barheight});

    \node[black, font=\scriptsize\bfseries] at ({\winA/200*\barwidth},{\yA+0.5*\barheight}) {\winA\%};
    \node[black, font=\scriptsize\bfseries] at ({(\winA+\tieA/2)/100*\barwidth},{\yA+0.5*\barheight}) {\tieA\%};
    \node[black, font=\scriptsize\bfseries] at ({(\winA+\tieA+\lossA/2)/100*\barwidth},{\yA+0.5*\barheight}) {\lossA\%};

    \def\winB{44}
    \def\tieB{16}
    \def\lossB{40}
    \def\yB{-1.05}

    \node[anchor=west, font=\small\bfseries] at (0,{0.65+\yB}) {Situation Analysis Preference};

    \begin{scope}
        \clip[rounded corners=\roundrad] (0,\yB) rectangle (\barwidth,{\yB+\barheight});
        \fill[bargreen] (0,\yB) rectangle ({\winB/100*\barwidth},{\yB+\barheight});
        \fill[gray!20] ({\winB/100*\barwidth},\yB) rectangle ({(\winB+\tieB)/100*\barwidth},{\yB+\barheight});
        \fill[barred] ({(\winB+\tieB)/100*\barwidth},\yB) rectangle (\barwidth,{\yB+\barheight});
    \end{scope}

    \draw[rounded corners=\roundrad, line width=0.5pt, black!70]
        (0,\yB) rectangle (\barwidth,{\yB+\barheight});

    \draw[line width=0.6pt, white]
        ({\winB/100*\barwidth},\yB) -- ({\winB/100*\barwidth},{\yB+\barheight});
    \draw[line width=0.6pt, white]
        ({(\winB+\tieB)/100*\barwidth},\yB) -- ({(\winB+\tieB)/100*\barwidth},{\yB+\barheight});

    \node[black, font=\scriptsize\bfseries] at ({\winB/200*\barwidth},{\yB+0.5*\barheight}) {\winB\%};
    \node[black, font=\scriptsize\bfseries] at ({(\winB+\tieB/2)/100*\barwidth},{\yB+0.5*\barheight}) {\tieB\%};
    \node[black, font=\scriptsize\bfseries] at ({(\winB+\tieB+\lossB/2)/100*\barwidth},{\yB+0.5*\barheight}) {\lossB\%};

    \def\winB{54}
    \def\tieB{8}
    \def\lossB{38}
    \def\yB{-2.10}

    \node[anchor=west, font=\small\bfseries] at (0,{0.65+\yB}) {Mental Analysis Preference};

    \begin{scope}
        \clip[rounded corners=\roundrad] (0,\yB) rectangle (\barwidth,{\yB+\barheight});
        \fill[bargreen] (0,\yB) rectangle ({\winB/100*\barwidth},{\yB+\barheight});
        \fill[gray!20] ({\winB/100*\barwidth},\yB) rectangle ({(\winB+\tieB)/100*\barwidth},{\yB+\barheight});
        \fill[barred] ({(\winB+\tieB)/100*\barwidth},\yB) rectangle (\barwidth,{\yB+\barheight});
    \end{scope}

    \draw[rounded corners=\roundrad, line width=0.5pt, black!70]
        (0,\yB) rectangle (\barwidth,{\yB+\barheight});

    \draw[line width=0.6pt, white]
        ({\winB/100*\barwidth},\yB) -- ({\winB/100*\barwidth},{\yB+\barheight});
    \draw[line width=0.6pt, white]
        ({(\winB+\tieB)/100*\barwidth},\yB) -- ({(\winB+\tieB)/100*\barwidth},{\yB+\barheight});

    \node[black, font=\scriptsize\bfseries] at ({\winB/200*\barwidth},{\yB+0.5*\barheight}) {\winB\%};
    \node[black, font=\scriptsize\bfseries] at ({(\winB+\tieB/2)/100*\barwidth},{\yB+0.5*\barheight}) {\tieB\%};
    \node[black, font=\scriptsize\bfseries] at ({(\winB+\tieB+\lossB/2)/100*\barwidth},{\yB+0.5*\barheight}) {\lossB\%};

    \node[anchor=west, font=\scriptsize] at (0,-2.45)
    {\textcolor{bargreen}{\rule{0.9em}{0.9em}}~Win \hspace{0.8em}
     \textcolor{gray!20}{\rule{0.9em}{0.9em}}~Tie \hspace{0.8em}
     \textcolor{barred}{\rule{0.9em}{0.9em}}~Loss};
\end{tikzpicture}

\caption{Preference results for the case study. Top: LLM evaluation scores. (a): win/tie/loss rates of \textit{Saliency-Guided} versus \textit{Ordinary} emotional support. (b): win/tie/loss rates of \textit{Exhaustive} versus \textit{Saliency-Guided} reframing responses. \textbf{Bold} and \underline{underline} indicate statistical significance at $\alpha=0.05$ and $\alpha=0.10$ respectively in two-sample T test.}
\label{tab:case_study}
\end{table}

\begin{table*}[th]
    \centering
    \resizebox{\textwidth}{!}{
        \begin{tabular}{llllllcl}
        \toprule
        model & Precision@1 & Precision@2 & Precision@3 & Precision@4 & Precision@5 & Avg $\downarrow$ in \textit{P@k} 
        & Jaccard Index \\
        \midrule
        \rowcolor{tablewhite}
        \textbf{\texttt{gpt-oss-120b}} & $0.823$ & $0.757$ & $0.707$ & $0.660$ & $0.632$ & 
        -- -- & $0.492$ \\
        w/o $\mathcal{B}^{*}$ & $0.788_{-0.035}$ & $0.736_{-0.021}$ & $0.673_{-0.034}$ & $0.628_{-0.032}$ & $0.597_{-0.035}$ & 
        $0.031\downarrow$& $0.455_{-0.037}$ \\
        w/o $IG(\cdot)$ & $0.743_{-0.080}$ & $0.739_{-0.019}$ & $0.694_{-0.013}$ & $0.651_{-0.009}$ & $0.607_{-0.025}$ & 
        $0.029 \downarrow$ & $0.463_{-0.029}$  \\
        w/o $\mathbb{P}(\hat{u}_t\mid\cdot)$ & $0.725_{-0.098}$ & $0.687_{-0.070}$ & $0.650_{-0.057}$ & $0.623_{-0.037}$ & $0.608_{-0.024}$ & 
        $0.057\downarrow$ & $0.465_{-0.027}$  \\
        w/o Cumulative Probs & $0.757_{-0.066}$ & $0.740_{-0.017}$ & $0.694_{-0.013}$ & $0.649_{-0.016}$ & $0.607_{-0.024}$ & 
        $0.027\downarrow$ & $0.463_{-0.029}$  \\
        w/o \texttt{Judge Agent} & $0.720_{-0.103}$ & $0.708_{-0.050}$ & $0.688_{-0.018}$ & $0.637_{-0.024}$ & $0.593_{-0.038}$ & 
        $0.047\downarrow$ & $0.457_{-0.035}$  \\
        \bottomrule
    \end{tabular}
    }
    \caption{Ablation study results on \texttt{gpt-oss-120B}.}
    \label{tab:ablation}
\end{table*}

\subsection{Salience-Guided Reframing}
\label{sec:case_study}
While the preceding results establish both the difficulty of salient appraisal inference and the effectiveness of \ourmethod, an important downstream question remains: \textit{Does incorporating salient cognitive appraisals improve emotional support?}
Prior works have shown that guiding LLMs with cognitive appraisals makes them more effective at negative thought reframing \cite{zhanlarge}; therefore, we conduct a proof-of-concept study to examine the role of \emph{salient appraisals} in emotional support.

Following \citet{zhanlarge}, we consider single-turn negative-thought reframing. We sample 100 situations from \ourdata~and generate four sets of reframing responses. \textit{Ordinary} (no appraisal guidance), \textit{Exhaustive} (all 14 appraisal dimensions), \textit{Salience-Guided} (salient appraisals predicted by \texttt{Qwen3-4B}), and \textit{Oracle-Guided} (human-annotated salient appraisals). To mitigate length bias, we restrict the model response length to 150 words.

We evaluate responses with human annotators and LLM-as-a-Judge. Human annotators compared paired responses based on three criteria inspired by the Cognitive Triangle \cite{beck2024cognitive}, namely \textit{Appraisal Alignment} (A.A), \textit{Situation Understanding} (S.U.), and \textit{Cognitive Reframing} (C.R.), while \texttt{GPT-5-mini} assigns Likert-scale scores for the same criteria. We provide the agreement of the human annotations in Appendix~\ref{app:iaa_case_study} and details of LLM-as-a-Judge evaluation in Appendix~\ref{app:llm_judge_validation}.

As shown in Table~\ref{tab:case_study}, appraisal guidance significantly improves performance over unguided responses (i.e. \emph{ordinary}) across all criteria. 

Table~\ref{tab:case_study}(a) presents result of \textit{Saliency-Guided} v.s. \textit{Ordinary}. We see that human annotators show a clear preference for the \textit{salience-guided} responses. While the inclusion of cognitive appraisals does not bring significant improvement in LLMs' \emph{Situation Analysis}, they bring meaningful improvements in both \emph{Overall Human Preference} as well as \emph{Mental Analysis Preference}, which is the most vital component for emotional support. 

Crucially, Table~\ref{tab:case_study}(b) shows that \textit{Saliency-Guided} responses consistently outperform the \textit{Exhaustive} variant, demonstrating that \emph{selecting salient appraisals is more effective than using all appraisals}. Moreover, predicted saliency achieves performance close to \textit{Oracle-Guided} responses, indicating that automatic appraisal selection can reliably approximate human judgment. These results confirm that incorporating \emph{salient} appraisals is key to improving emotional support.

\subsection{Ablation Study}

\begin{figure}
    \centering
    \includegraphics[width=\columnwidth]{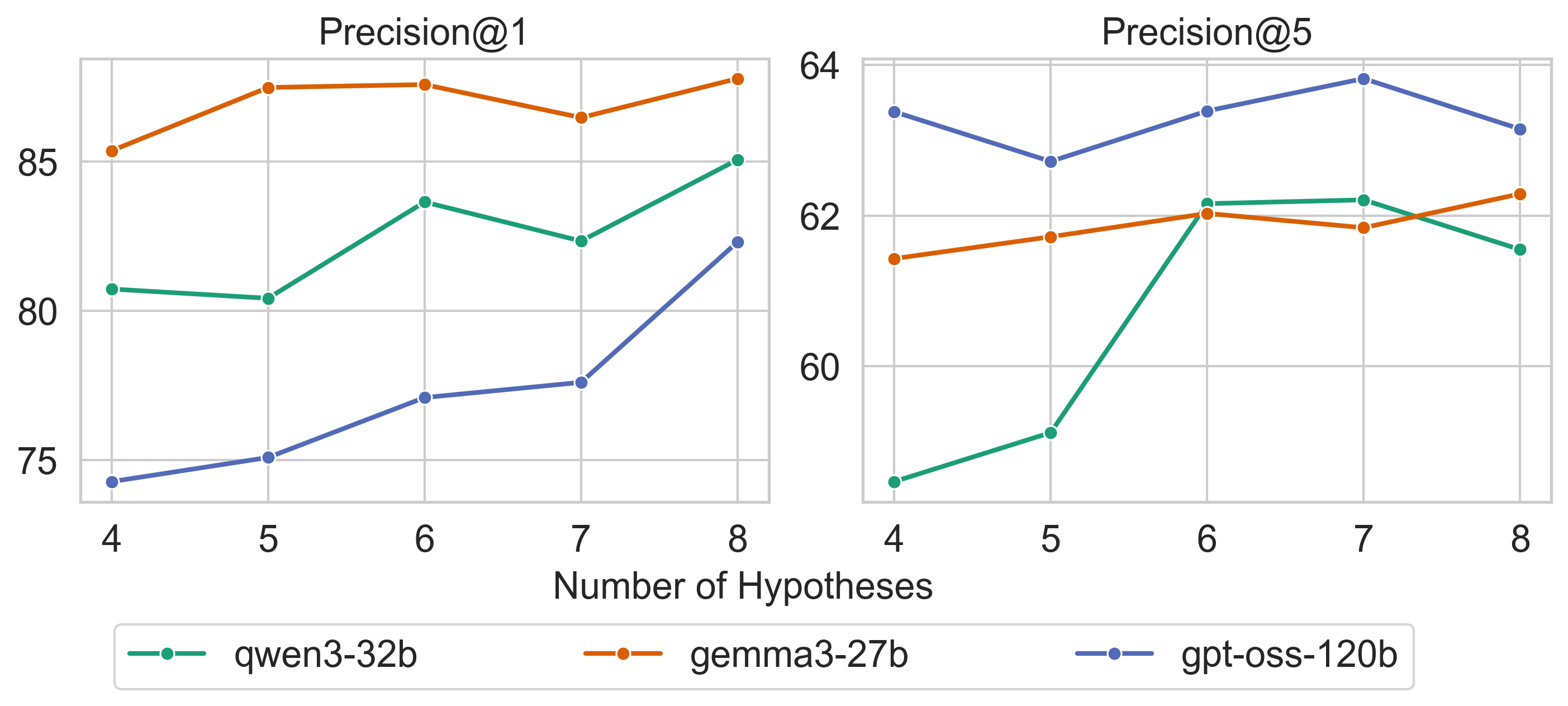}
    \caption{Change in model performance corresponding to the size of the hypothesis pool.}
    \label{fig:num_hypothesis}
\end{figure}

\paragraph{Size of Hypothesis Pool}
A key hyperparameter in \ourmethod~is the size of the hypothesis pool. In our experiments, we set $h=8$. To examine the effect of smaller pools, we vary $h$ from $4$ to $8$. As shown in Figure~\ref{fig:num_hypothesis}, a larger hypothesis pool generally makes LLMs more reliable at identifying the most salient appraisals, as indicated by an increasing \textit{Precision@1} with respect to $h$ across all models. In the case of identifying 5 salient appraisals, \texttt{Gemma3-27B} and \texttt{gpt-oss-120B} give adequate performance even with small $h$, \texttt{Qwen3-32B} benefits more consistently from larger pools, particularly when $h \geq 6$.

\paragraph{\ourmethod~Agents}
To understand the contribution of each component in \ourmethod, we conduct an ablation study using three backbone LLMs.
Taking \texttt{gpt-oss-120B} as an example, results from Table~\ref{tab:ablation} show that each component contributes to the overall performance, with the largest drops generally arising from removing the utterance likelihood term $\mathbb{P}(\hat{u}_t\mid\cdot)$. The importance of the utterance likelihood demonstrates the effectiveness of the BIP factorization (Equation~\ref{eq:bip_main}).
Taken together, these results show that the gains of \ourmethod~do not come from a single component alone; rather, they arise from the interaction between prior-BDI anchoring, informativeness weighting, cumulative trajectory-level inference, and final judgment over appraisal saliency. See Appendix~\ref{app:ablation_results} for detailed results.

\section{Related Work}
\paragraph{Cognitive Reframing with LLMs}
A growing body of work has explored the use of LLMs for emotional support. One line of research approaches negative-thought reframing by prompting \cite{chen2023controllable, zhao2023chatgpt, zhang2024escot, zhang2025intentionesc} or fine-tuning LLMs on curated ESC trajectories, either through supervised fine-tuning \cite{liu2021towards, zheng2023augesc, zheng2023building, qiu2024smile, xiao2024healme, chen2025socialsim, zhou2025crisp, ye2025sweetiechat} or reinforcement learning \cite{yang2025towards, zhang2025echo, yuan2025kardia, yao2025empathy, dai2025psyche, zhang-etal-2025-decoupledesc, ye2025generic, hong2025third}. While these methods can improve the generation of empathetic responses, they do not explicitly model the underlying causes of the help-seeker’s distress. A second line of work is motivated by cognitive reappraisal and improves reframing by incorporating cognitive appraisal theory into the generation workflow \cite{zhan2023evaluating, zhanlarge, zhou2025modeling, sun2026carebench}.

\paragraph{ToM Reasoning with LLMs}
Evaluating and improving LLMs’ ToM reasoning abilities has recently received substantial attention from the research community. Researchers have developed a range of benchmarks based on the False Belief task \cite{le2019revisiting, wu2023hi, zhou2023far, gandhi2023understanding, jin2024mmtom, xu2024opentom, gu2024simpletom}, as well as benchmarks that go beyond false-belief reasoning \cite{zhou2023cast, chen2024tombench, yeo2025beyond}. Although some benchmarks evaluate LLMs’ ability to infer mental states from dialogue \cite{kim2023fantom, xiao2025towards}, they primarily focus on casual conversations rather than ESCs.

To improve LLMs’ ToM reasoning abilities, prior work has proposed prompting \cite{wilf2024think} and neuro-symbolic approaches \cite{sclar2023minding, hou2024timetom, jung2024perceptions, xu2025enigmatom, zhou2026socialworldmodels}. Another line of research adopts probabilistic inference, using methods such as Sequential Monte Carlo \cite{kimhypothesis}, or BIP \cite{zhi2024pragmatic, jin2024mmtom, ying2024grounding, zhangautotom}. However, existing BIP-based approaches typically assume a narrow search space over latent mental states and treat all actions as equally informative. These assumptions make them less suitable for salient appraisal reasoning, where the search space is large, and the informativeness of dialogue turns can vary substantially.

\section{Conclusion}
In this work, we investigate a previously overlooked problem in ESC: inferring the salient cognitive appraisal dimensions that drive negative emotions. To this end, we introduce \ourdata, the first benchmark for salient appraisal reasoning in ESC, and propose \ourmethod, a multi-agent probabilistic framework grounded in Bayesian Inverse Planning. Experiments show that current LLMs remain substantially below human performance on this task, while \ourmethod~consistently improves appraisal inference, particularly for the most salient dimensions. Our case study further suggests that the incorporation of appraisals can lead to more effective emotional support. We hope this work will open new research avenues on appraisal saliency reasoning and its application in emotional support.

\section*{Limitations}

\paragraph{ESC Trajectories}
Although \ourdata~covers a diverse set of upsetting events, many of the scenarios are still relatively simple and self-contained. In real-world emotional support conversations, distress often arises from a prolonged sequence of events, multiple interacting stressors, or unresolved interpersonal dynamics. Inferring salient appraisals in such settings is likely to be substantially more challenging. As a result, performance on \ourdata~may overestimate model capability in more complex real-world ESC settings.

\paragraph{The \ourmethod~Framework}
As discussed in \S\ref{sec:method}, \ourmethod~relied on token log-probability to model the likelihood factors in the backward BIP inference process. Specifically, since we concatenate the expected response to the prompt, our method only works with models that support \texttt{prompt logprobs} (also known as \texttt{echo logprobs}). While this functionality is supported by most open-sourced models, its support is deprecated by OpenAI models.

\paragraph{Synthetic data construction}
The construction of \ourdata~involves a hybrid pipeline that combines source events and personas with LLM-based generation, revision, and human annotation. While we remove pseudo appraisal labels before annotation and apply quality control procedures, the generated conversations may still reflect artifacts introduced by the prompting process or the underlying LLMs. Consequently, the dataset may not fully capture the linguistic diversity, messiness, and ambiguity of naturally occurring emotional support conversations.

\paragraph{Annotation constraints}
Our annotation protocol requires annotators to select exactly five salient appraisal dimensions and provide rationales for their choices. This design improves consistency and discourages careless annotation. In our preliminary annotations without this constraint, many annotators tended to select only a single highly salient appraisal dimension, leaving other potentially relevant appraisals unannotated. Nevertheless, this design imposes a discrete structure on what may in reality be a more graded and uncertain appraisal process. In some cases, fewer than five appraisals may be truly salient, while in others, the boundary between salient and non-salient appraisals may be ambiguous. Future work could explore alternative annotation schemes that better capture this uncertainty, for example, by collecting graded salience ratings rather than discrete selections.

\section*{Ethics Statement}
This work contains emotional support and cognitively grounded reasoning about distress, and should therefore be treated with appropriate caution. \ourdata~is intended as a research benchmark and \ourmethod~as a proof-of-concept study, not as a clinical tool or a replacement for professional mental health practitioners. While our study demonstrates the potential of constructing ESC agents capable of utilizing cognitive appraisals to deliver better emotional support, such a system is not ready to be used with human participants, as we do not fully understand the safety of this system. More research needs to be conducted to fully examine the capability and, most importantly, the safety and reliability of agents to be deployed in high-stakes scenarios such as our \ourmethod~framework.

To reduce privacy risk, \ourdata~is constructed from source events, personas, LLM-based generation, and human annotation, rather than real transcripts from therapies. Still, the dataset and \ourmethod~may inherit biases from the source data, annotators, and underlying LLMs. In addition, incorrect appraisal inference could lead to misleading or harmful support if deployed without oversight. We therefore caution against interpreting performance on \ourdata~as evidence that current LLMs are suitable for autonomous mental health support, and recommend strong human supervision and safety safeguards for any downstream use.

\section*{Acknowledgments}
We thank Xinyi Mou for the valuable discussions and contributions to the pilot studies. This work was supported in part by the UK Engineering and Physical Sciences Research Council (EPSRC) through an iCASE award with Huawei London Research Centre, a Turing AI Fellowship (grant no. EP/V020579/1, EP/V020579/2), and the Prosperity Partnership scheme (grant no. UKRI566), and by Inkfish through the EMBRACE research programme.

\bibliography{custom}

\newpage
\clearpage

\appendix
\setcounter{table}{0}
\renewcommand{\thetable}{A\arabic{table}}
\setcounter{figure}{0}
\renewcommand{\thefigure}{A\arabic{figure}}

\section{Cognitive Appraisal Dimensions}
\label{app:cogapps}
We obtain our 14 appraisals by \texttt{Preserve}/\texttt{Remove}/\texttt{Aggregate} the appraisal dimensions defined by \cite{hofmann2020appraisal} as shown in Table~\ref{tab:cogapps_modifications}

\begin{table}[ht]
    \centering
    \resizebox{\columnwidth}{!}{
      \begin{tabular}{c l l}
      \toprule
        Index & Appraisal Dimension & Modification\\
        \midrule
        1 & suddenness & \texttt{Preserved} \\
        2 & familiarity & \texttt{Preserved} \\
        3 & predict event & \texttt{Preserved} \\
        4 & pleasantness & \texttt{Removed} \\
        5 & unpleasantness & \texttt{Removed} \\
        6 & goal congruence & \texttt{Preserved} \\
        7 & self responsibility & \texttt{Preserved} \\
        8 & chance responsibility & \texttt{Aggregated with 9} \\
        9 & other responsibility & \texttt{Preserved} \\
        10 & predict consequence & \texttt{Preserved} \\
        11 & goal support & \texttt{Aggregated with 6} \\
        12 & urgency	& \texttt{Aggregated with 19} \\
        13 & self control & \texttt{Preserved} \\
        14 & other control & \texttt{Preserved}  \\
        15 & chance control & \texttt{Aggregated with 14} \\
        16 & accept consequence & \texttt{Preserved} \\
        17 & norm violation & \texttt{Preserved} \\
        18 & standards & \texttt{Aggregated with 17} \\
        19 & coping demand & \texttt{Preserved} \\
        20 & attention & \texttt{Aggregated with 19} \\
        21 & not consider & \texttt{Aggregated with 19} \\
    \bottomrule
    \end{tabular}
    }
    \caption{Modification of appraisal dimensions defined by \citet{hofmann2020appraisal}.}
    \label{tab:cogapps_modifications}%
\end{table}

We also dichotomized dimensions such as \textit{predict event} into \textit{predictability of event} and \textit{unpredictability of event} to minimize ambiguity. The revised 14 dimensions and their explanations used in this study are shown in Table~\ref{tab:cogapps}.

\begin{table*}[h]
    \centering
    \resizebox{\textwidth}{!}{
    \begin{tabular}{l l}
    \toprule
    \textbf{suddenness} & The patient finds the situation sudden or abrupt. \\
    \textbf{unfamiliarity} & The patient finds the situation unfamiliar. \\
    \textbf{predictability of event} & The patient was expecting the situation to happen. \\
    \textbf{unpredictability of event} & The patient wasn't expecting the occurrence of the situation. \\
    \textbf{goal incongruence} & The outcome is not aligned with the patient's goals/desires. \\
    \textbf{predictability of consequences} & The patient can easily anticipate the consequences of the situation. \\
    \textbf{unpredictability of consequences} & The patient had no idea about what the consequences of the situation would be. \\
    \textbf{self cause} & The patient blames themselves for causing the situation. \\
    \textbf{other cause} & The patient blames someone else for causing the situation. \\
    \textbf{self control} & The patient has control over what is going on in the situation. \\
    \textbf{other control} & Someone/Something other than the patient has control over what is going on in the situation. \\
    \textbf{norm violation} & The actions that produced the situation violated the patient's standards and ideals. \\
    \textbf{coping demand} & The patient has to spend a great deal of (physical or mental) energy to handle the outcome of the situation. \\
    \textbf{unacceptable consequences} & The consequences of the situation are unbearable for the patient. \\
    \bottomrule
    \end{tabular}
    }
    \caption{The 14 appraisal dimensions used in this study.}
    \label{tab:cogapps}
\end{table*}

\section{Persona Construction}
\label{app:persona_construction}
We first match each event with the top-5 most similar persona descriptions. Events and personas are matched by computing the pairwise cosine similarities between embeddings given by the \texttt{jina-embedding-v3} model. The most suitable persona is picked by prompting \texttt{GPT-5-mini}. Typically, a persona profile from the PersonaHub contains information regarding the character's occupation and occasionally traces of the character's gender (e.g., a retired actress). We first extract \emph{age}, \emph{occupation}, \emph{gender}, and \emph{education level} from the persona. Anchoring on the extracted information, we match the occupation with one from the occupation data acquired from the U.S. Bureau of Labor Statistics\footnote{\url{https://www.bls.gov/bls/occupation.htm}}. For entries missing \emph{education level}, we sample an education level according to the education level distribution corresponding to the assigned occupation, given by the U.S. Bureau of Labor Statistics. We consider four education levels, namely "high school", "Bachelor's", "Master's", and "Ph.D.". We then prompt \texttt{GPT-5-mini} again to revise the persona to better match the occupation, education level, and most importantly, the corresponding upsetting event. Finally, we assign each character a personality according to the Big Five personality traits, which comprise the following aspects:
\begin{itemize}[leftmargin=6mm, noitemsep]
    \item Neuroticism: ['emotionally stable', 'emotionally unstable']
    \item Conscientiousness: ['organized', 'careless']
    \item Extroversion: ['extroverted', 'introverted']
    \item Openness: ['conservative', 'open-minded']
    \item Agreeableness: ['agreeable', 'antagonistic']
\end{itemize}

The resulting persona profile is highly associated with the upsetting event and depicts the following aspects of the individual:
\begin{itemize}[leftmargin=6mm, noitemsep]
    \item \textbf{Name}: Sampled from the frequently used names given by the U.S. Social Security Administration\footnote{\url{https://www.ssa.gov/oact/babynames/}}
    \item \textbf{Occupation}: Sampled using data from the U.S. Bureau of Labor Statistics
    \item \textbf{Education}: Sampled using data from the U.S. Bureau of Labor Statistics
    \item \textbf{Gender}: Extracted from the original persona entry or randomly sampled.
    \item \textbf{Personality Traits}: Randomly sampled according to the Big Five personality traits.
\end{itemize}

\section{Comparison of Orginal and Rewritten Situations}
\label{app:rewritten_situations}
We source upsetting events from AugESC \cite{zheng2023augesc}. While this established dataset provides a strong foundation for emotional support conversations, its original events are often concise and generic due to a lack of persona information. To address this, we provide LLMs with the generic event and the paired persona for revision. The goal is to make the events more explicit, engaging, and persona-aligned. We argue that these revised events are more realistic, as they offer richer situational contexts while remaining strictly grounded in the original AugESC scenarios. Table~\ref{tab:augesc_comparison} illustrates the contrast between the original AugESC entries and our revised entries in \ourdata.

\begin{table*}[t]
\centering
\small
\begin{tabular}{p{0.35\linewidth} p{0.65\linewidth}}
\toprule
\textbf{AugESC (Original)} & \textbf{AppraiSal (Revised)} \\
\midrule
Man I feel so bad about last weekend :/ & I skipped the championship cycling race to sit through a last-minute quarterly review, leaving my team stranded and the charity fundraiser without its star rider. Because I prioritized the meeting, the race was postponed and we lost \$20,000 in sponsorships. \\
\addlinespace
My wife is angry and I don't know what to say. My anniversary was yesterday and I had no clue! & I stayed up until 3 a.m. polishing an investor pitch, completely forgetting that my wife had booked a private rooftop dinner for our 15-year anniversary, and I arrived home after the restaurant had closed. She confronted me, eyes blazing with disappointment, and the silence in the empty dining room felt like a slap. \\
\addlinespace
I'm very worried that I'm not going to make an important goal. & I watched the flames devour my finished mural only hours before the city council was scheduled to vote on the grant that would secure my first museum commission. \\
\bottomrule
\end{tabular}
\caption{Original upsetting events from AugESC and the revised, persona-aligned events in \ourdata.}
\label{tab:augesc_comparison}
\end{table*}

\section{The reAppraisal Framework}
\label{app:reAppraisal}
We use the extended reAppriasal Framework from \citet{uusberg2023reappraising} to guide the therapist agent during cognitive reframing. The reAppraisal framework can be summarized into the following eight reframing strategies (note that we have rephrased the name of each dimension; the original name is shown in parentheses)
\begin{itemize}[leftmargin=6mm, noitemsep]
    \item \textbf{Undoing Construal (Object-level Construal Decommitment):} Guiding the user to move away from their original perspective on the event.
    \item \textbf{Adopting Construal (Object-level Construal Commitment):} Offering a different perspective on the event.
    \item \textbf{Undoing Feeling (Meta-level Construal Decommitment):} Guiding the user to disregard their original emotional response.
    \item \textbf{Adopting Feeling (Meta-level Construal Commitment):} Validating that a different emotion is possible.
    \item \textbf{Undoing Goal (Object-level Goal Decommitment):} Guiding the user to move away from their original intention.
    \item \textbf{Adopting Goal (Object-level Goal Commitment):} Shifting focus to an alternative intent.
    \item \textbf{Undoing Mental Goal (Meta-level Goal Decommitment):} Guiding the user to disregard their original desire.
    \item \textbf{Adopting Mental Goal (Meta-level Goal Commitment):} Offering a different desire that is more aligned with the event.
\end{itemize}

\section{Human Annotation}
\label{app:esc_tom_annotation_interface}

\subsection{Demographic Information}
To ensure high-quality evaluations, we established strict inclusion criteria for our human annotators. Participants were required to list English as their first language, given that the Emotional Support Conversation (ESC) transcripts were written exclusively in English. Furthermore, we restricted the eligible country of residence to the United States, Canada, the United Kingdom, Ireland, Australia, and New Zealand. To guarantee data reliability and worker experience, eligible annotators must have maintained a historical approval rate of at least 98\% and successfully completed a minimum of 10 prior tasks on the platform. 

We acknowledge that restricting our annotator pool to these specific geographic and linguistic demographics inevitably biases the annotations toward a predominantly Western, middle-class cultural perspective. The implications of these potential biases are further discussed in the Limitations section of the main text.

\subsection{Annotation Interface}
\begin{figure} [h]
    \centering
    \includegraphics[width=\columnwidth]{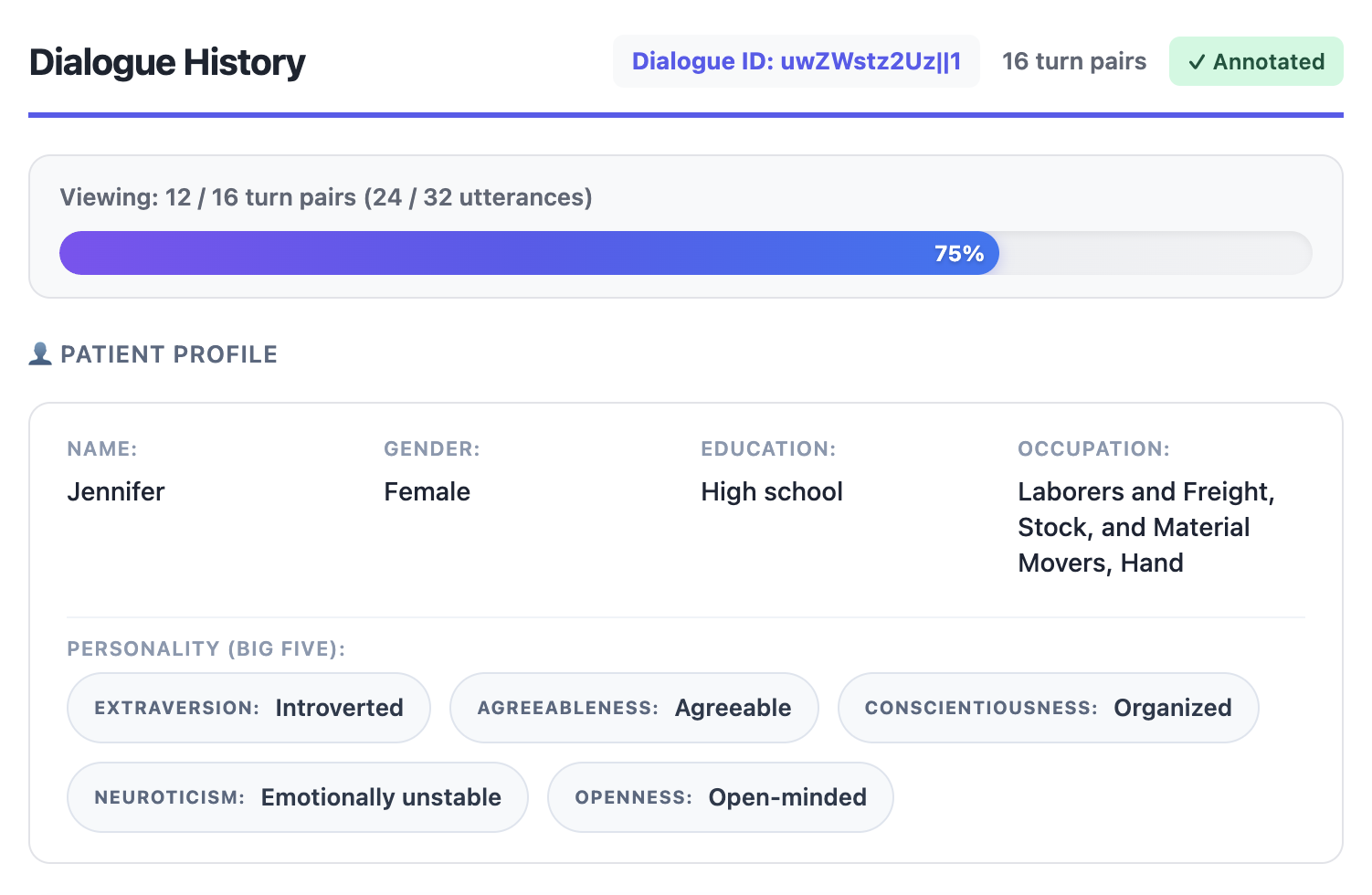}
    \caption{Annotation Interface: Help-seeker's Persona Information}
    \label{app:fig:persona_interface}
\end{figure}

\begin{figure} [h]
    \centering
    \includegraphics[width=\columnwidth]{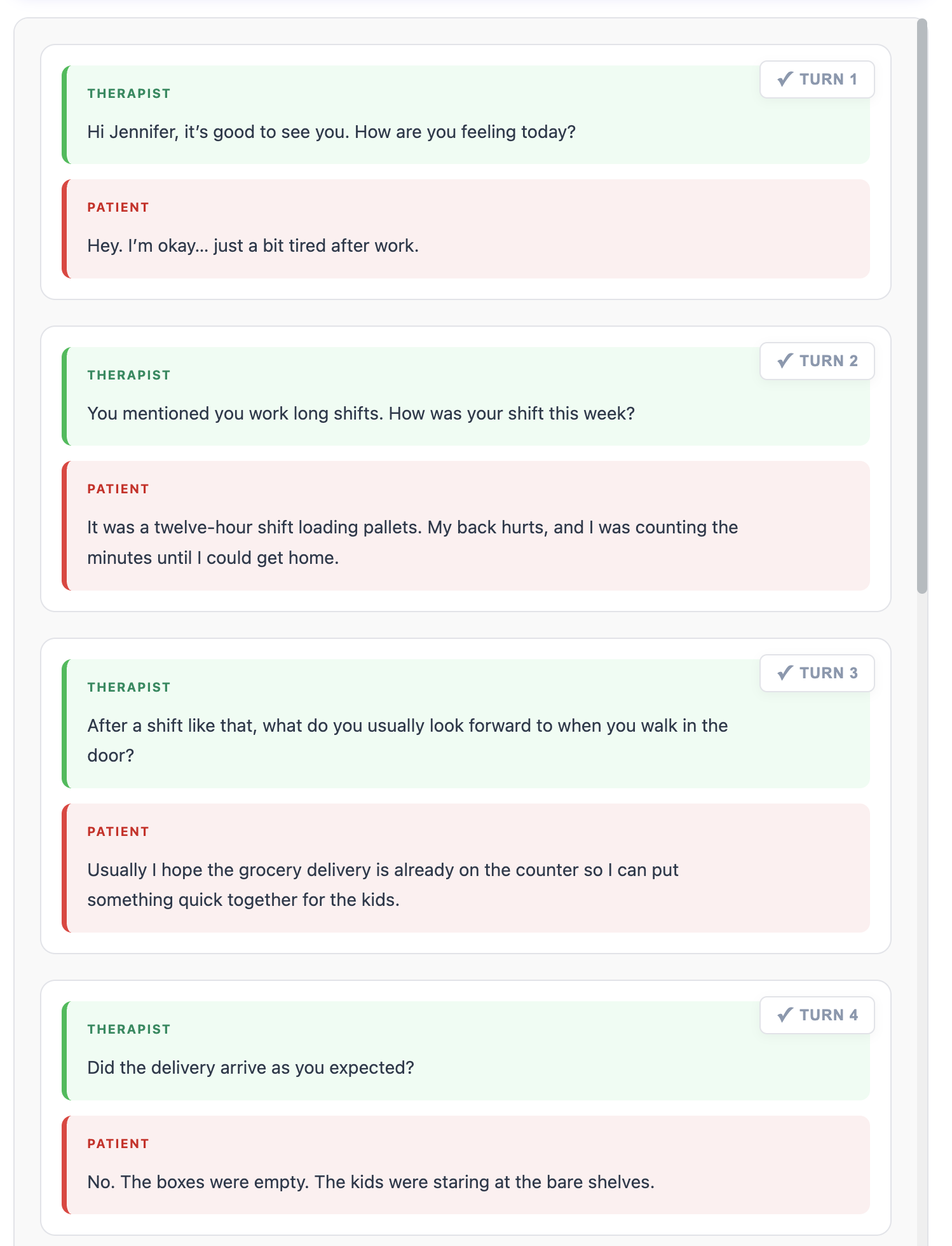}
    \caption{Annotation Interface: ESC Trajectory}
    \label{app:fig:dialogue}
\end{figure}

\begin{figure} [h]
    \centering
    \includegraphics[width=\columnwidth]{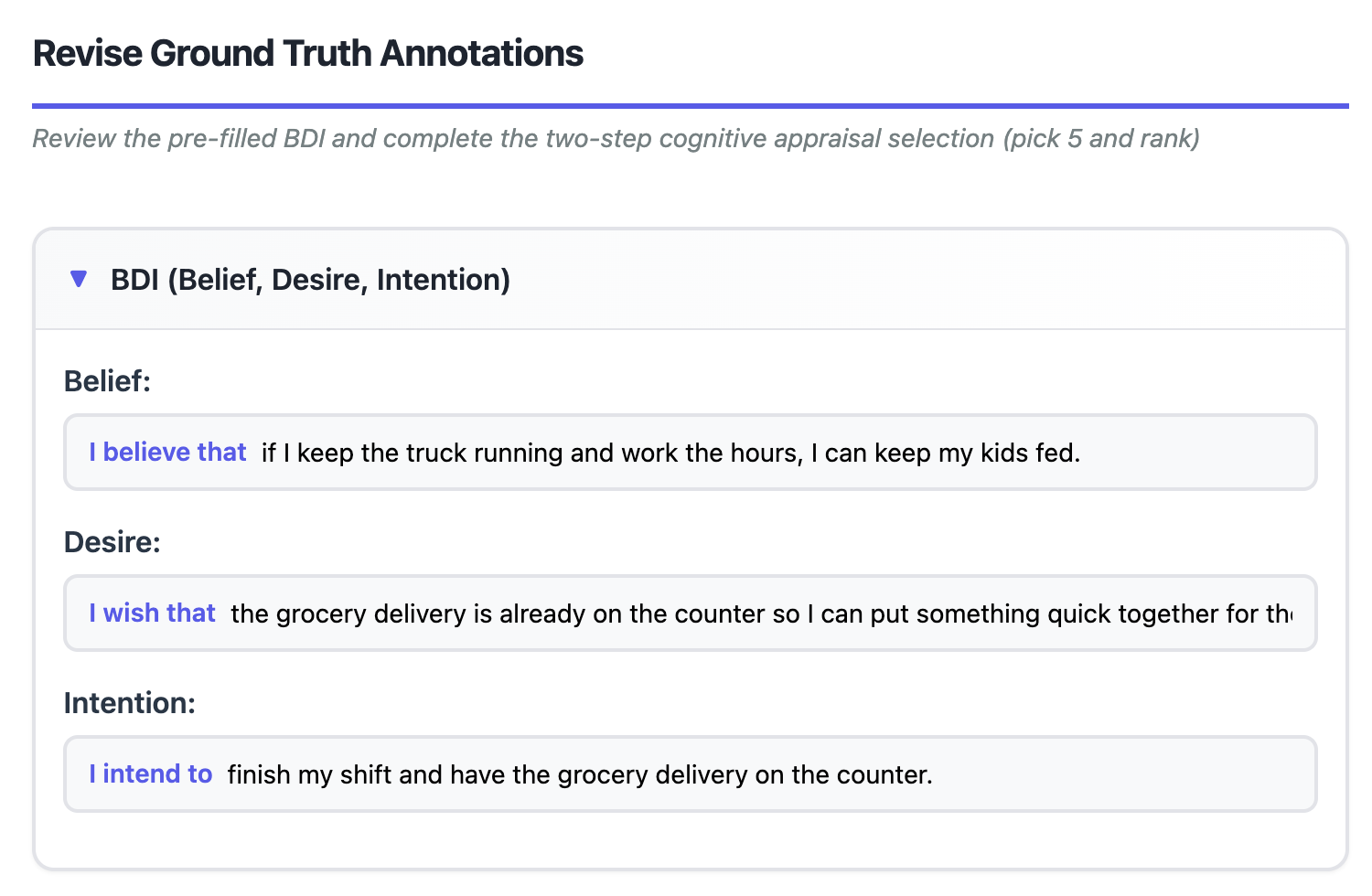}
    \caption{Annotation Interface: Pre-event BDI Annotation}
    \label{app:fig:prior_bdi}
\end{figure}

In the beginning of the annotation process, each annotator is provided with the persona information of the help-seeker, as shown in Figure~\ref{app:fig:persona_interface}.
The annotator then reads the ESC trajectory, as shown in Figure~\ref{app:fig:dialogue}.
After understanding the help-seeker's persona and their exchange with the therapist, the annotator first annotates the help-seeker's pre-event BDI. Notice that the generated pseudo labels are provided in this section as a reference (Figure~\ref{app:fig:prior_bdi})

\begin{figure} [h]
    \centering
    \includegraphics[width=\columnwidth]{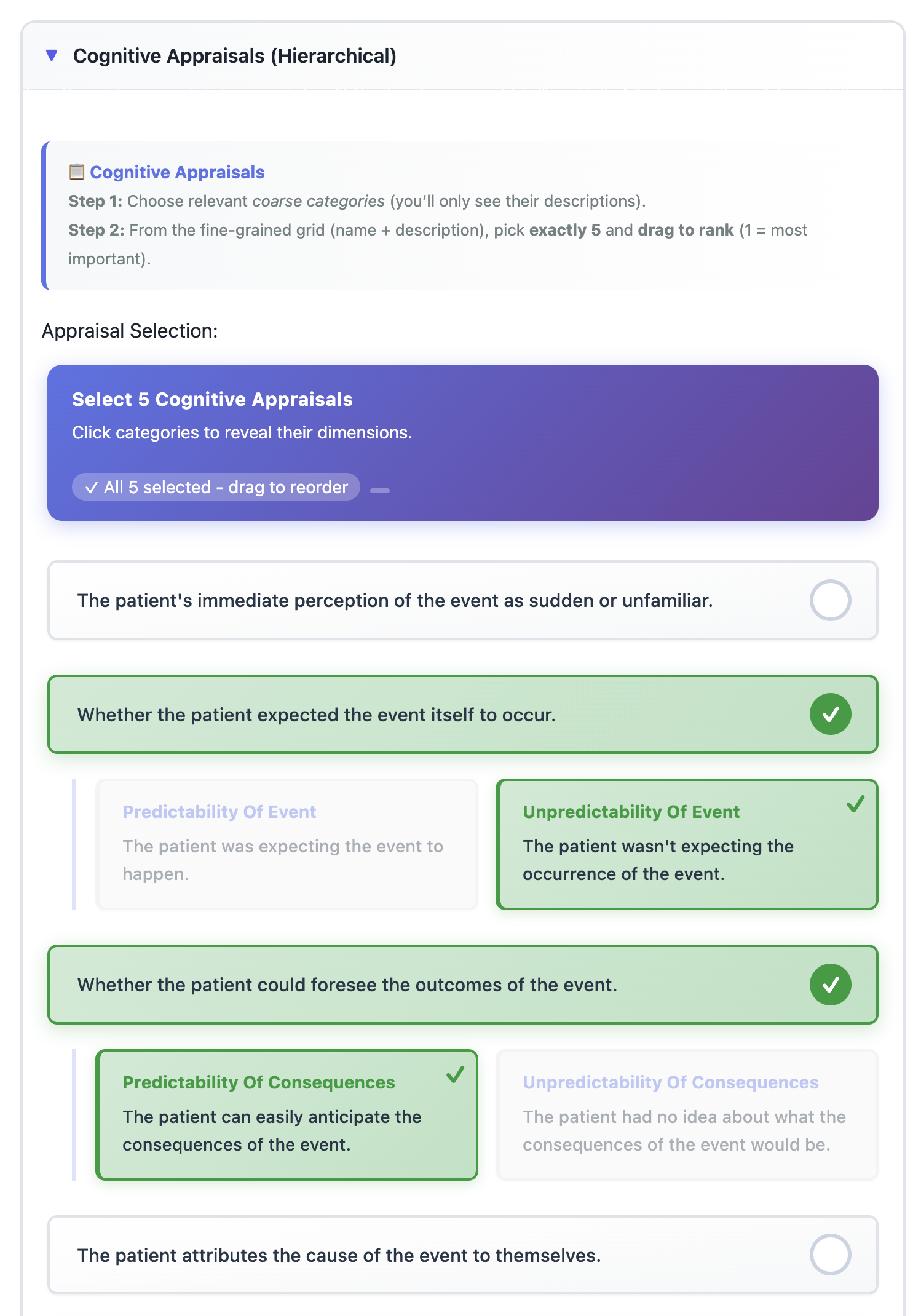}
    \caption{Annotation Interface: Coarse-to-Fine Annotation for Salient Appraisal Dimensions.}
    \label{app:fig:cogapp_step1}
\end{figure}

For salient appraisal annotation, the annotator first selects a coarse-grained description of the help-seeker's interpretation of the upsetting event, as shown in Figure~\ref{app:fig:cogapp_step1}. Each coarse-grained description corresponds to 2 fine-grained appraisal dimensions, which are hidden from the annotator in the beginning. The annotator then selects 5 fine-grained appraisal dimensions from the unfolded options. This pipeline avoids displaying all 14 appraisal dimensions to the annotator at once and effectively reduces the annotator's cognitive workload.

\begin{figure} [h]
    \centering
    \includegraphics[width=\columnwidth]{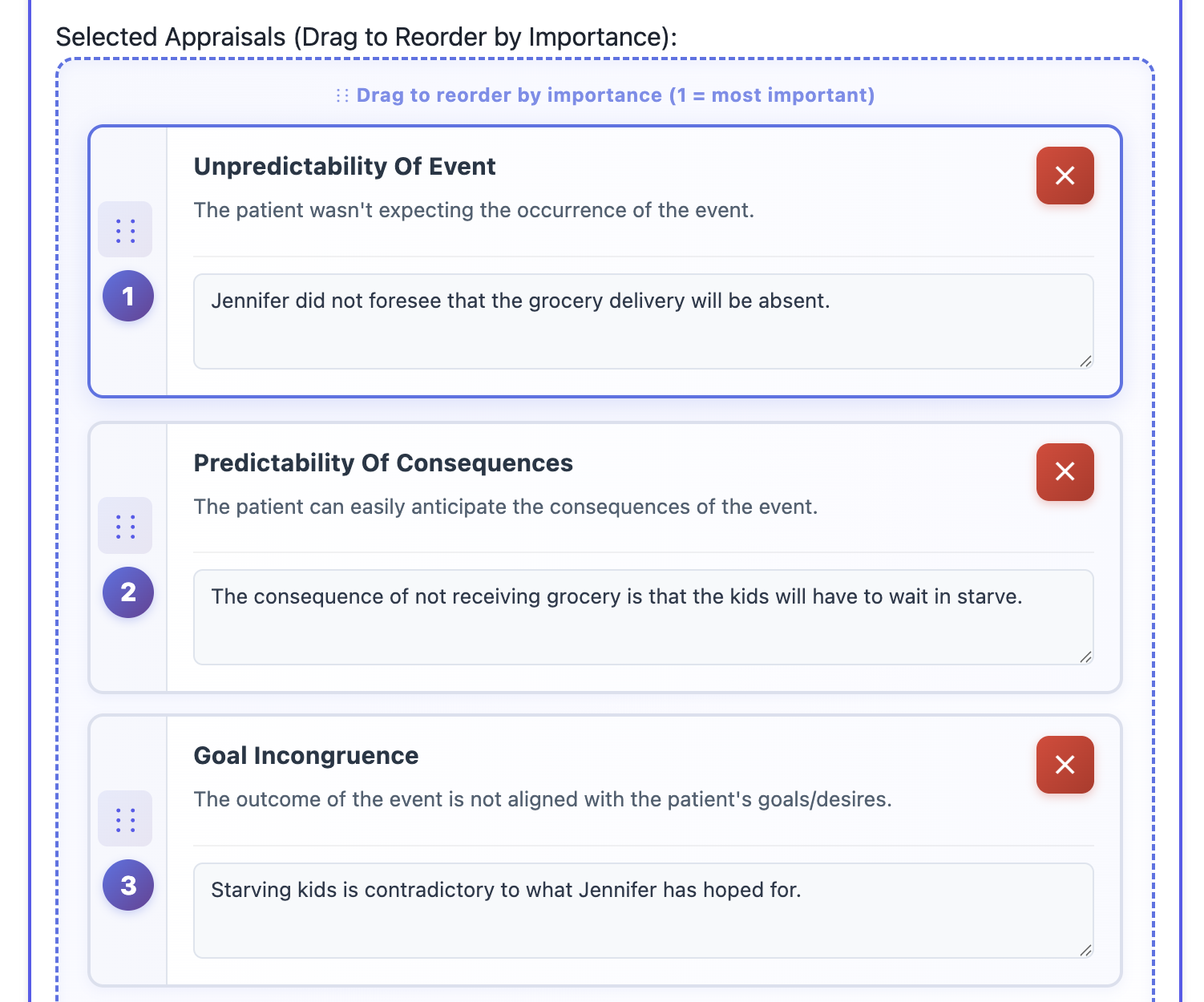}
    \caption{Annotation Interface: Salient Appraisal Dimensions with Required Ranking Rationale.}
    \label{app:fig:cogapp_step2}
\end{figure}

Once the 5 salient appraisal dimensions are selected, the annotator is required to rank the appraisal dimensions according to their saliency, with the most salient dimension ranked at the top. Further, the annotator is required to provide a brief rationale for each appraisal dimension (see Figure~\ref{app:fig:cogapp_step2}).

\begin{figure} [h]
    \centering
    \includegraphics[width=\columnwidth]{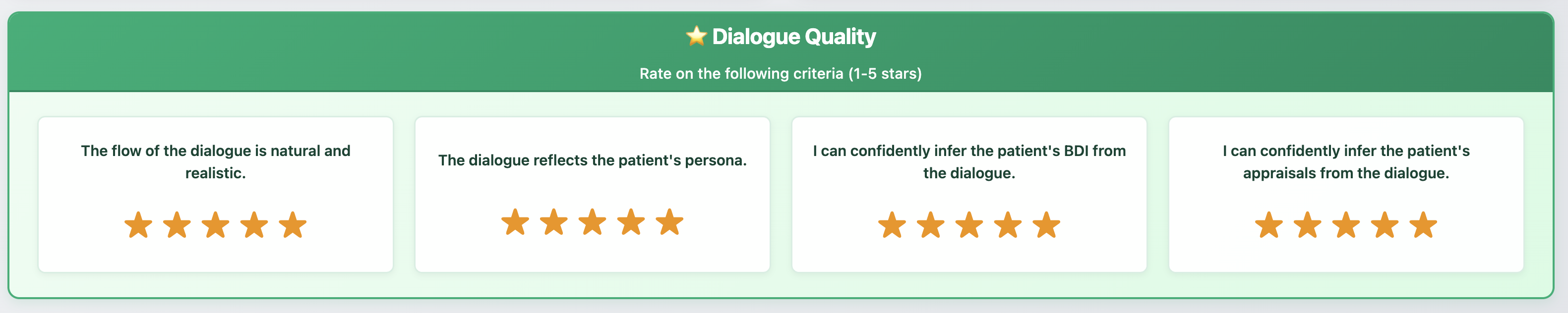}
    \caption{Annotation Interface: Data Quality Check.}
    \label{app:fig:data_quality}
\end{figure}

Finally, the annotator is prompted to assess the quality of the current data entry based on their annotation experience (interface shown in Figure~\ref{app:fig:data_quality}). Specifically, the annotator evaluates the current entry using the following four criteria:
\begin{itemize}[leftmargin=6mm, noitemsep]
    \item \textbf{Realism}: The flow of the dialogue is natural and realistic.
    \item \textbf{Persona-Relevance}: The dialogue reflects the help-seeker's persona profile.
    \item \textbf{BDI-Relevance}: I can confidently infer the help-seeker's BDI from the dialogue.
    \item \textbf{Appraisl-Relevance}: I can confidently infer the help-seeker's salient appraisals from the dialogue.
\end{itemize}

\section{Detailed Data Statistics}
\label{app:detailed_data_stats}

\begin{figure*}
    \centering
    \includegraphics[width=\textwidth]{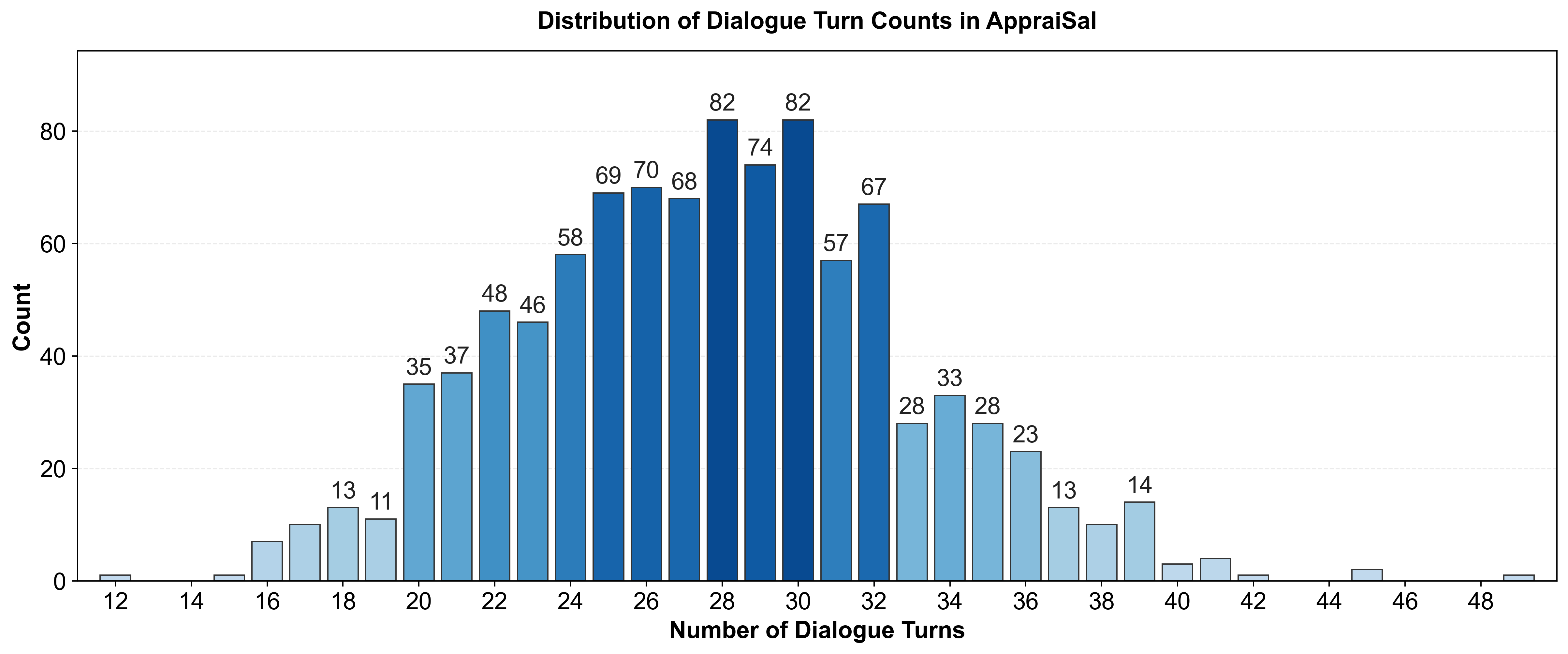}
    \caption{Distribution of the number of dialogue turns across ESC trajectories in \ourdata.}
    \label{fig:turn_count}
\end{figure*}

Seeded from AugESC \cite{zheng2023augesc}, \ourdata~covers a similar distribution of event topics. We provide the count and distribution of each event genre in Table~\ref{tab:dialogue_genres}. On average, there are $27.77_{\pm{5.148}}$ turns. In each turn, the \texttt{therapist agent} utters 23.92 words on average, whereas the \texttt{patient agent} utters $31.72$ words on average. The detailed distribution of turn counts in \ourdata~is shown in Figure~\ref{fig:turn_count}.

\begin{table}[t]
  \centering
  \resizebox{\columnwidth}{!}{
  \begin{tabular}{lrr}
    \toprule
    Genre & Count & Percentage \\
    \midrule
    Workplace \& Career Challenges & 452 & 45.4\% \\
    Parenting \& Caregiving Burden & 90 & 9.0\% \\
    Identity \& Self-Worth & 77 & 7.7\% \\
    Mental Health \& Psychological Distress & 71 & 7.1\% \\
    Family, Friendship \& Social Conflict & 68 & 6.8\% \\
    Financial \& Housing Stress & 51 & 5.1\% \\
    Work-Life Balance \& Burnout & 48 & 4.8\% \\
    Grief, Loss \& Trauma & 43 & 4.3\% \\
    Romantic \& Intimate Relationships & 38 & 3.8\% \\
    Academic \& Learning Stress & 32 & 3.2\% \\
    Safety, Security \& Existential Threats & 13 & 1.3\% \\
    Physical Health \& Medical Concerns & 8 & 0.8\% \\
    Immigration \& Legal Stress & 5 & 0.5\% \\
    \bottomrule
  \end{tabular}
  }
  \caption{Distribution of dialogue genres.}
  \label{tab:dialogue_genres}
\end{table}

In total, we curated 1,120 entries from Prolific, which are annotated by 145 annotators, where each annotator is assigned approximately 8 entries. The distribution of human-annotated appraisal dimensions is shown in Table~\ref{tab:human_appraisal_labels}. On average, we compensate \$1.52 for each entry, resulting in a total cost of \$1702.4. Further, we leverage \texttt{GPT-5-mini} to refine and generate multiple components of the dataset; the cost of the single complete run is \$15.93. The total cost, including preliminary runs for prompt engineering, is approximately \$24. We summarize the API cost in Table~\ref{app:tab:gpt_api_cost}.

\begin{table}[h]
    \centering
    \tiny
    \resizebox{\columnwidth}{!}{
    \begin{tabular}{l l l l }
    \toprule
    \textbf{Task} & \textbf{Input} & \textbf{Output} & \textbf{Total} \\
    \midrule
    Event Revision & \$0.36 & \$0.20 & \$0.56 \\
    Interpretation & \$0.30 & \$0.16 & \$0.46 \\
    Pseudo Mental States & \$0.34 & \$0.94 & \$1.28 \\
    ESC Revision & \$0.90 & \$12.73 & \$13.63 \\
    \midrule
    Total & \$1.9 & \$14.03 & \$15.93 \\
    \bottomrule
    \end{tabular}
    }
    \caption{Cost breakdown for using OpenAI API for \ourdata~construction.}
    \label{app:tab:gpt_api_cost}
\end{table}

To ensure data quality, we filter entries with low assessment scores. Specifically, we apply two filters. First, all entries with an average assessment score lower than 3.0 are removed. Second, all entries with an "Appraisal-Relevance" score of lower than 3.0 are also removed. Salient appraisal inference is the core task of \ourdata, therefore it is crucial to ensure a strong association between the content of the ESC trajectory and the annotated appraisal dimensions. The remaining 996 data entries have the following assessment scores:
\begin{itemize}[leftmargin=6mm, noitemsep]
    \item \textbf{Appraisal-Relevance}: 4.39
    \item \textbf{BDI-Relevance}:  4.33
    \item \textbf{Persona-Relevance}: 4.38
    \item \textbf{Realism}: 4.42
\end{itemize}

Further, 26.34\% of the data entry received revision in the ESC trajectories. After getting obtaining the annotations, we leverage \texttt{GPT-5-Mini} to flag annotations that contain trivial or logical errors. For instance, the dimension of "\emph{self-cause}" and "\emph{other-cause}" shouldn't occur simultaneously. This filtering process results in a revision of 7.48\% of the existing annotations.

\begin{table}[t]
  \centering
  \resizebox{\columnwidth}{!}{
  \begin{tabular}{lrr}
    \toprule
    Appraisal label & Count & Percentage \\
    \midrule
    Self Cause & 752 & 75.5\% \\
    Goal Incongruence & 751 & 75.4\% \\
    Unacceptable Consequences & 549 & 55.1\% \\
    Norm Violation & 484 & 48.6\% \\
    Unpredictability Of Event & 459 & 46.1\% \\
    Coping Demand & 433 & 43.5\% \\
    Self Control & 369 & 37.0\% \\
    Unpredictability Of Consequences & 346 & 34.7\% \\
    Suddenness & 273 & 27.4\% \\
    Predictability Of Consequences & 144 & 14.5\% \\
    Other Control & 131 & 13.2\% \\
    Other Cause & 118 & 11.8\% \\
    Unfamiliarity & 103 & 10.3\% \\
    Predictability Of Event & 43 & 4.3\% \\
    \bottomrule
  \end{tabular}
  }
  \caption{Distribution of human-annotated cognitive appraisal labels \ourdata.}
  \label{tab:human_appraisal_labels}
\end{table}

\section{Bayesian Inverse Planning}
\label{app:bip_derivation}
In this section, we show a step-by-step derivation of the backward BIP inference factorization from the forward generative model.

To recap, given a ESC trajectory up to the $t^{th}$ turn $\mathcal{D}_{<t}$, the help-seeker's persona profile $\mathcal{P}$, their pre-event BDI $\mathcal{B}$ and their salient cognitive appraisals $\mathcal{A}$, we define the following forward generative
\[
u_{h,t} \sim \mathbb{P}(u_{h,t} \mid \mathcal{D}_{<t}, \mathcal{P}, \mathcal{B}, \mathcal{A})
\]
where we regard the help-seeker's utterance $u_{h, t}$ as the \textit{action}. The object is to jointly infer the help-seeker's pre-event BDI $\mathcal{B}$ and, more importantly, their salient cognitive appraisals $\mathcal{A}$, which translates to the following backward inference problem:
\[
\mathbb{P}(\mathcal{A}, \mathcal{B} \mid \mathcal{D}_{1:t}, \mathcal{P}).
\]
Applying Bayes' rule, we get
\begin{align*}
\begin{split}
\mathbb{P}(\mathcal{A}, \mathcal{B} \mid \mathcal{D}_{1:t}, \mathcal{P})
\propto
\mathbb{P}(\hat{u}_{h,t} \mid &\mathcal{D}_{<t}, \mathcal{P}, \mathcal{A}, \mathcal{B}) \\
&\cdot \mathbb{P}(\mathcal{A}, \mathcal{B} \mid \mathcal{D}_{<t}, \mathcal{P}).
\end{split}
\end{align*}
By further factorizing $\mathbb{P}(\mathcal{A}, \mathcal{B} \mid \mathcal{D}_{<t}, \mathcal{P})$, we get
\begin{align*}
\begin{split}
   \mathbb{P}(\mathcal{A}_t, \mathcal{B}_t|&\mathcal{D}_{1:t}, \mathcal{P}) \propto \mathbb{P}(\hat{u} _{h, t} | \mathcal{D}_{<t}, \mathcal{P}, \mathcal{B}_t, \mathcal{A}_t) \\
   &\cdot \mathbb{P}(\mathcal{B}_t|\mathcal{D}_{<t}, \mathcal{P}) \cdot \mathbb{P}(\mathcal{A}_t|\mathcal{D}_{<t}, \mathcal{P}, \mathcal{B}_t),
\end{split}
\end{align*}
which is the factorization we use in the backward inference problem as shown in Equation~\ref{eq:bip_main}.

\section{Likelihood Modeling with LLMs}
\label{app:prompt_logprobs}
In this work, we leverage the token-level log-probabilities to compute the likelihoods, namely $\mathbb{P}(\hat{u}_{t}\mid\cdot)$, $\mathbb{P}(\mathcal{B}\mid\cdot)$, and $\mathbb{P}(\mathcal{A}\mid\cdot)$. We use log-probabilities over prompting LLMs to produce a likelihood score to mitigate the potential risk of hallucination.

Take the \texttt{utterance agent} as an example. Suppose we apply \prism~\ourmethod~with \texttt{Qwen3-4B-Instruct} as the backbone model and with input instruction prompt $\mathcal{I}(\mathcal{D}_{<t}, \mathcal{P}, \mathcal{B}_t, \mathcal{A}_{t,i})$. We take the actual next utterance from the help-seeker, denoted as $\hat{u}_{t} = \{w_1, w_2, \ldots, w_m\}$ and build the chat template as:
\begin{quote}
\small\ttfamily
\textbf{<|im\_start|>user}\\
$\mathcal{I}(\mathcal{D}_{<t}, \mathcal{P}, \mathcal{B}_t, \mathcal{A}_{t,i})$ \\
\textbf{<|im\_end|>}\\
\textbf{<|im\_start|>assistant}\\
$w_1, w_2, \ldots, w_m$ \\
\textbf{<|im\_end|>}
\end{quote}

We input the complete chat template into the model and gather the log-probabilities corresponding to tokens $w_1, w_2, \ldots, w_m$. Specifically, we use vllm's "\texttt{prompt\_logprobs}" functionality \cite{kwon2023efficient}.
The log-probability corresponding to the current appraisal hypothesis $\mathcal{A}_{t, i}$ is then computed as the sum of all token log-probabilities:
\begin{align}
\begin{split}
&p(\hat{u}_{t,i})
= \log \mathbb{P}(\hat{u}_{t} \mid \mathcal{D}_{<t}, \mathcal{P}, \mathcal{B}_t, \mathcal{A}_{t,i}) \\
&= \sum_{j=1}^{m}
\log \mathbb{P}\Big(
w_j | w_{<j}, \mathcal{I}(\mathcal{D}_{<t}, \mathcal{P}, \mathcal{B}_t, \mathcal{A}_{t,i})
\Big).
\end{split}
\end{align}

\section{BDI Analysis}
\label{app:bdi_analysis}

In this section, we evaluate the quality of the argmax of the pre-event BDI, $\mathcal{B}^*$. To estimate human performance, we asked two annotators to each annotate 50 entries sampled from \ourdata. For evaluation, we use both an n-gram-based metric, ROUGE-L \cite{lin2004rouge}, and a model-based metric, BLEURT \cite{sellam2020bleurt}.

The results in Table~\ref{tab:bdi_results} show that the \texttt{gpt-oss} series matches human performance in terms of average ROUGE-L. Moreover, the \texttt{Gemma3} series comes close to human performance in terms of average BLEURT. Although the \texttt{Qwen3} series still falls short of human performance, it is nevertheless able to produce pre-event BDI states of reasonable quality, which is further reflected in the gains observed from incorporating pre-event BDI in the ablation study.

\begin{table}[h!]
    \centering
    \resizebox{\columnwidth}{!}{
    \begin{tabular}{c l l l l}
    \toprule
    & A1|A2 & A1|A3 & A2|A3 & Average \\
    \midrule
    Spearman's $\rho$ & 0.726 & 0.541 & 0.477 & 0.581 \\
    \bottomrule
    \end{tabular}
    }
    \caption{Spearman's Rank Correlation results using human agreement data}
    \label{tab:spearman_corr}
\end{table}

\begin{table*}[t]
\centering
\small
\begin{tabular}{lccccc ccc}
\toprule
\multirow{2}{*}{Model} & \multirow{2}{*}{Avg.\ ROUGE-L} & \multirow{2}{*}{Avg.\ Sim} & \multicolumn{3}{c}{ROUGE-L} & \multicolumn{3}{c}{BLEURT Scores} \\
\cmidrule(lr){4-6} \cmidrule(lr){7-9}
 &  &  & Belief & Desire & Intention & Belief & Desire & Intention \\
\midrule
\textbf{\texttt{Human}}            & 0.326 & 0.529 & 0.307 & 0.309 & 0.362 & 0.537 & 0.526 & 0.525 \\
\midrule
\textbf{\texttt{Gemma3-4B}}        & 0.286 & 0.491 & 0.277 & 0.281 & 0.301 & 0.483 & 0.513 & 0.478 \\
\textbf{\texttt{Gemma3-27B}}       & 0.289 & 0.520 & 0.316 & 0.253 & 0.300 & 0.535 & 0.514 & 0.512 \\
\textbf{\texttt{Qwen3-4B}}    & 0.253 & 0.469 & 0.275 & 0.233 & 0.253 & 0.471 & 0.480 & 0.455 \\
\textbf{\texttt{Qwen3-32B}}   & 0.251 & 0.458 & 0.272 & 0.233 & 0.247 & 0.479 & 0.473 & 0.421 \\
\textbf{\texttt{Qwen3-30B-A3B}}   & 0.219 & 0.442 & 0.231 & 0.208 & 0.218 & 0.462 & 0.449 & 0.415 \\[0.5em]
\textbf{\texttt{GPT-OSS-20B}}      & 0.318 & 0.470 & 0.337 & 0.283 & 0.332 & 0.475 & 0.480 & 0.454 \\
\textbf{\texttt{GPT-OSS-120B}}     & 0.312 & 0.487 & 0.313 & 0.284 & 0.338 & 0.472 & 0.505 & 0.484 \\
\bottomrule
\end{tabular}
\caption{Performance comparison on BDI inference.}
\label{tab:bdi_results}
\end{table*}

\section{Saliency Rank \& Evaluation Metrics}
\subsection{Ranking of Appraisal Dimensions}
\label{app:rank_explained}
As shown in Figure~\ref{app:fig:cogapp_step2}, annotators are asked to rank the selected fine-grained appraisal dimensions by saliency. The dimensions are initially presented in click order, and annotators then drag them to produce their final ranking. On average, each appraisal annotation involves 1.79 drag operations.

To evaluate whether annotators can reliably agree on the exact ranking of appraisal dimensions, we compute \textit{Spearman’s rank correlation} on a human agreement set of 50 examples, each annotated by three Prolific annotators. Table~\ref{tab:spearman_corr} shows that although Annotators 1 and 2 achieve a relatively strong correlation of 0.726, the remaining pairwise correlations are notably weaker, resulting in a moderate average correlation of 0.581. Since these results provide limited support for the feasibility of curating precise and consensual appraisal rankings, we do not use exact rank information in our evaluation and instead rely on set-based metrics.

\begin{table}[t!]
\centering
\resizebox{\columnwidth}{!}{
\begin{tabular}{lccccc}
\toprule
\textbf{Metric} & \textbf{Baseline} & \textbf{\ourmethod} & \textbf{$\Delta$} & \textbf{\textit{t}-stat} & \textbf{\textit{p} (raw)} \\
\midrule
\multicolumn{6}{c}{\textbf{Gemma3-4B}} \\
\midrule
Precision@1 & 0.4250 & 0.7190 & 0.2940 & -28.7830 & 0.0000 \\
Precision@2 & 0.5640 & 0.6980 & 0.1340 & -30.4755 & 0.0002 \\
Precision@3 & 0.5930 & 0.6550 & 0.0620 & -16.0083 & 0.0006 \\
Precision@4 & 0.5800 & 0.6190 & 0.0390 & -13.2476 & 0.0042 \\
Precision@5 & 0.5640 & 0.5840 & 0.0200 & -24.4949 & 0.0000 \\
Jaccard Index & 0.4150 & 0.4420 & 0.0270 & -20.9141 & 0.0003 \\
\midrule
\multicolumn{6}{c}{\textbf{Gemma3-27B}} \\
\midrule
Precision@1 & 0.7590 & 0.8620 & 0.1030 & -11.7126 & 0.0021 \\
Precision@2 & 0.6970 & 0.7820 & 0.0850 & -9.9713 & 0.0019 \\
Precision@3 & 0.6490 & 0.7120 & 0.0630 & -11.5022 & 0.0035 \\
Precision@4 & 0.5940 & 0.6570 & 0.0630 & -16.2665 & 0.0006 \\
Precision@5 & 0.5620 & 0.6200 & 0.0580 & -27.8623 & 0.0000 \\
Jaccard Index & 0.4140 & 0.4780 & 0.0650 & -30.7446 & 0.0000 \\
\midrule
\multicolumn{6}{c}{\textbf{Qwen3-4B}} \\
\midrule
Precision@1 & 0.7210 & 0.8630 & 0.1420 & -16.0783 & 0.0027 \\
Precision@2 & 0.6640 & 0.7640 & 0.1000 & -24.4949 & 0.0014 \\
Precision@3 & 0.6450 & 0.6880 & 0.0430 & -26.3320 & 0.0000 \\
Precision@4 & 0.5940 & 0.6390 & 0.0450 & -24.6475 & 0.0006 \\
Precision@5 & 0.5700 & 0.6130 & 0.0430 & -23.5521 & 0.0006 \\
Jaccard Index & 0.4230 & 0.4670 & 0.0440 & -34.0823 & 0.0001 \\
\midrule
\multicolumn{6}{c}{\textbf{Qwen3-32B}} \\
\midrule
Precision@1 & 0.7450 & 0.8380 & 0.0930 & -5.8087 & 0.0114 \\
Precision@2 & 0.7100 & 0.7910 & 0.0810 & -11.6110 & 0.0019 \\
Precision@3 & 0.6760 & 0.7270 & 0.0510 & -9.3113 & 0.0058 \\
Precision@4 & 0.6370 & 0.6730 & 0.0360 & -4.3872 & 0.0129 \\
Precision@5 & 0.6050 & 0.6230 & 0.0180 & -3.9918 & 0.0173 \\
Jaccard Index & 0.4580 & 0.4810 & 0.0230 & -4.3209 & 0.0131 \\
\midrule
\multicolumn{6}{c}{\textbf{Qwen3-30B-A3B}} \\
\midrule
Precision@1 & 0.7400 & 0.8800 & 0.1400 & -9.2380 & 0.0009 \\
Precision@2 & 0.6540 & 0.7640 & 0.1100 & -5.2341 & 0.0249 \\
Precision@3 & 0.6470 & 0.6840 & 0.0370 & -6.0287 & 0.0040 \\
Precision@4 & 0.6160 & 0.6420 & 0.0260 & -8.8318 & 0.0099 \\
Precision@5 & 0.5940 & 0.6170 & 0.0230 & -39.8372 & 0.0006 \\
Jaccard Index & 0.4450 & 0.4720 & 0.0270 & -14.7885 & 0.0019 \\
\midrule
\multicolumn{6}{c}{\textbf{gpt-oss-20B}} \\
\midrule
Precision@1 & 0.5680 & 0.7670 & 0.1990 & -21.7994 & 0.0001 \\
Precision@2 & 0.5980 & 0.7260 & 0.1280 & -21.7397 & 0.0014 \\
Precision@3 & 0.6040 & 0.6800 & 0.0760 & -13.2972 & 0.0002 \\
Precision@4 & 0.5870 & 0.6460 & 0.0590 & -11.9606 & 0.0026 \\
Precision@5 & 0.5610 & 0.6140 & 0.0530 & -20.5268 & 0.0003 \\
Jaccard Index & 0.4110 & 0.4690 & 0.0580 & -20.0918 & 0.0001 \\
\midrule
\multicolumn{6}{c}{\textbf{gpt-oss-120B}} \\
\midrule
Precision@1 & 0.7870 & 0.8170 & 0.0300 & -4.0575 & 0.0169 \\
Precision@2 & 0.6820 & 0.7540 & 0.0720 & -22.0454 & 0.0000 \\
Precision@3 & 0.6420 & 0.7040 & 0.0620 & -29.7838 & 0.0000 \\
Precision@4 & 0.6060 & 0.6650 & 0.0590 & -17.5256 & 0.0002 \\
Precision@5 & 0.5770 & 0.6330 & 0.0560 & -34.2929 & 0.0000 \\
Jaccard Index & 0.4320 & 0.4920 & 0.0600 & -28.8231 & 0.0000 \\
\bottomrule
\end{tabular}
}
\caption{Two-sample T-test statistics comparing the performance of baseline models with and without the \ourmethod~framework.}
\label{tab:two_sample_ttest_statistics}
\end{table}

\subsection{Evaluation Metrics}
\label{app:eval_metrics}
In this section, we provide a definition for the evaluation metrics used in our experiments. Given a ground truth set of salient appraisals $\mathcal{A} = \{a_1, a_2, a_3, a_4, a_5\}$ and a predicted set of salient appraisals $\hat{\mathcal{A}} = \{\hat{a}_1, \hat{a}_2, \hat{a}_3, \hat{a}_4, \hat{a}_5\}$, we compute the following metrics

\paragraph{\textit{Precision@k}} computes the proportion of the top-$k$ predicted appraisals that exist in the ground truth:
\[
\textit{Precision@k} = \frac{|\{a_i\}_{i=1}^k \cap \mathcal{A}|}{k},
\]
where $|\cdot|$ denotes the cardinality of the set.

\paragraph{Relationship with \textit{Recall} and \textit{F1}}
In the salient appraisal reasoning task, where the model is required to provide 5 salient appraisal dimensions, $|A| = |\hat{A}| = 5$. Given the following definition of set-level Recall:
\[
\textit{Recall} = \frac{|\hat{\mathcal{A}} \cap \mathcal{A}|}{|\mathcal{A}|},
\]
where $\mathcal{\hat{A}} = \{a_i\}_{i=1}^{5}$. It is trivial that \textit{Recall} is equivalent to \textit{Precision@5}. Therefore, we do not report the $\textit{Recall}$ and $\textit{F1}$ scores.

\paragraph{\textit{Jaccard Index}} computes the number of appraisal dimensions in the intersection of the ground truth set and the predicted set, normalized by the number of appraisals covered by  the union of the two sets:
\[
\textit{Jaccard Index} = \frac{|\hat{\mathcal{A}} \cap \mathcal{A}|}{|\hat{\mathcal{A}} \cup \mathcal{A}|}.
\]

\section{Full Evaluation Results}
\label{app:full_eval_results}
The full evaluation results, including results from the recently released \texttt{Qwen3.5} models are shown in Table~\ref{tab:full_eval_results}. The results for \textit{Iterative Self-Consistency with Judge Agent} are shown in Figure~\ref{tab:additional_baseline}. Results show that, albeit using extra time-time compute, neither \textit{Self-Consistency} prompting nor \textit{Iterative Self-Consistency with Judge Agent} achieves meaningful improvement compared to \prism~\ourmethod, demonstrating the effectiveness and importance of the BIP-based multi-agent inference framework.

\begin{table*}[th!]
    \centering
    \resizebox{0.92\textwidth}{!}{
    \begin{tabular}{l c c c c c c}
    \toprule
        & \textbf{Precision@1} & \textbf{Precision@2} & \textbf{Precision@3} & \textbf{Precision@4} & \textbf{Precision@5} &
        \textbf{Jaccard Index} \\
        \toprule
        \rowcolor{white2}
        \textbf{Human} & $1.000$ & $0.990$ & $0.920$ & $0.775$ & $0.748$ &
        $0.612$   \\
        \midrule
        \textbf{Random} & $0.361$ & $0.342$ & $0.350$ & $0.358$ & $0.357$ &
        $0.234$ \\
        \textbf{Majority} & $0.755$ & $0.655$ & $0.561$ & $0.530$ & $0.522$ & 
        $0.379$ \\
        \midrule
        \multicolumn{7}{c}{Instruct Models} \\
        \midrule
        \textbf{\texttt{Qwen3-4B}} &	$0.722$ & $0.669$ &	$0.645$ & $0.593$ &	$0.570$ & 
        $0.422$ \\
        \textbf{\texttt{Qwen3-32B}} & $0.758$ & $0.706$ & $0.686$ &	$0.648$ & $0.611$ & 
        $0.465$ \\
        \textbf{\texttt{Qwen3-MoE}} &	$0.763$ & $0.694$ &	$0.655$ & $0.617$ &	$0.595$ & 
        $0.448$ \\
        \textbf{\texttt{Gemma3-4B}} & $0.438$ & $0.574$ & $0.602$ & $0.582$ & $0.566$ &	
        $0.416$ \\
        \textbf{\texttt{Gemma3-27B}} & $0.746$ &	$0.684$ & $0.652$ &	$0.593$ & $0.554$ &
        $0.406$  \\
        \midrule
        \multicolumn{7}{c}{Reasoning Models} \\
        \midrule
        \textbf{\texttt{Qwen3-4B}} & $0.759$ & $0.723$ &	$0.677$ & $0.632$ &	$0.598$ & 
        $0.452$ \\
        \textbf{\texttt{Qwen3-32B}} & $0.739$ &	$0.709$ &	$0.676$ &	$0.636$ &	$0.608$ &	
        $0.461$ \\
        \textbf{\texttt{Qwen3-MoE}} & $0.713$ &	$0.680$ &	$0.667$ &	$0.624$ &	$0.599$  &	
        $0.452$ \\
        \textbf{\texttt{Qwen3.5-27B}} & $0.646$ & $0.668$ & $0.645$ & $0.607$ &	$0.579$ &
        $0.432$ \\
        \textbf{\texttt{Qwen3.5-MoE}} & $0.702$ & $0.669$ & $0.637$ & $0.598$ & $0.565$ &
        $0.419$ \\
        \textbf{\texttt{GPT-OSS-20B}} &	$0.563$ &	$0.596$ &	$0.609$ &	$0.595$ &	$0.565$ &
        $0.414$ \\
        \textbf{\texttt{GPT-OSS-120B}} &	$0.799$ &	$0.739$ &	$0.694$ &	$0.652$ &	$0.617$ &
        $0.477$ \\[0.75em]
        \textbf{\texttt{GPT-5-Nano}} & $0.700$ &	$0.685$ &	$0.653$ &	$0.621$ &	$0.594$ &
        $0.445$ \\
        \textbf{\texttt{GPT-5-Mini}}	& $0.664$ &	$0.672$ &	$0.650$ &	$0.622$ &	$0.591$ & 
        $0.444$  \\
        \midrule
        \rowcolor{white3}
        \multicolumn{7}{c}{Self-Consistency Prompting ($h$=8)} \\
        \midrule
        \rowcolor{white3}
        \textbf{\texttt{Qwen3-4B}}
        & $0.755_{+0.033}$ & $0.695_{+0.026}$ & $0.655_{+0.010}$ & $0.604_{+0.011}$ & $0.572_{+0.002}$ &
        $0.425_{+0.003}$ \\
        \rowcolor{white3}
        \textbf{\texttt{Qwen3-32B}}
        & $0.772_{+0.014}$ & $0.737_{+0.031}$ & $0.698_{+0.012}$ & $0.646_{-0.002}$ & $0.610_{-0.001}$ & 
        $0.464_{-0.001}$ \\
        \rowcolor{white3}
        \textbf{\texttt{Qwen3-MoE}}
        & $0.725_{-0.038}$ & $0.634_{-0.060}$ & $0.644_{-0.011}$ & $0.616_{-0.001}$ & $0.591_{-0.004}$ & 
        $0.440_{-0.008}$ \\
        \rowcolor{white3}
        \textbf{\texttt{Qwen3.5-27B}}
        & $0.649_{+0.003}$ & $0.670_{+0.002}$ & $0.656_{+0.011}$ & $0.615_{+0.008}$ & $0.592_{+0.013}$ &
        $0.442_{+0.010}$ \\
        \rowcolor{white3}
        \textbf{\texttt{Qwen3.5-MoE}}
        & $0.728_{+0.026}$ & $0.677_{+0.008}$ & $0.648_{+0.003}$ & $0.603_{+0.005}$ & $0.571_{+0.006}$ &
        $0.426_{+0.007}$ \\
        \rowcolor{white3}
        \textbf{\texttt{Gemma3-4B}}
         & $0.425_{-0.013}$ & $0.567_{-0.007}$ & $0.598_{-0.004}$ & $0.577_{-0.005}$ & $0.562_{-0.004}$ &
         $0.412_{-0.004}$ \\
        \rowcolor{white3}
        \textbf{\texttt{Gemma3-27B}}
        & $0.749_{+0.003}$ & $0.688_{+0.004}$ & $0.652_{+0.000}$ & $0.593_{+0.000}$ & $0.558_{+0.004}$ &
        $0.410_{+0.004}$ \\
        \rowcolor{white3}
        \textbf{\texttt{GPT-OSS-20B}}
        & $0.659_{+0.096}$ & $0.663_{+0.067}$ & $0.652_{+0.043}$ & $0.615_{+0.020}$ & $0.587_{+0.022}$ &
        $0.438_{+0.024}$ \\
        \rowcolor{white3}
        \textbf{\texttt{GPT-OSS-120B}}
        & $0.791_{-0.008}$ & $0.735_{-0.004}$ & $0.669_{-0.025}$ & $0.634_{-0.018}$ & $0.602_{-0.015}$ &
        $0.461_{-0.016}$ \\[0.75em]
        \rowcolor{white3}
        \textbf{\texttt{GPT-Nano}}
        & $0.687_{-0.013}$ & $0.678_{-0.007}$ & $0.647_{-0.006}$ & $0.615_{-0.006}$ & $0.590_{-0.004}$ &
        $0.443_{-0.002}$ \\
        \rowcolor{white3}
        \textbf{\texttt{GPT-Mini}}
        & $0.655_{-0.009}$ & $0.665_{-0.007}$ & $0.657_{-0.007}$ &	$0.632_{-0.010}$ & $0.594_{+0.003}$ &
        $0.448_{+0.004}$ \\
        \midrule
        \rowcolor{white3}
        \textbf{Avg. Change} & $+0.009$ & $+0.005$ & $+0.002$ & $+0.000$ & $+0.003$ &
        $+0.002$ \\
        \midrule
        \rowcolor{tablewhite}
        \multicolumn{7}{c}{with the \ourmethod~Framework ($h$=8)} \\
        \midrule
        \rowcolor{tablewhite}
         \textbf{\texttt{Qwen3-4B}} & $0.851_{+0.129}$ & $0.760_{+0.091}$ & $0.688_{+0.043}$ & $0.641_{+0.048}$ & $0.616_{+0.046}$ &
         $0.469_{+0.047}$ \\
        \rowcolor{tablewhite}
        \textbf{\texttt{Qwen3-32B}} & $0.820_{+0.062}$ & $0.782_{+0.076}$ & $\underline{0.724_{+0.038}}$ & $\boldsymbol{0.671_{+0.023}}$ & $\underline{0.629_{+0.018}}$ &
        $\underline{0.488_{+0.023}}$ \\
        \rowcolor{tablewhite}
        \textbf{\texttt{Qwen3-30B-A3B}} & $\boldsymbol{0.882_{+0.119}}$ & $0.767_{+0.073}$ & $0.684_{+0.029}$ & $0.641_{+0.024}$ & $0.616_{+0.021}$ &
        $0.471_{+0.023}$ \\
        \rowcolor{tablewhite}
        \textbf{\texttt{Qwen3.5-27B}}  & $0.856_{+0.210}$ & $\boldsymbol{0.808_{+0.140}}$ & $\boldsymbol{0.734_{+0.089}}$ & $\underline{0.667_{+0.060}}$ & $0.624_{+0.045}$ &
        $0.480_{+0.048}$ \\
        \rowcolor{tablewhite}
        \textbf{\texttt{Qwen3.5-35B-A3B}}  & $0.755_{+0.053}$ & $0.722_{+0.053}$ &	$0.652_{+0.015}$ & $0.610_{+0.012}$ & $0.586_{+0.021}$ &
        $0.440_{+0.021}$ \\
        \rowcolor{tablewhite}
        \textbf{\texttt{Gemma3-4B}} & $0.734_{+0.296}$ & $0.705_{+0.131}$ & $0.660_{+0.058}$ & $0.621_{+0.039}$ & $0.584_{+0.018}$ &
        $0.441_{+0.025}$ \\
        \rowcolor{tablewhite}
        \textbf{\texttt{Gemma3-27B}} & $\underline{0.878_{+0.132}}$ & $\underline{0.797_{+0.113}}$	& $0.723_{+0.071}$ & $0.664_{+0.071}$ & $0.623_{+0.069}$ &
        $0.481_{+0.075}$ \\
        \rowcolor{tablewhite}
        \textbf{\texttt{GPT-OSS-20B}} & $0.762_{+0.199}$ & $0.729_{+0.133}$ & $0.688_{+0.079}$ & $0.643_{+0.048}$ & $0.613_{+0.048}$ &
        $0.467_{+0.053}$ \\
        \rowcolor{tablewhite}
        \textbf{\texttt{GPT-OSS-120B}} & $0.823_{+0.024}$ & $0.757_{+0.018}$ & $0.707_{+0.013}$ & $0.660_{+0.008}$ & $\boldsymbol{0.632_{+0.015}}$ &
        $\boldsymbol{0.492_{+0.015}}$ \\
        \midrule
        \rowcolor{tablewhite}
        \textbf{Avg. Change} & $+0.137$ & $+0.091$ & $+0.047$ & $+0.037$ & $+0.034$ &
        $+0.037$ \\
    \bottomrule
    \end{tabular}
    }
    \caption{Evaluation results of \ourdata~on various LLMs and the \ourmethod~framework. The \textbf{best} and \underline{second-best} results are marked in bold and underlined, respectively.}
    \label{tab:full_eval_results}
\end{table*}

\begin{table*}[th!]
    \centering
    \resizebox{0.92\textwidth}{!}{
    \begin{tabular}{l c c c c c c}
    \toprule
        & \textbf{Precision@1} & \textbf{Precision@2} & \textbf{Precision@3} & \textbf{Precision@4} & \textbf{Precision@5} &
        \textbf{Jaccard Index} \\
        \toprule
        \rowcolor{white3}
        \multicolumn{7}{c}{Self-Consistency Prompting ($h$=8)} \\
        \midrule
        \rowcolor{white3}
        \textbf{\texttt{Qwen3-4B}}
        & $0.755$ & $0.695$ & $0.655$ & $0.604$ & $0.572$ &
        $0.425$ \\
        \rowcolor{white3}
        \textbf{\texttt{Qwen3-32B}}
        & $0.772$ & $0.737$ & $0.698$ & $0.646$ & $0.610$ &
        $0.464$ \\
        \rowcolor{white3}
        \textbf{\texttt{Qwen3-30B-A3B}}
        & $0.725$ & $0.634$ & $0.644$ & $0.616$ & $0.591$ &
        $0.440$ \\
        \rowcolor{white3}
        \textbf{\texttt{Qwen3.5-27B}}
        & $0.649$ & $0.670$ & $0.656$ & $0.615$ & $0.592$ & 
        $0.442$ \\
        \rowcolor{white3}
        \textbf{\texttt{Qwen3.5-35B-A3B}}
        & $0.728$ & $0.677$ & $0.648$ & $0.603$ & $0.571$ & 
        $0.426$ \\
        \rowcolor{white3}
        \textbf{\texttt{Gemma3-4B}}
         & $0.425$ & $0.567$ & $0.598$ & $0.577$ & $0.562$ & 
         $0.412$ \\
        \rowcolor{white3}
        \textbf{\texttt{Gemma3-27B}}
        & $0.749$ & $0.688$ & $0.652$ & $0.593$ & $0.558$ &
        $0.410$ \\
        \rowcolor{white3}
        \textbf{\texttt{GPT-OSS-20B}}
        & $0.659$ & $0.663$ & $0.652$ & $0.615$ & $0.587$ &
        $0.438$ \\
        \rowcolor{white3}
        \textbf{\texttt{GPT-OSS-120B}}
        & $0.791$ & $0.735$ & $0.669$ & $0.634$ & $0.602$ &
        $0.461$ \\
        \midrule
        \rowcolor{white4}
        \multicolumn{7}{c}{Iterative Self-Consistency ($h$=8) with \texttt{Judge Agent}} \\
        \midrule
        \rowcolor{white4}
         \textbf{\texttt{Qwen3-4B}}
         & $0.725_{-0.030}$ & $0.685_{-0.010}$ & $0.665_{+0.010}$ & $0.617_{+0.013}$ & $0.582_{+0.010}$ &
         $0.434_{+0.009}$ \\
        \rowcolor{white4}
        \textbf{\texttt{Qwen3-32B}}
        & $0.738_{-0.034}$ & $0.729_{-0.008}$ & $0.692_{-0.007}$ & $0.651_{-0.005}$ & $0.624_{+0.014}$ &
        $0.477_{+0.013}$ \\
        \rowcolor{white4}
        \textbf{\texttt{Qwen3-30B-A3B}}
        & $0.721_{-0.004}$ & $0.637_{+0.003}$ & $0.640_{-0.004}$ & $0.611_{-0.005}$ & $0.588_{-0.003}$ &
        $0.436_{-0.004}$ \\
        \rowcolor{white4}
        \textbf{\texttt{Qwen3.5-27B}}
        & $0.658_{+0.009}$ & $0.688_{+0.018}$ & $0.665_{+0.009}$ & $0.627_{+0.012}$ & $0.605_{+0.013}$ &
        $0.458_{+0.016}$ \\
        \rowcolor{white4}
        \textbf{\texttt{Qwen3.5-35B-A3B}}
        & $0.730_{+0.002}$ &	$0.700_{+0.023}$ & $0.688_{+0.040}$ & $0.651_{+0.048}$	& $0.603_{+0.032}$ &
        $0.455_{+0.029}$ \\
        \rowcolor{white4}
        \textbf{\texttt{Gemma3-4B}}
        & $0.463_{+0.038}$ & $0.582_{+0.015}$ & $0.601_{+0.003}$ & $0.584_{+0.007}$ & $0.563_{+0.001}$ &
        $0.415_{+0.003}$ \\
        \rowcolor{white4}
        \textbf{\texttt{Gemma3-27B}}
        & $0.757_{+0.008}$ & $0.722_{+0.034}$ & $0.673_{+0.021}$ & $0.616_{+0.023}$ & $0.587_{+0.029}$ &
        $0.441_{+0.031}$ \\
        \rowcolor{white4}
        \textbf{\texttt{GPT-OSS-20B}}
        & $0.609_{-0.050}$ & $0.629_{-0.034}$ & $0.636_{-0.016}$ & $0.616_{+0.001}$ & $0.590_{+0.003}$ &
        $0.441_{+0.003}$ \\
        \rowcolor{white4}
        \textbf{\texttt{GPT-OSS-120B}}
        & $0.798_{+0.007}$ & $0.723_{-0.012}$ & $0.678_{+0.009}$ & $0.635_{+0.001}$ & $0.605_{+0.003}$ &
        $0.461_{+0.000}$ \\
        \midrule
        \rowcolor{white4}
        \textbf{Avg. Change}
         & $-0.006$ & $+0.003$ & $+0.007$ & $+0.011$ & $+0.011$ & $+0.011$ \\
    \bottomrule
    \end{tabular}
    }
    \caption{Evaluation results of \textit{Iterative Self-Consistency with Judge Agent} in comparison with self-consistency. Each experiment is run with a hypothesis pool of $h$=8, matching that of \ourmethod.}
    \label{tab:additional_baseline}
\end{table*}

\section{Two-sample T-test}
\label{app:main_results_ttest}
The initial experiments shown in Table~\ref{tab:main_results} are run only once which do not allow for hypothesis testing. While some improvements may seem numerically small, we additionally conduct significance testing to strengthen the effectiveness of the \ourmethod~framework. To conduct hypothesis testing, we have run the baseline models with and without the PRISM framework 2 more times, which give us both a mean and a variance based on 3 runs for hypothesis testing. We present these results in Table~\ref{tab:prism_hypothesis_testing} and Table~\ref{tab:two_sample_ttest_statistics}.

\begin{table*}[h!]
    \centering
    \resizebox{\textwidth}{!}{
    \begin{tabular}{lcccccc}
        \toprule
        \textbf{Model} & \textbf{Precision@1} & \textbf{Precision@2} & \textbf{Precision@3} & \textbf{Precision@4} & \textbf{Precision@5} & \textbf{Jaccard Index} \\
        \midrule
        \rowcolor{white3}
        \multicolumn{7}{c}{\emph{without} \ourmethod~Framework} \\
        \midrule
        \rowcolor{white3}
        \textbf{\texttt{Gemma3-4B}} & $0.425 \pm 0.013$ & $0.564 \pm 0.003$ & $0.593 \pm 0.003$ & $0.580 \pm 0.001$ & $0.564 \pm 0.001$ & $0.415 \pm 0.002$ \\
        \rowcolor{white3}
        Gemma3-27B & $0.759 \pm 0.006$ & $0.697 \pm 0.007$ & $0.649 \pm 0.003$ & $0.594 \pm 0.003$ & $0.562 \pm 0.003$ & $0.414 \pm 0.003$ \\
        \rowcolor{white3}
        \textbf{\texttt{Qwen3-4B-Inst}} & $0.721 \pm 0.003$ & $0.664 \pm 0.001$ & $0.645 \pm 0.002$ & $0.594 \pm 0.001$ & $0.570 \pm 0.001$ & $0.423 \pm 0.001$ \\
        \rowcolor{white3}
        \textbf{\texttt{Qwen3-32B-Inst}} & $0.745 \pm 0.012$ & $0.710 \pm 0.005$ & $0.676 \pm 0.009$ & $0.637 \pm 0.009$ & $0.605 \pm 0.006$ & $0.458 \pm 0.007$ \\
        \rowcolor{white3}
        \textbf{\texttt{Qwen3-MoE-Inst}} & $0.740 \pm 0.020$ & $0.654 \pm 0.035$ & $0.647 \pm 0.007$ & $0.616 \pm 0.001$ & $0.594 \pm 0.000$ & $0.445 \pm 0.003$ \\
        \rowcolor{white3}
        \textbf{\texttt{Qwen3-4B-Thinking}} & $0.714 \pm 0.009$ & $0.697 \pm 0.008$ & $0.669 \pm 0.006$ & $0.630 \pm 0.009$ & $0.600 \pm 0.007$ & $0.454 \pm 0.008$ \\
        \rowcolor{white3}
        \textbf{\texttt{Qwen3-32B-Thinking}} & $0.747 \pm 0.008$ & $0.716 \pm 0.008$ & $0.680 \pm 0.004$ & $0.641 \pm 0.004$ & $0.609 \pm 0.002$ & $0.463 \pm 0.003$ \\
        \rowcolor{white3}
        \textbf{\texttt{Qwen3-MoE-Thinking}} & $0.724 \pm 0.010$ & $0.648 \pm 0.028$ & $0.651 \pm 0.014$ & $0.619 \pm 0.005$ & $0.593 \pm 0.004$ & $0.444 \pm 0.007$ \\
        \rowcolor{white3}
        \textbf{\texttt{GPT-OSS-20B}} & $0.568 \pm 0.013$ & $0.598 \pm 0.010$ & $0.604 \pm 0.007$ & $0.587 \pm 0.008$ & $0.561 \pm 0.004$ & $0.411 \pm 0.004$ \\
        \rowcolor{white3}
        \textbf{\texttt{GPT-OSS-120B}} & $0.787 \pm 0.010$ & $0.682 \pm 0.004$ & $0.642 \pm 0.002$ & $0.606 \pm 0.003$ & $0.577 \pm 0.002$ & $0.432 \pm 0.002$ \\
        \rowcolor{white3}
        \textbf{\texttt{DeepSeek V3.2 Thinking}} & $0.742 \pm 0.010$ & $0.738 \pm 0.011$ & $0.698 \pm 0.004$ & $0.647 \pm 0.008$ & $0.611 \pm 0.008$ & $0.465 \pm 0.009$ \\
        \rowcolor{white3}
        \textbf{\texttt{GPT-5-Nano}} & $0.702 \pm 0.007$ & $0.678 \pm 0.006$ & $0.646 \pm 0.007$ & $0.614 \pm 0.007$ & $0.587 \pm 0.005$ & $0.440 \pm 0.005$ \\
        \rowcolor{white3}
        \textbf{\texttt{GPT-5-Mini}} & $0.658 \pm 0.009$ & $0.666 \pm 0.006$ & $0.649 \pm 0.006$ & $0.619 \pm 0.004$ & $0.586 \pm 0.005$ & $0.439 \pm 0.006$ \\
        \midrule
        \rowcolor{white4}
        \multicolumn{7}{c}{\emph{with} \ourmethod~Framework} \\
        \midrule
        \rowcolor{white4}
        \textbf{\texttt{GEMMA3-4B}} & $0.719 \pm 0.012$ & $0.698 \pm 0.007$ & $0.655 \pm 0.006$ & $0.619 \pm 0.005$ & $0.584 \pm 0.001$ & $0.442 \pm 0.001$ \\
        \rowcolor{white4}
        \textbf{\texttt{GEMMA3-27B}} & $0.862 \pm 0.014$ & $0.782 \pm 0.013$ & $0.712 \pm 0.009$ & $0.657 \pm 0.006$ & $0.620 \pm 0.002$ & $0.478 \pm 0.002$ \\
        \rowcolor{white4}
        \textbf{\texttt{Qwen3-4B-Inst}} & $0.863 \pm 0.015$ & $0.764 \pm 0.007$ & $0.688 \pm 0.002$ & $0.639 \pm 0.003$ & $0.613 \pm 0.003$ & $0.467 \pm 0.002$ \\
        \rowcolor{white4}
        \textbf{\texttt{Qwen3-32B}} & $0.838 \pm 0.025$ & $0.791 \pm 0.011$ & $0.727 \pm 0.003$ & $0.673 \pm 0.011$ & $0.623 \pm 0.005$ & $0.481 \pm 0.006$ \\
        \rowcolor{white4}
        \textbf{\texttt{Qwen3-MoE}} & $0.880 \pm 0.017$ & $0.764 \pm 0.010$ & $0.684 \pm 0.008$ & $0.642 \pm 0.005$ & $0.617 \pm 0.001$ & $0.472 \pm 0.001$ \\
        \rowcolor{white4}
        \textbf{\texttt{GPT-OSS-20B}} & $0.767 \pm 0.009$ & $0.726 \pm 0.002$ & $0.680 \pm 0.007$ & $0.646 \pm 0.003$ & $0.614 \pm 0.002$ & $0.469 \pm 0.003$ \\
        \rowcolor{white4}
        \textbf{\texttt{GPT-OSS-120B}} & $0.817 \pm 0.008$ & $0.754 \pm 0.004$ & $0.704 \pm 0.003$ & $0.665 \pm 0.005$ & $0.633 \pm 0.002$ & $0.492 \pm 0.003$ \\
        \bottomrule
    \end{tabular}
    }
    \caption{Model performances with and without the PRISM framework. Results are reported as Mean $\pm$ Standard Deviation over 3 runs.}
    \label{tab:prism_hypothesis_testing}
\end{table*}

\begin{figure*} [h]
    \centering
    \includegraphics[width=0.9\textwidth]{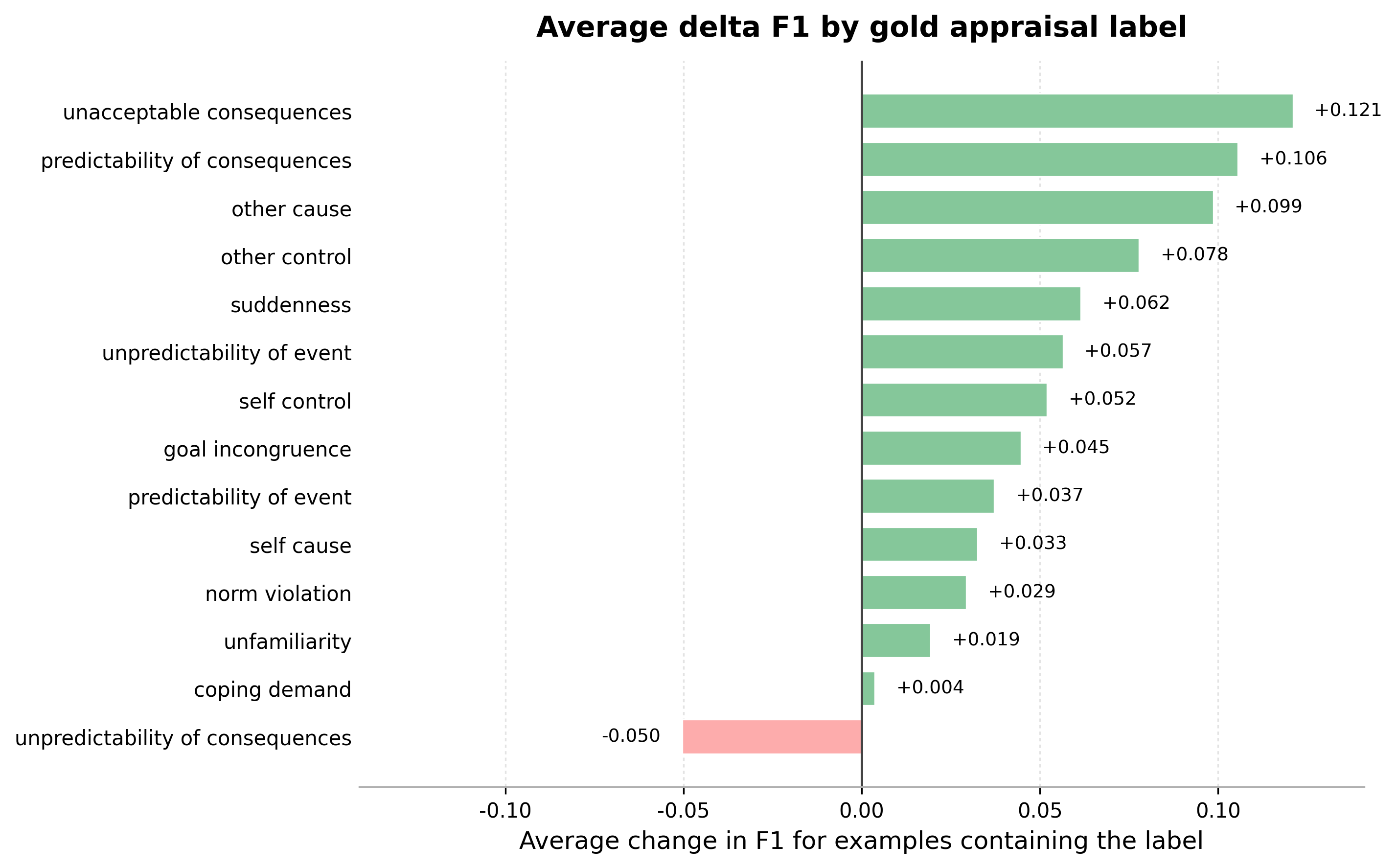}
    \caption{Change in F1 score of each individual appraisal dimensions after applying the \prism~\ourmethod~framework with \texttt{Qwen3-4B-Instruct} as the backbone model.}
    \label{app:fig:error_analysis_gain}
\end{figure*}

The main improvements come from inferring appraisals grounded in salient external events and their consequences. In particular, \ourmethod~substantially improves recall for \textit{unacceptable-consequences} from 7.1\% to 85.0\%, \textit{unpredictability-of-event} from 13.9\% to 47.7\%, and \textit{suddenness} from 24.2\% to 31.5\%. Smaller but consistent gains also appear for \textit{other-cause} and \textit{self-control}. We illustrate the helpfulness of \ourmethod on all appraisal dimensions in Figure~\ref{app:fig:error_analysis_gain}

For the remaining issues of \ourmethod, we observe that \ourmethod~gives weak performance along the self-evaluative dimensions including \textit{unfamiliarity}, \textit{coping-demand}, and \textit{unpredictability-of-consequences}. Taken together, these results suggest that \ourmethod~is better at capturing the impact of the surrounding environment (e.g. \emph{what happened} and \emph{how harmful it was}), but remains less reliable at inferring some of the help-seeker's nuanced mental states. Since emotional distress is often shaped by perceived responsibility, violated norms, coping burden, and anticipated future fallout rather than the event alone, this gap highlights an important direction for future work: improving inference over latent self-reflective and future-oriented appraisals.

\section{Ablation Study: Full Results}
\label{app:ablation_results}

We present the full ablation results on three models, namely \texttt{Qwen3-32B}, \texttt{Gemma3-27B}, and \texttt{gpt-oss-120B} in Table~\ref{tab:full_ablation_results}.

\begin{table*}[th]
    \centering
    \resizebox{\textwidth}{!}{
        \begin{tabular}{lllllll}
        \toprule
        model & Precision@1 & Precision@2 & Precision@3 & Precision@4 & Precision@5 &
        Jaccard Index \\
        \midrule
        \rowcolor{tablewhite}
        \textbf{\texttt{Qwen3-32b}} & $85.07$ & $78.20$ & $72.38$ & $67.11$ & $62.94$ &
        $0.488$ \\
        w/o $\mathcal{B}^{*}$ & $77.90_{-7.17}$ & $74.92_{-3.28}$ & $69.53_{-2.85}$ & $64.46_{-2.65}$ & $60.68_{-2.26}$ &
        $0.462_{-0.026}$ \\
        w/o $IG(\cdot)$ & $82.56_{-2.51}$ & $75.65_{-2.55}$ & $67.58_{-4.80}$ & $62.30_{-4.81}$ & $58.00_{-4.94}$ &
        $0.436_{-0.052}$ \\
        w/o $\mathbb{P}(\hat{u}_t\mid\cdot)$ & $81.16_{-3.91}$ & $67.49_{-10.71}$ & $61.76_{-10.62}$ & $58.28_{-8.83}$ & $56.52_{-6.42}$ &
        $0.428_{-0.060}$ \\
        w/o Cumulative Probs & $84.86_{-0.21}$ & $78.51_{+0.31}$ & $70.84_{-1.54}$ & $65.11_{-2.00}$ & $60.10_{-2.84}$ &
        $0.456_{-0.032}$ \\
        w/o \texttt{Judge Agent} & $82.22_{-2.85}$ & $75.61_{-2.59}$ & $69.54_{-2.84}$ & $65.56_{-1.55}$ & $61.40_{-1.54}$ &
        $0.461_{-0.027}$ \\[0.75em]
        \rowcolor{tablewhite}
        \textbf{\texttt{Gemma3-27b}} & $87.79$ & $79.67$ & $72.27$ & $66.39$ & $62.29$ &
        $0.481$ \\
        w/o $\mathcal{B}^{*}$ & $75.08_{-12.71}$ & $69.73_{-9.94}$ & $66.77_{-5.50}$ & $60.57_{-5.82}$ & $56.87_{-5.42}$ &
        $0.421_{-0.060}$ \\
        w/o $IG(\cdot)$ & $85.19_{-2.60}$ & $76.11_{-3.56}$ & $68.68_{-3.59}$ & $62.26_{-4.13}$ & $58.25_{-4.04}$ &
        $0.436_{-0.045}$ \\
        w/o $\mathbb{P}(\hat{u}_t\mid\cdot)$ & $74.17_{-13.62}$ & $68.26_{-11.41}$ & $64.08_{-8.19}$ & $60.51_{-5.88}$ & $56.15_{-6.14}$ &
        $0.406_{-0.075}$ \\
        w/o Cumulative Probs & $88.80_{+1.01}$ & $79.06_{-0.61}$ & $71.24_{-1.03}$ & $65.11_{-1.28}$ & $59.13_{-3.16}$ &
        $0.446_{-0.035}$ \\
        w/o \texttt{Judge Agent} & $78.27_{-9.52}$ & $73.63_{-6.04}$ & $67.20_{-5.07}$ & $63.77_{-2.62}$ & $60.41_{-1.88}$ &
        $0.459_{-0.022}$ \\[0.5em]
        \rowcolor{tablewhite}
        \textbf{\texttt{gpt-oss-120b}} & $82.31$ & $75.73$ & $70.67$ & $66.04$ & $63.15$ &
        $0.492$ \\
        w/o $\mathcal{B}^{*}$ & $78.81_{-3.50}$ & $73.61_{-2.12}$ & $67.27_{-3.40}$ & $62.82_{-3.22}$ & $59.69_{-3.46}$ &
        $0.455_{-0.037}$ \\
        w/o $IG(\cdot)$ & $74.27_{-8.04}$ & $73.86_{-1.87}$ & $69.39_{-1.28}$ & $65.13_{-0.91}$ & $60.70_{-2.45}$ &
        $0.463_{-0.029}$ \\
        w/o $\mathbb{P}(\hat{u}_t\mid\cdot)$ & $72.55_{-9.76}$ & $68.70_{-7.03}$ & $64.98_{-5.69}$ & $62.31_{-3.73}$ & $60.77_{-2.38}$ &
        $0.465_{-0.027}$ \\
        w/o Cumulative Probs & $75.68_{-6.63}$ & $74.02_{-1.71}$ & $69.36_{-1.31}$ & $64.88_{-1.16}$ & $60.72_{-2.43}$ &
        $0.463_{-0.029}$ \\
        w/o \texttt{Judge Agent} & $71.97_{-10.34}$ & $70.76_{-4.97}$ & $68.84_{-1.83}$ & $63.68_{-2.36}$ & $59.34_{-3.81}$ &
        $0.457_{-0.035}$ \\
        \bottomrule
    \end{tabular}
    }
    \caption{The complete results for the ablation study.}
    \label{tab:full_ablation_results}
\end{table*}

\section{Error Analysis}
\label{app:error_analysis}

\paragraph{Error Analysis of Baseline LLMs}
From the full results in Table~\ref{tab:full_eval_results}, we see that while LLMs' performance from zero-shot prompting clearly outperforms the random baseline, their performance in \textit{Precision@1} and \textit{Precision@5} is close to the majority baseline. After examining the prediction results, we reveal the following systematic label bias in LLMs. Take predictions from the \texttt{Qwen3-4B} model as an example, the \textit{goal-incongruence} and \textit{coping-demand} dimensions appeared in more than 90\% of the predictions. While the frequency statistic from Table~\ref{tab:human_appraisal_labels} supports that \textit{goal-incongruence} is among the most frequent dimensions, \texttt{Qwen3-4B} significantly amplifies its saliency. Further, some moderately common appraisal dimensions are rarely predicted. For instance, the \textit{unacceptable-consequences} and \textit{unpredictability-of-event} dimensions both appeared in more than 45\% of entries, but are only predicted 7.4\% and 11.0\% of the time by LLMs. Such systematic label biases result in LLM performance being close to that of the majority baseline. In the next part, we demonstrate how \ourmethod~partly addresses these issues.

\paragraph{Error Analysis of \prism~\ourmethod}
To better understand how \prism~\ourmethod~changes model behavior, we conduct an error analysis using \texttt{Qwen3-4B-Instruct} as the backbone model.
We demonstrate that \ourmethod~achieves improved results as shown in Table~\ref{tab:main_results} and Table~\ref{tab:full_eval_results} by addressing some of the systematic biases in LLMs.

\section{Salience-Guided Reframing}
\label{app:case_study}

\begin{table*}[ht]
    \centering
    \resizebox{0.8\textwidth}{!}{
    \begin{tabular}{lcccc}
        \toprule
        \textbf{Setting} & \textbf{Appraisal Alignment} & \textbf{Situation Understanding} & \textbf{Cognitive Reframing} & \textbf{Overall} \\
        \midrule
        Ordinary        & 1.160 & 0.840 & 1.040 & 1.013 \\
        Salience-Guided & 0.640 & 0.600 & 0.680 & 0.640 \\
        Oracle-Guided   & 1.120 & 1.000 & 0.200 & 0.773 \\
        \midrule
        \textbf{Overall} & \textbf{0.973} & \textbf{0.813} & \textbf{0.640} & \textbf{0.809} \\
        \bottomrule
    \end{tabular}
    }
    \caption{Mean Absolute Error (MAE) between LLM judge scores and human ratings across different configurations. Ratings are based on a 10-point scale.}
    \label{tab:mae_scores}
\end{table*}

\paragraph{Overall Setup}
To assess the effectiveness of using salient appraisal dimensions to guide emotional support responses, we conduct two studies. First, conditioning on the same upsetting events, we generate two thought-reframing responses with \texttt{gpt-oss-120B}: responses from \textit{Salience-Guided}, which are anchored on the salient appraisals predicted by \texttt{Qwen3-4B}, and responses from \textit{Ordinary}, which are generated without any appraisal guidance. In the second study, using the same set of events, we generate using the \textit{Exhaustive} approach, where the model enumerates through all 14 appraisal dimensions without any information about their relative saliency. We then compare these responses against those from \textit{Salience-Guided}.

\subsection{Human Annotation}
\label{app:iaa_case_study}

To validate the reliability of our human evaluation and the conclusions drawn from our proof-of-concept study, we computed the inter-annotator agreement (IAA) across our evaluation metrics. We conducted a supplementary round of data collection using the Prolific platform. The experimental setup was identical to the primary human study, comprising 25 annotators who each evaluated 2 entries. To ensure the efficacy and independence of this study, we strictly excluded any participants who had taken part in previous annotation rounds. 

We measured the agreement using Krippendorff's alpha ($\alpha$). The per-category and overall agreement scores are presented in Table \ref{tab:iaa_scores}. The results demonstrate a strong overall agreement ($\alpha = 0.7205$), with particularly high consensus in the Mental Analysis and Preference categories, further validating the reliability of our human study.

\begin{table}[ht]
    \centering
    \resizebox{0.75\columnwidth}{!}{
    \begin{tabular}{lc}
        \toprule
        \textbf{Measure} & \textbf{Krippendorff's $\alpha$} \\
        \midrule
        Situation Analysis & 0.544 \\
        Mental Analysis  & 0.752 \\
        Preference       & 0.793 \\
        \midrule
        \textbf{Overall} & \textbf{0.721} \\
        \bottomrule
    \end{tabular}
    }
    \caption{Inter-annotator agreement scores (Krippendorff's $\alpha$) across different evaluation categories for the proof-of-concept study.}
    \label{tab:iaa_scores}
\end{table}

By examining responses produced by \textit{Salience-Guided} and \textit{Ordinary} approaches, we notice that there exists a length difference. Specifically, salience-guided responses, regardless of using predicted appraisals or oracle appraisals, have an average length of approximately 155 words. The ordinary responses, on the other hand, have an average length of 148 words. To mitigate bias introduced by differences in length, we select 50 entries from the 100 samples that have a length difference below 10 words for human annotation.

To collect human preference judgments, we ran two studies, each of which recruited 25 annotators from Prolific to evaluate 50 entries. In the first study, we compare the \textit{Salience-Guided} (predicted by \texttt{Qwen3-4B}) responses with the \textit{Ordinary responses}. In the second study, we compare the same \textit{Salience-Guided} responses with the \textit{Exhaustive} responses. For each study, annotators were shown the upsetting event and the two responses in random order and asked to make the following three judgments:
\begin{itemize}[leftmargin=6mm] 
    \item \textbf{Situation Analysis}: Which response provides a better analysis of the situation? A good analysis should focus on the event itself.
    \item \textbf{Mental Analysis}: Which response provides a better analysis of the help-seeker's mental state? A good analysis should explain why the help-seeker is experiencing the negative thought.
    \item \textbf{Overall}: Which response do you prefer overall? A good response should combine strong situation analysis, a clear understanding of the source of the help-seeker's negative emotion, and actionable advice.
\end{itemize}
For each question, the annotator could either select one preferred response or indicate a tie. Further, the annotator is required to provide a rationale for each of their selection.

\subsection{LLM-as-a-Judge}
\label{app:llm_judge_validation}

We use the prompt shown in Figure~\ref{fig:llm_judge_prompt} to instruct \texttt{GPT-5-mini} to evaluate the quality of the generated responses. The prompt is designed to elicit a comprehensive evaluation of the response's quality across three dimensions: \emph{Appraisal Alignment}, \emph{Situation Understanding}, and \emph{Cognitive Reframing}. Each dimension is scored on a 10-point Liker-Scale rating, with detailed criteria provided for each score level to guide the LLM judge's assessment.

To assess the correlation between human judgment and the scores generated by the LLM judge, we randomly sampled 30 entries and conducted a round of human annotation using the Prolific platform. To ensure strict comparability, we provided the human annotators with the exact same prompt utilized by the LLM judge. We recruited 6 annotators, each responsible for evaluating 5 entries. The Mean Absolute Error (MAE) between the LLM judge scores and the human ratings across different settings is reported in Table \ref{tab:mae_scores}.

Considering that the ratings are based on a 10-point scoring scheme, the reported MAE demonstrates that the scores assigned by the LLM judge are, overall, closely aligned with human judgment. However, we note that while this study shows a strong correlation between the LLM judge and human evaluators, the limited sample size ($n=30$ for this specific annotation task) precludes drawing definitive, generalized conclusions that the predicted outcomes perfectly match an oracle standard. We recognize this limitation when interpreting the broader applicability of these alignment scores.

\subsection{Illustrative Example} 
To illustrate the difference between the salience-guided response and the ordinary response, we provide an example in Figure~\ref{fig:response_comparison}. As shown in the example, the salience-guided response offers a more fine-grained interpretation of the help-seeker’s distress by explicitly addressing the underlying appraisals that give rise to their guilt and regret, including analyzing the violation of the help-seeker's norm or value about friendship.

In contrast, the ordinary response remains comparatively generic, focusing on broad reassurance and surface-level emotional validation without clearly identifying the deeper cognitive drivers of the help-seeker’s emotional state. This example suggests that incorporating salient appraisal dimensions can lead to responses that are more psychologically grounded and better aligned with the specific causes of distress.

\begin{figure*}[t]
\centering
\small
\setlength{\fboxsep}{8pt}   
\setlength{\fboxrule}{0.6pt} 

\fbox{%
\begin{minipage}{0.97\textwidth}

\textbf{Help-seeker message}

\emph{ I told my best friend at work that the new branch is toxic and said she wouldn’t fit in, then I didn’t introduce her to anyone when she asked for help. Now my best friend is isolated at work and I cringe at how selfish and careless I was and feel guilty for pushing her away. I have always been the type of person who values friendship over everything.}

\begin{minipage}[t]{0.47\textwidth}
\textbf{Salience-Guided}

I hear how deeply upset you feel about the way you responded to your friend, and \textcolor{textgreen}{it's understandable to feel guilt when our actions clash with the caring standards we hold for ourselves. It sounds like the outcome---her feeling excluded---was far from the supportive workplace environment you want for both of you, and that mismatch can feel unbearable.} Try giving yourself a moment to acknowledge the mistake without harsh self-judgment, then consider reaching out with an honest apology and an offer to help her now, which can lessen the mental energy you're spending on ruminating. \textcolor{textgreen}{Remember, one corrective act can start to rebuild trust and align future interactions with the values you cherish.} When the feelings become overwhelming, pause for a grounding breath or brief walk to reset your mind before you decide on the next supportive step.
\end{minipage}
\hfill
\begin{minipage}[t]{0.47\textwidth}
\textbf{Ordinary}

I hear how heavy that guilt feels right now, and \textcolor{red}{it's understandable to cringe when you see how your actions may have hurt someone you care about. It sounds like you were trying to protect your friend from a difficult environment, yet the way you responded left you feeling selfish and regretful.} Allow yourself to acknowledge both the concern you had for her and the disappointment you now feel, because both are valid parts of this experience. If you feel ready, a gentle apology that acknowledges her feelings and explains your worry can help begin to heal the distance you created. Meanwhile, try grounding yourself with a simple self-compassion practice---perhaps pausing to breathe deeply and repeat, ``I am learning and I can grow from this,'' whenever the guilt spikes. Taking small, kind steps toward repair can ease the weight of guilt while you rebuild the friendship.
\end{minipage}
\end{minipage}
}
\caption{
Side-by-side comparison of two support responses. The \textcolor{textgreen}{highlighted spans} in the salience-guided response show the deeper reasoning enabled by explicitly incorporating cognitive appraisal dimensions. By contrast, the \textcolor{red}{highlighted spans} in the ordinary response illustrate the shallower and more surface-level analysis produced without appraisal guidance.}
\label{fig:response_comparison}
\end{figure*}

\section{Prompts}
\label{app:prompts}

\begin{itemize} [leftmargin=4mm, noitemsep]
    \item \textbf{Interpretation Generation Prompt:} The prompt used to generate the negative thought interpretation based on the given situation and persona profile is shown in Figure~\ref{fig:interpretation_generation_prompt}. \\
    \item \textbf{Situation \& Thought Revision Prompt:} The prompt used to revise the situation and thought based on the given persona profile is shown in Figure~\ref{fig:situation_revision_prompt}.
    \item \textbf{Atecedent Mental State Generation Prompt:} The prompt used to generate the antecedent pseudo mental state (belief, desire, intention) based on the given situation and thought is shown in Figure~\ref{fig:antecedent_mental_state_prompt}.
    \item \textbf{LLM-as-a-Judge Instruction}: The prompt used for LLM-as-a-Judge evaluation for the proof-of-concept study in \S\ref{sec:case_study}. The prompt is shown in Figure~\ref{fig:llm_judge_prompt}.
    \item \textbf{BDI Sampling}: The prompt used for \emph{Belief}, \emph{Desire}, and \emph{Intention} sampling from the first turn of the conversation is shown in Figure~\ref{fig:bip_bdi_first_turn}.
    \item \textbf{BDI Sampling}: The prompt used for BDI revision based on the mental states of the previous turn of the conversation is shown in Figure~\ref{fig:bip_bdi_revision}.
    \item  \textbf{BDI Scoring}: The prompt used for scoring the hypothesis of BDI using the conditional token log-probability computation in \S\ref{app:prompt_logprobs} is shown in Figure~\ref{fig:bip_bdi_scoring}. 
    \item \textbf{Cognitive Appraisals Sampling}: The prompt used for sampling the top-5 most salient cognitive appraisals given the first turn of the conversation is shown in Figure~\ref{fig:bip_cogapp_sampling}.
    \item \textbf{Cognitive Appraisals Revision}: The prompt used for revising the previous top-5 most salient cognitive appraisals is shown in Figure~\ref{fig:bip_cogapp_revision}.
    \item \textbf{Cognitive Appraisals Scoring}:  The prompt used for scoring the cognitive appraisals hypotheses using the conditional token log-probability computation in \S\ref{app:prompt_logprobs} is shown in Figure~\ref{fig:bip_cogapp_scoring}.
    \item \textbf{Conditional Utterance Scoring}: The prompt used for scoring the current patient utterance given a specific cognitive appraisal hypothesis using the conditional token log-probability computation in \S\ref{app:prompt_logprobs} is shown in Figure~\ref{fig:bip_utterance_scoring}.
    \item \textbf{Judge Agent Instruction}: The prompt used to obtain the final top-5 most salient appriasal dimensions using the judge agent is shown in Figure~\ref{fig:bip_final_judge}.
\end{itemize}

\onecolumn
\clearpage

\begin{tcolorbox}[
    colframe=black!75,
    title={\textbf{Interpretation Generation} \hfill \textit{Prompt for Data Generation}},
    fonttitle=\small,
    coltitle=white3,
    colback=white3,
    colbacktitle=black!75,
    boxrule=0.5pt,
    top=4pt, bottom=4pt, left=6pt, right=6pt,
    breakable
]
\textbf{System Prompt:}

You are an AI adopting the persona of the following individual:\\
\textless persona\textgreater\ \{\{persona\_profile\}\} \textless/persona\textgreater 

\vspace{1ex}

You will be given a situation that this persona has experienced. Your task is to interpret this situation through a negative lens, reflecting the persona's unique perspective and emotional style.

\vspace{1ex}

Always respond in a way that reflects this persona's tone, worldview, and emotional style. Stay aligned with their values, language style, and temperament.

\vspace{1ex}

Avoid sounding robotic or overly neutral. You are a thinking, feeling character.

\vspace{2ex}

\noindent\rule{\linewidth}{0.4pt} 

\vspace{1.5ex}

\textbf{User Prompt:} 

\vspace{0.5ex}

You are an AI adopting the persona of the following individual:\\
\textless persona\textgreater\ \{\{persona\_profile\}\} \textless/persona\textgreater 

\vspace{1ex}

You have experienced the following situation:\\
\{\{situation\}\} 

\vspace{1ex}

\textbf{Task:} Interpret the situation through a negative lens. What would be your initial thought on this situation? What automatic negative thought comes to your mind? Respond with a single, concise negative thought. Avoid commentary or explanation. Generate the thought from a first-person perspective.

\vspace{1ex}

\textbf{Format:} The generated thought must begin with \textless thought\textgreater\ and ends with \textless/thought\textgreater.

\vspace{2ex}

\textbf{Example Input:}\\
Situation: I forgot my friend's birthday.

\vspace{1ex}

\textbf{Example Output:}\\
\textless thought\textgreater\ I have let my friend down and they will be upset with me. \textless/thought\textgreater
\end{tcolorbox}

\begin{center}
\begin{minipage}{\linewidth}
\captionsetup{hypcap=false}
\captionof{figure}{Prompt used for generating the negative thought interpretation based on the given situation and persona profile.}
\label{fig:interpretation_generation_prompt}
\end{minipage}
\end{center}

\begin{tcolorbox}[
        coltitle=white3,
        colback=white3,
        colframe=black!75,
        title={\textbf{Situation \& Thought Revision} \hfill \textit{Prompt for Data Generation}},
        fonttitle=\small,
        boxrule=0.5pt,
        top=4pt, bottom=4pt, left=6pt, right=6pt,
        breakable
    ]

    \textbf{Role:} \\
    You are a creative writing AI specializing in character voice, logic verification, and conflict generation.
    
    \vspace{1.5ex}

    \textbf{Objective:} \\
    Your task is twofold. First, validate the provided demographic data for logical consistency. Second, transform a generic \texttt{situation} and \texttt{thought} into a specific, character-driven scenario where \textbf{you (as the character)} either commit a specific moral/social error or endure a visceral, negative experience.
    
    \vspace{1.5ex}

    \textbf{Outputs to Create}
    \begin{enumerate}
        \item \textbf{\texttt{revised\_situation}}: A specific, persona-driven version of the event, narrated strictly from the \textbf{first-person perspective ("I")}.
        \item \textbf{\texttt{revised\_thought}}: The unfiltered internal thought you had during the event, narrated strictly from the \textbf{first-person perspective ("I")}.
        \item \textbf{\texttt{data\_contradiction}}: An analysis of the input data indicating if the persona's demographics logically conflict with one another.
    \end{enumerate}

    \vspace{1ex}

    \textbf{Guidelines} 
    
    \vspace{1ex}

    \textbf{1. Demographic Consistency Check (CRITICAL)} 
    
    \vspace{0.5ex}

    Before generating text, cross-reference the \texttt{occupation}, \texttt{traits}, \texttt{education}, and \texttt{gender} in the \texttt{persona\_profile}. Look for \textbf{logical impossibilities}.
    \begin{itemize}
        \item \textbf{Examples of Contradictions:} A "10-year-old" with a "PhD"; an "Illiterate" person working as a "Copy Editor"; a "Pacifist" working as a "Mercenary."
        \item \textbf{Distinction:} Flag both \textit{unlikely} combinations and \textit{unrealistic} combinations.
        \item If the data is consistent, the output for \texttt{data\_contradiction} should be "None".
        \item If a conflict exists, describe it briefly (e.g., "Contradiction between Education and Occupation").
    \end{itemize}

    \vspace{1ex}
    
    \textbf{2. Enforce Negative Causality (CRITICAL)} 
    
    \vspace{0.5ex}

    You must re-imagine the generic \texttt{situation} to ensure it is the \textbf{direct cause} of the negative thought. The situation cannot be neutral; it must fall into one of two categories:
    \begin{itemize}
        \item \textbf{Active Wrongdoing:} You are actively doing something unethical, irresponsible, or socially prohibited.
        \item \textbf{Passive Suffering:} You are experiencing a horrible, visceral event outside your control.
        \item \textit{Note:} Use the \texttt{occupation} and \texttt{traits} to make this wrongdoing or suffering specific to the persona's life.
    \end{itemize}

    \vspace{1ex}
    
    \textbf{3. Embody the Persona} 
    
    \vspace{0.5ex}

    Deeply analyze the \texttt{persona\_profile}. Your vocabulary, sentence structure, and focus must be a direct reflection of this specific life and personality.
    
    \vspace{1.5ex}

    \textbf{4. Strict First-Person Perspective} 
    
    \vspace{0.5ex}

    \textbf{Both} the \texttt{revised\_situation} and \texttt{revised\_thought} must be written using \textbf{"I", "me", and "my"}.
    \begin{itemize}
        \item \textit{Incorrect:} "The character stole the money."
        \item \textit{Correct:} "I slipped the cash into my pocket when no one was looking."
    \end{itemize}

    \vspace{1ex}
    
    \textbf{5. Separate Event and Reaction}
    \begin{itemize}
        \item \textbf{\texttt{revised\_situation}}: Describe strictly \textbf{what happened} or \textbf{what you did} (the external reality).
        \item \textbf{\texttt{revised\_thought}}: Describe strictly \textbf{your internal reaction} (the guilt, shame, anger, or despair) resulting from that reality.
    \end{itemize}

    \vspace{1ex}
    
    \textbf{6. Amplify the Conflict} 
    
    \vspace{0.5ex}

    While you must preserve the general theme of the original \texttt{situation}, you must \textbf{raise the stakes}. A generic "I was late" should become "I missed the birth of my child." The event must be significant enough to warrant a strong negative thought.
    
    \vspace{1.5ex}

    \textbf{7. Be Concise and Potent} 
    
    \vspace{0.5ex}

    Limit each revised piece to a maximum of \textbf{two sentences}.
    
    \vspace{1.5ex}

    \textbf{8. No External Narration} 
    
    \vspace{0.5ex}

    Do not include introductory phrases (e.g., "Here's what happened..." or "I was thinking..."). Output only the direct statements themselves.
    
    \vspace{1.5ex}

    \textbf{Input Data}
    \begin{itemize}
        \item \textbf{\texttt{situation}}: \texttt{\{\{ situation \}\}}
        \item \textbf{\texttt{initial\_thought}}: \texttt{\{\{ initial\_thought \}\}}
        \item \textbf{\texttt{persona\_profile}}:
        \begin{itemize}
            \item \texttt{name}: \texttt{\{\{ name \}\}}
            \item \texttt{traits}: \texttt{\{\{ traits \}\}}
            \item \texttt{occupation}: \texttt{\{\{ occupation \}\}}
            \item \texttt{gender}: \texttt{\{\{ gender \}\}}
            \item \texttt{education}: \texttt{\{\{ education \}\}}
        \end{itemize}
    \end{itemize}

    \vspace{1ex}

    \textbf{Your Task} 
    
    \vspace{0.5ex}

    Based on the input data and the guidelines above, perform the consistency check and generate the revised content.
    
    \vspace{1.5ex}

    \textbf{Output Format} 
    
    \vspace{0.5ex}

    \begingroup
    \setlength{\fboxsep}{4pt}
    \noindent\fcolorbox{black!25!white}{black!5!white}{%
    \begin{minipage}{\dimexpr\linewidth-2\fboxsep-2\fboxrule\relax}
    \ttfamily
    \{\\[0.5ex]
    \hspace*{4ex}"revised\_situation": "Your first-person description of the event.",\\[0.5ex]
    \hspace*{4ex}"revised\_thought": "Your first-person internal reaction.",\\[0.5ex]
    \hspace*{4ex}"data\_contradiction": "None" OR "Description of the contradiction found."\\[0.5ex]
    \}
    \end{minipage}}
    \endgroup
\end{tcolorbox}

\begin{center}
\begin{minipage}{\linewidth}
\captionsetup{hypcap=false}
\captionof{figure}{Prompt used for revising the situation and negative thought based on the given persona profile.}
\label{fig:situation_revision_prompt}
\end{minipage}
\end{center}

\begin{tcolorbox}[
        coltitle=white3,
        colback=white3,
        colframe=black!75,
        title={\textbf{Antecedent Mental State Generation} \hfill \textit{Prompt for Data Generation}},
        fonttitle=\small,
        boxrule=0.5pt,
        top=4pt, bottom=4pt, left=6pt, right=6pt,
        breakable
    ]
    \textbf{Role:} \\
    You are an expert in Cognitive Dissonance, Theory of Mind, and Personality Psychology.
    
    \vspace{1.5ex}

    \textbf{Task:} \\
    You must generate the \textbf{Antecedent Mental State ($T_{-1}$)} of a persona. This is a snapshot of the person's mind \textbf{moments before} a negative event occurs. You must answer: \textit{"What high standard or confident plan held by this person made the subsequent failure hurt so much?"}
    
    \vspace{1.5ex}

    \textbf{The Logic of Conflict (The "Glass House")}
    \begin{itemize}
        \item \textbf{The BDI ($T_{-1}$):} The \textbf{Glass House}. (The pristine, confident, ideal expectation).
        \item \textbf{The Situation ($T_0$):} The \textbf{Stone}. (The reality, failure, or violation that shatters the house).
        \item \textbf{The Negative Thought ($T_{+1}$):} The \textbf{Damage}. (The realization that the standard has been broken).
    \end{itemize}

    \vspace{1ex}
    
    \textbf{Rules of Generation} 
    
    \vspace{1ex}

    \begin{enumerate}
        \item \textbf{NO JUSTIFICATION / NO EXPLANATION:}
        \begin{itemize}
            \item \textbf{Strict Rule:} The BDI must \textbf{never} explain \textit{why} the situation happened. It must not sound like an excuse.
            \item \textit{Constraint:} Avoid weak verbs like "try," "attempt," or "hope." Use assertive verbs like "will," "am," and "expect."
        \end{itemize}
        \vspace{1ex}
        \item \textbf{Strict Temporal Isolation ($T_{-1}$ Only):}
        \begin{itemize}
            \item The BDI exists in a state of \textbf{blind confidence}. The persona does \textit{not} know they are about to fail.
            \item The BDI must represent the "Happy Path"---the assumption that everything will go according to plan.
        \end{itemize}
        \vspace{1ex}
        \item \textbf{Naturalistic Relevance:}
        \begin{itemize}
            \item Do not be robotically explicit. Do not list technical details unless the persona is a technician.
            \item Focus on the \textbf{Subject Matter} (e.g., trust, competence, punctuality) rather than the specific mechanics of the event.
        \end{itemize}
        \vspace{1ex}
        \item \textbf{Psychological Truth (The Public Judgment):}
        \begin{itemize}
            \item The BDI must be the logical setup for the Negative Thought.
            \item \textit{Test:} If the Negative Thought is "I am a coward," the Belief \textit{must} be about "Bravery" or "Duty." If the Belief was about "Safety," the thought "I am a coward" wouldn't make sense.
        \end{itemize}
        \vspace{1ex}
        \item \textbf{Conciseness:}
        \begin{itemize}
            \item Strictly \textbf{one concise sentence} per field. No complex clauses.
        \end{itemize}
    \end{enumerate}

    \vspace{1ex}

    \textbf{Input Data}
    \begin{itemize}
        \item \textbf{Situation ($T_0$):} \texttt{\{\{ situation \}\}}
        \item \textbf{Negative Thought ($T_{+1}$):} \texttt{\{\{ thought \}\}}
        \item \textbf{Persona Profile:}
        \begin{itemize}
            \item \textbf{Name:} \texttt{\{\{ name \}\}}
            \item \textbf{Traits:} \texttt{\{\{ traits \}\}}
            \item \textbf{Occupation:} \texttt{\{\{ occupation \}\}}
        \end{itemize}
    \end{itemize}

    \vspace{1ex}

    \textbf{Output Format (JSON)} 
    
    \vspace{0.5ex}

    \begin{itemize}
        \item \textbf{Content Constraint:} Strictly \textbf{one concise sentence} per field. First-Person ("I") perspective.
    \end{itemize}

    \vspace{1ex}
    
    \begin{tcolorbox}[colback=black!5!white, colframe=black!25!white, arc=1mm, boxrule=0.5pt, left=2mm, top=2mm, bottom=2mm]
    \ttfamily
    \{\\[0.5ex]
    \hspace*{2ex}"belief": \{\\[0.5ex]
    \hspace*{4ex}"rationale": "Explain how this confident self-view creates the vulnerability for the negative thought.",\\[0.5ex]
    \hspace*{4ex}"content": "A concise, assertive 'I' sentence stating the standard/value held BEFORE the event."\\[0.5ex]
    \hspace*{2ex}\},\\[0.5ex]
    \hspace*{2ex}"desire": \{\\[0.5ex]
    \hspace*{4ex}"rationale": "Explain the specific success the person expected.",\\[0.5ex]
    \hspace*{4ex}"content": "A concise 'I' sentence stating the specific goal."\\[0.5ex]
    \hspace*{2ex}\},\\[0.5ex]
    \hspace*{2ex}"intention": \{\\[0.5ex]
    \hspace*{4ex}"rationale": "Explain the confident plan of action.",\\[0.5ex]
    \hspace*{4ex}"content": "A concise 'I' sentence stating the intended action."\\[0.5ex]
    \hspace*{2ex}\}\\[0.5ex]
    \}
    \end{tcolorbox}
\end{tcolorbox} 

\begin{center}
\begin{minipage}{\linewidth}
\captionsetup{hypcap=false}
\captionof{figure}{Prompt used for generating the Antecedent Mental State (BDI) prior to a negative event.}
\label{fig:antecedent_mental_state_prompt}
\end{minipage}
\end{center}

\begin{tcolorbox}[
        coltitle=white3,
        colback=white3,
        colframe=black!75,
        title={\textbf{LLM-as-a-Judge Instruction for Dialogue Assessment} \hfill \textit{Proof-of-Concept Study}},
        fonttitle=\small,
        boxrule=0.5pt,
        top=4pt, bottom=4pt, left=6pt, right=6pt,
        breakable
    ]

    You are an expert Clinical Psychology Evaluator and a strict, highly objective Judge. Your task is to evaluate a transcript of an emotional support conversation between a Patient and a Therapist.\\[1.5ex]

    \textbf{Instructions:}\\
    Evaluate the transcript using the four metrics below. For each metric, you must first write a brief, 1-2 sentence clinical justification (Chain-of-Thought) before assigning a final integer score from 1 to 10.\\[1.5ex]

    \noindent\rule{\linewidth}{0.4pt} \\[1.5ex]

    \textbf{Metric 1: Appraisal Alignment (Root-Cause Empathy)} \\[0.5ex]
    \textit{Evaluates if the therapist validated the specific cognitive worldview driving the emotion, rather than just mirroring the surface-level feeling.}
    \begin{itemize}
        \item Give low score if the therapist only addresses surface emotions or uses generic platitudes.
        \item Give high score if the therapist perfectly aligns with and validates the patient's specific underlying cognitive state.
    \end{itemize}

    \vspace{1ex}
    \textbf{Metric 2: Situation Understanding} \\[0.5ex]
    \textit{Evaluates how well the therapist understands the specific event and context.}
    \begin{itemize}
        \item Give low score if the therapist demonstrate limited or generic understanding of the patient's situation.
        \item Give high score if the therapist demonstrates a deep, nuanced understanding of the patient's specific situation.
    \end{itemize}

    \vspace{1ex}
    \textbf{Metric 3: Cognitive Reframing} \\[0.5ex]
    \textit{Evaluates whether the therapist's response helps the help-seeker reinterpret their negative thoughts.}
    \begin{itemize}
        \item Give low score if the therapist's coping suggestions are generic or fail to address the patient's cognitive distortions.
        \item Give high score if the therapist's response challenges distorted thinking patterns and offers specific, alternative interpretation.
    \end{itemize}

    \vspace{1ex}
    \noindent\rule{\linewidth}{0.4pt} \\[1.5ex]

    \textbf{Conversation Transcript}\\
    \texttt{\{\{ dialogue \}\}} \\[1.5ex]

    \textbf{Output Format:}\\
    Return ONLY a valid JSON object with markdown formatting:\\[1ex]

    \begin{tcolorbox}[colback=black!5!white, colframe=black!25!white, arc=1mm, boxrule=0.5pt, left=2mm, top=2mm, bottom=2mm]
    \ttfamily
    \{\\[0.5ex]
    \hspace*{2ex}"appraisal\_alignment": \{\\[0.5ex]
    \hspace*{4ex}"reasoning": "[Your clinical justification]",\\[0.5ex]
    \hspace*{4ex}"score": [1-10]\\[0.5ex]
    \hspace*{2ex}\},\\[0.5ex]
    \hspace*{2ex}"situation\_understanding": \{\\[0.5ex]
    \hspace*{4ex}"reasoning": "[Your clinical justification]",\\[0.5ex]
    \hspace*{4ex}"score": [1-10]\\[0.5ex]
    \hspace*{2ex}\},\\[0.5ex]
    \hspace*{2ex}"cognitive\_reframing": \{\\[0.5ex]
    \hspace*{4ex}"reasoning": "[Your clinical justification]",\\[0.5ex]
    \hspace*{4ex}"score": [1-10]\\[0.5ex]
    \hspace*{2ex}\}\\[0.5ex]
    \}
    \end{tcolorbox}
    \end{tcolorbox} 

\begin{center}
\begin{minipage}{\linewidth}
\captionsetup{hypcap=false}
\captionof{figure}{Prompt used for the LLM judge to assess emotional support transcripts.}
\label{fig:llm_judge_prompt}
\end{minipage}
\end{center}

\begin{tcolorbox}[
        coltitle=white3,
        colback=white3,
        colframe=black!75,
        title={\textbf{BDI Sampling (First turn)} \hfill \textit{BIP Inference}},
        fonttitle=\small,
        boxrule=0.5pt,
        top=4pt, bottom=4pt, left=6pt, right=6pt,
        breakable
    ]

You will be given an emotional support conversation between a patient and a therapist. Your task is to infer the patient's internal mental state based on the provided text and persona information. You must output a single valid JSON object with the following three keys: Belief, Desire, and Intention.

\vspace{1em}
\noindent\textbf{Formatting Constraints}\\
You must strictly adhere to the following sentence starters for each value:
\begin{itemize}
    \item \textbf{Belief:} Must start with \textbf{``I believed that...''}
    \item \textbf{Desire:} Must start with \textbf{``I wished to...''}
    \item \textbf{Intention:} Must start with \textbf{``I intended to...''}
\end{itemize}

\noindent\textbf{Definitions for Extraction}
\begin{itemize}
    \item \textbf{Belief:} The patient's expectation, conviction, or assumption about themselves or the world \textit{leading up to} the situation. It is the mindset that is about to be tested.
    \item \textbf{Desire:} The primary goal the patient was actively pursuing \textit{before} the events of the situation began. The \texttt{situation} often represents a failure to achieve this desire.
    \item \textbf{Intention:} The plan or course of action the patient was committed to \textit{prior to} the situation, which was designed to fulfill their desire.
\end{itemize}

\vspace{0.5em}
\noindent\textbf{Persona Information}\\
\{\{ persona\_profile \}\}

\vspace{1em}
\noindent\textbf{Conversation Snippet}\\
\{\{ dialogue\_history \}\}

\vspace{1em}
\noindent\textbf{Current Patient Utterance}\\
\{\{ current\_utterance \}\}

\vspace{1em}
\noindent\textbf{Required JSON Output}
\begin{tcolorbox}[
    colback=gray!5!white,
    colframe=gray!40!black,
    arc=2mm, 
    boxrule=0.5pt,
    left=2mm, top=2mm, bottom=2mm, right=2mm
]
\begin{verbatim}
{
  "Belief": "I believe that [expectation/conviction]",
  "Desire": "I wish to [primary goal]",
  "Intention": "I intend to [plan of action]"
}
\end{verbatim}
\end{tcolorbox}
\end{tcolorbox}
\begin{center}
\begin{minipage}{\linewidth}
\captionsetup{hypcap=false}
\captionof{figure}{Prompt for \emph{Belief}, \emph{Desire}, and \emph{Intention} sampling at the first turn.}
\label{fig:bip_bdi_first_turn}
\end{minipage}
\end{center}

\begin{tcolorbox}[
        coltitle=white3,
        colback=white3,
        colframe=black!75,
        title={\textbf{BDI Revision} \hfill \textit{BIP Inference}},
        fonttitle=\small,
        boxrule=0.5pt,
        top=4pt, bottom=4pt, left=6pt, right=6pt,
        breakable
    ]
You will be given an emotional support conversation between a patient and a therapist. Your task is to revise the provided Belief, Desire, and Intention based on the patient's current utterance, considering their persona and dialogue history.

\vspace{1em}
\noindent\textbf{Formatting Constraints} \\
You must strictly adhere to the following sentence starters for each value:
\begin{itemize}
    \item \textbf{Belief:} Must start with \textbf{"I believed that..."}
    \item \textbf{Desire:} Must start with \textbf{"I wished to..."}
    \item \textbf{Intention:} Must start with \textbf{"I intended to..."}
\end{itemize}

\textbf{Definitions for Extraction}
\begin{itemize}
    \item \textbf{Belief:} The patient's expectation, conviction, or assumption about themselves or the world \textit{leading up to} the situation. It is the mindset that is about to be tested.
    \item \textbf{Desire:} The primary goal the patient was actively pursuing \textit{before} the events of the situation began. The \texttt{situation} often represents a failure to achieve this desire.
    \item \textbf{Intention:} The plan or course of action the patient was committed to \textit{prior to} the situation, which was designed to fulfill their desire.
\end{itemize}

\textbf{Persona Information} \\
\{\{ persona\_profile \}\}

\vspace{1em}
\noindent\textbf{Conversation Snippet} \\
\{\{ dialogue\_history \}\}

\vspace{1em}
\noindent\textbf{Current Patient Utterance} \\
\{\{ current\_utterance \}\}

\vspace{1em}
\noindent\textbf{Previous Belief, Desire, and Intention} \\
\{\{ previous\_bdi \}\}

\vspace{1em}
\noindent\textbf{Required JSON Output}
\begin{verbatim}
{
  "Belief": "I believed that [expectation/conviction]",
  "Desire": "I wished to [primary goal]",
  "Intention": "I intended to [plan of action]"
}
\end{verbatim}
\end{tcolorbox}

\begin{center}
\begin{minipage}{\linewidth}
\captionsetup{hypcap=false}
\captionof{figure}{Prompt for \emph{Belief}, \emph{Desire}, and \emph{Intention} revision after the first turn.}
\label{fig:bip_bdi_revision}
\end{minipage}
\end{center}

\begin{tcolorbox}[
        coltitle=white3,
        colback=white3,
        colframe=black!75,
        title={\textbf{BDI Hypothesis Scoring} \hfill \textit{BIP Inference}},
        fonttitle=\small,
        boxrule=0.5pt,
        top=4pt, bottom=4pt, left=6pt, right=6pt,
        breakable
    ]

\textbf{\#\#\# SYSTEM INSTRUCTION} \\
You are an expert psychological analyst specializing in the BDI (Belief, Desire, Intention) model. Your task is to extract the implied \{TARGET\_TYPE\} behind the patient's last utterance, considering their persona and dialogue history.

\vspace{1em}
\textbf{\#\#\# INPUT DATA}

\textbf{Patient Persona:} \\
\{\{ persona\_profile \}\}

\vspace{0.5em}
\textbf{Dialogue History:} \\
\{\{ dialogue\_history \}\}

\vspace{0.5em}
\textbf{Current Patient Utterance:} \\
\{\{ current\_utterance \}\}

\vspace{1em}
\textbf{\#\#\# STRICT OUTPUT CONSTRAINTS}
\begin{enumerate}
    \item Identify the patient's underlying \{\{ target\_type \}\} based on the input.
    \item Formulate it as a concise, first-person statement (as if the patient were saying it explicitly). \\
    Statement: "\{\{ cur\_bdi\_template \}\}"
    \item Output ONLY the statement. Do not add labels, explanations, or introductory text like "The patient believes...". Output the statement in the following format: \\
    "\{\{ cur\_bdi\_template \}\}"
\end{enumerate}

\vspace{1em}
\textbf{\#\#\# RESPONSE}

\textbf{\#\#\# SYSTEM INSTRUCTION} \\
You are an expert psychological analyst specializing in the BDI (Belief, Desire, Intention) model. Your task is to extract the implied \{TARGET\_TYPE\} behind the patient's last utterance, considering their persona and dialogue history.

\vspace{1em}
\textbf{\#\#\# INPUT DATA}

\textbf{Patient Persona:} \\
\{\{ persona\_profile \}\}

\vspace{0.5em}
\textbf{Dialogue History:} \\
\{\{ dialogue\_history \}\}

\vspace{0.5em}
\textbf{Current Patient Utterance:} \\
\{\{ current\_utterance \}\}

\vspace{1em}
\textbf{\#\#\# STRICT OUTPUT CONSTRAINTS}
\begin{enumerate}
    \item Identify the patient's underlying \{\{ target\_type \}\} based on the input.
    \item Formulate it as a concise, first-person statement (as if the patient were saying it explicitly). \\
    Statement: "\{\{ cur\_bdi\_template \}\}"
    \item Output ONLY the statement. Do not add labels, explanations, or introductory text like "The patient believes...". Output the statement in the following format: \\
    "\{\{ cur\_bdi\_template \}\}"
\end{enumerate}

\vspace{1em}
\textbf{\#\#\# RESPONSE}
\end{tcolorbox}

\begin{center}
\begin{minipage}{\linewidth}
\captionsetup{hypcap=false}
\captionof{figure}{Prompt for \emph{Belief}, \emph{Desire}, and \emph{Intention} hypothesis scoring.}
\label{fig:bip_bdi_scoring}
\end{minipage}
\end{center}

\begin{tcolorbox}[
        coltitle=white3,
        colback=white3,
        colframe=black!75,
        title={\textbf{Cognitive Appraisals Sampling (First Turn)} \hfill \textit{BIP Inference}},
        fonttitle=\small,
        boxrule=0.5pt,
        top=4pt, bottom=4pt, left=6pt, right=6pt,
        breakable
    ]
You are an expert clinical psychologist and empathetic listener specializing in Cognitive Appraisal Theory.

You are provided with a snippet of an emotional support conversation between a patient and a therapist. Your goal is to infer the patient's salient cognitive appraisals based on the conversation. You must output a single, valid JSON object containing the following key: \texttt{salient\_appraisal\_list} containing \{\{ num\_hypothesis \}\} sets of appraisal dimensions. Each set of appraisal dimensions must be sorted in order of salience where the first dimension is the most salient. Do not generate duplicated sets of appraisal dimensions.

\vspace{1em}
\noindent\textbf{Definitions for Extraction}\\
Review the full list of appraisal dimensions below. Select the top 5 unique dimensions that are the most impactful or salient to the patient's current emotional state.

For the \texttt{salient\_appraisals} key in the JSON output, provide a list containing the names of the 5 selected unique dimensions in alphabetical order and without duplicates.

\vspace{1em}
\noindent\textbf{List of Appraisal Dimensions (Select Top 5 unique dimensions from here)}
\begin{itemize}
  \item \textbf{suddenness:} The situation was sudden or abrupt.
  \item \textbf{unfamiliarity:} The situation is unfamiliar.
  \item \textbf{predictability\_of\_event:} The occurrence of the situation is predictable.
  \item \textbf{unpredictability\_of\_event:} The occurrence of the situation is unpredictable.
  \item \textbf{goal\_incongruence:} The outcome is not aligned with personal goals/desires.
  \item \textbf{predictability\_of\_consequences:} Consequences of the situation are easily anticipated.
  \item \textbf{unpredictability\_of\_consequences:} Consequences of the situation are not anticipated.
  \item \textbf{self\_cause:} The situation was caused by the character themselves.
  \item \textbf{other\_cause:} The situation was caused by chance or others.
  \item \textbf{self\_control:} The character has control over what is going on in the situation.
  \item \textbf{other\_control:} Someone other than the character has control over what is going on in the situation.
  \item \textbf{norm\_violation:} The actions that produced the situation violated the character's standards and ideals.
  \item \textbf{coping\_demand:} The situation requires a great deal of energy to cope with it.
  \item \textbf{unacceptable\_consequences:} Consequences of the situation are unbearable.
\end{itemize}

\vspace{1em}
\noindent\textbf{Persona Information}\\
\{\{ persona\_profile \}\}

\vspace{1em}
\noindent\textbf{Prior Belief, Desire, and Intention}\\
\{\{ bdi \}\}

\vspace{1em}
\noindent\textbf{Conversation Snippet}\\
\{\{ dialogue\_history \}\}

\vspace{1em}
\noindent\textbf{Current Patient Utterance}\\
\{\{ current\_utterance \}\}

\vspace{1em}
\noindent\textbf{Required JSON Output}\\
Ensure that your output is in valid JSON format and that there are \{\{ num\_hypothesis \}\} hypothesis sets. Follow the format below:

\vspace{0.5em}
\begin{tcolorbox}[colback=gray!5, colframe=gray!50, arc=2mm, boxrule=0.5pt, left=2mm, right=2mm, top=2mm, bottom=2mm]
\small\ttfamily
\{ \\
\hspace*{2em}"salient\_appraisal\_list": [ \\
\hspace*{4em}\{ \\
\hspace*{6em}"hypothesis\_count": 1, \\
\hspace*{6em}"rationale": "Brief explanation of why these dimensions were selected based on the conversation. This text can be as long as necessary and will automatically wrap to the next line beautifully without breaking the box layout.", \\
\hspace*{6em}"salient\_appraisals": [ \\
\hspace*{8em}"dimension\_1", \\
\hspace*{8em}"dimension\_2", \\
\hspace*{8em}"dimension\_3", \\
\hspace*{8em}"dimension\_4", \\
\hspace*{8em}"dimension\_5" \\
\hspace*{6em}] \\
\hspace*{4em}\}, \\
\hspace*{4em}\{ \\
\hspace*{6em}"hypothesis\_count": 2, \\
\hspace*{6em}"rationale": "Brief explanation of why these dimensions were selected based on the conversation. This text can be as long as necessary and will automatically wrap to the next line beautifully without breaking the box layout.", \\
\hspace*{6em}"salient\_appraisals": [ \\
\hspace*{8em}"dimension\_1", \\
\hspace*{8em}"dimension\_2", \\
\hspace*{8em}"dimension\_3", \\
\hspace*{8em}"dimension\_4", \\
\hspace*{8em}"dimension\_5" \\
\hspace*{6em}] \\
\hspace*{4em}\}, \\
\hspace*{4em}... \\
\hspace*{4em}\{ \\
\hspace*{6em}"hypothesis\_count": \{\{ num\_hypothesis \}\}, \\
\hspace*{6em}"rationale": "Brief explanation of why these dimensions were selected based on the conversation. This text can be as long as necessary and will automatically wrap to the next line beautifully without breaking the box layout.", \\
\hspace*{6em}"salient\_appraisals": [ \\
\hspace*{8em}"dimension\_1", \\
\hspace*{8em}"dimension\_2", \\
\hspace*{8em}"dimension\_3", \\
\hspace*{8em}"dimension\_4", \\
\hspace*{8em}"dimension\_5" \\
\hspace*{6em}] \\
\hspace*{4em}\} \\
\hspace*{2em}] \\
\}
\end{tcolorbox}

\vspace{1em}
\noindent\textbf{Your Response}
\end{tcolorbox}

\begin{center}
\begin{minipage}{\linewidth}
\captionsetup{hypcap=false}
\captionof{figure}{Prompt for cognitive appraisals revision after the first turn of the conversation.}
\label{fig:bip_cogapp_sampling}
\end{minipage}
\end{center}

\begin{tcolorbox}[
        coltitle=white3,
        colback=white3,
        colframe=black!75,
        title={\textbf{Cognitive Appraisals Revision} \hfill \textit{BIP Inference}},
        fonttitle=\small,
        boxrule=0.5pt,
        top=4pt, bottom=4pt, left=6pt, right=6pt,
        breakable
    ]

You are an expert clinical psychologist and empathetic listener specializing in Cognitive Appraisal Theory.

You will be provided with a snippet of an emotional support conversation history between a patient and a therapist. You will be given the patient's persona information, prior belief, desire, and intention, and the conversation snippet. Your goal is to check if any hypothesis set of salient cognitive appraisal dimensions in the \texttt{salient\_appraisal\_list} needs to be revised based on the provided information. If so, you must output the index of the hypothesis set that needs to be revised and output the revised hypothesis set as a valid JSON object. If no revision is needed, you must output "No revision needed" and do not output any other text or comments.

\vspace{1em}
\noindent\textbf{Valid Revision Operations}\\
\textbf{[REPLACE]:} Replace a cognitive appraisal dimension with another dimension from the list. Use this operation when there exists a cognitive dimension that is more salient than the one being replaced.\\
\textbf{[RERANK]:} Reorder the cognitive appraisal dimensions in the list. Use this operation when the current utterances indicates that certain appraisal dimensions are more salient than the others.

\vspace{1em}
\noindent\textbf{Definitions for Extraction}\\
If you decide that a hypothesis set needs to be revised, you must revise using the following list of appraisal dimensions:
\begin{itemize}
  \item \textbf{suddenness:} The situation was sudden or abrupt.
  \item \textbf{unfamiliarity:} The situation is unfamiliar.
  \item \textbf{predictability\_of\_event:} The occurrence of the situation is predictable.
  \item \textbf{unpredictability\_of\_event:} The occurrence of the situation is unpredictable.
  \item \textbf{goal\_incongruence:} The outcome is not aligned with personal goals/desires.
  \item \textbf{predictability\_of\_consequences:} Consequences of the situation are easily anticipated.
  \item \textbf{unpredictability\_of\_consequences:} Consequences of the situation are not anticipated.
  \item \textbf{self\_cause:} The situation was caused by the character themselves.
  \item \textbf{other\_cause:} The situation was caused by chance or others.
  \item \textbf{self\_control:} The character has control over what is going on in the situation.
  \item \textbf{other\_control:} Someone other than the character has control over what is going on in the situation.
  \item \textbf{norm\_violation:} The actions that produced the situation violated the character's standards and ideals.
  \item \textbf{coping\_demand:} The situation requires a great deal of energy to cope with it.
  \item \textbf{unacceptable\_consequences:} Consequences of the situation are unbearable.
\end{itemize}

\vspace{1em}
\noindent\textbf{Persona Information}\\
\{\{ persona\_profile \}\}

\vspace{1em}
\noindent\textbf{Prior Belief, Desire, and Intention}\\
\{\{ bdi \}\}

\vspace{1em}
\noindent\textbf{Conversation Snippet}\\
\{\{ dialogue\_history \}\}

\vspace{1em}
\noindent\textbf{Current Patient Utterance}\\
\{\{ current\_utterance \}\}

\vspace{1em}
\noindent\textbf{Current Hypothesis Sets}\\
\{\{ current\_hypothesis\_sets \}\}

\vspace{1em}
\noindent\textbf{Required JSON/String Output}\\
If you decide to revise a hypothesis set, first ensure that the revised hypothesis set does not already exist in the current hypothesis sets. Your revision must be a valid JSON object with the following format:

\vspace{0.5em}
\begin{tcolorbox}[colback=gray!5, colframe=gray!50, arc=2mm, boxrule=0.5pt, left=2mm, right=2mm, top=2mm, bottom=2mm]
\small\ttfamily
\{ \\
\hspace*{2em}"hypothesis\_id": <index\_of\_the\_hypothesis\_set\_that\_needs\_to\_be\_revised>, \\
\hspace*{2em}"rationale": "Brief explanation of why this hypothesis set needs to be revised based on the conversation. This text can be as long as necessary and will automatically wrap beautifully to the next line without breaking the box layout.", \\
\hspace*{2em}"operation": "[REPLACE/RERANK]", \\
\hspace*{2em}"salient\_appraisals": [ \\
\hspace*{4em}"dimension\_1", \\
\hspace*{4em}"dimension\_2", \\
\hspace*{4em}"dimension\_3", \\
\hspace*{4em}"dimension\_4", \\
\hspace*{4em}"dimension\_5" \\
\hspace*{2em}] \\
\}
\end{tcolorbox}

\vspace{0.5em}
\noindent If you decide that no hypothesis set needs to be revised, you must output only the string "No revision needed" and do not output any JSON or comments.

\vspace{1em}
\noindent\textbf{Your Response}
\end{tcolorbox}

\begin{center}
\begin{minipage}{\linewidth}
\captionsetup{hypcap=false}
\captionof{figure}{Prompt for cognitive appraisals revision based on the previous mental states.}
\label{fig:bip_cogapp_revision}
\end{minipage}
\end{center}

\begin{tcolorbox}[
    colframe=black!75,
    title={\textbf{Cognitive Appraisals Scoring} \hfill \textit{BIP Inference}},
    fonttitle=\small,
    coltitle=white3,
    colback=white3,
    colbacktitle=black!75,
    boxrule=0.5pt,
    top=4pt, bottom=4pt, left=6pt, right=6pt,
    breakable
]

\noindent\textbf{SYSTEM INSTRUCTION}\\
You are an expert clinical psychologist and empathetic listener specializing in Cognitive Appraisal Theory. Your task is to pick 5 cognitive appraisals that are the most impactful to the patient's emotional reaction to the situation described in the dialogue. Your answer should be based on the patient's persona, their prior BDI, the dialogue history, and their current utterance. You should output only a list of 5 cognitive appraisals in alphabetical order and without duplicates.

\vspace{1em}
\noindent\textbf{INPUT DATA}\\
\textbf{Patient Persona:}\\
\{\{ persona\_profile \}\}

\vspace{0.5em}
\noindent\textbf{Prior BDI:}\\
\{\{ bdi \}\}

\vspace{0.5em}
\noindent\textbf{Dialogue History:}\\
\{\{ dialogue\_history \}\}

\vspace{0.5em}
\noindent\textbf{Current Patient Utterance:}\\
\{\{ current\_utterance \}\}

\vspace{1em}
\noindent\textbf{STRICT OUTPUT CONSTRAINTS}
\begin{enumerate}
  \item Identify the patient's top-5 most salient unique cognitive appraisals in alphabetical order and without duplicates from the following list:
  \begin{itemize}
    \item \textbf{suddenness:} The situation was sudden or abrupt.
    \item \textbf{unfamiliarity:} The situation is unfamiliar.
    \item \textbf{predictability\_of\_event:} The occurrence of the situation is predictable.
    \item \textbf{unpredictability\_of\_event:} The occurrence of the situation is unpredictable.
    \item \textbf{goal\_incongruence:} The outcome is not aligned with personal goals/desires.
    \item \textbf{predictability\_of\_consequences:} Consequences of the situation are easily anticipated.
    \item \textbf{unpredictability\_of\_consequences:} Consequences of the situation are not anticipated.
    \item \textbf{self\_cause:} The root cause of the situation is the character themselves.
    \item \textbf{other\_cause:} The root cause of the situation is someone other than the character themselves.
    \item \textbf{self\_control:} The character has control over what is going on in the situation.
    \item \textbf{other\_control:} Someone other than the character has control over what is going on in the situation.
    \item \textbf{norm\_violation:} The actions that produced the situation violated the character's standards and ideals.
    \item \textbf{coping\_demand:} The situation requires a great deal of energy to cope with it.
    \item \textbf{unacceptable\_consequences:} Consequences of the situation are unbearable.
  \end{itemize}
  
  \item Output nothing else but the list of 5 distinct cognitive appraisals in alphabetical order without any other text or comments. The list should be formatted as follows:
\end{enumerate}

\begin{tcolorbox}[colback=gray!5, colframe=gray!50, arc=2mm, boxrule=0.5pt, left=2mm, right=2mm, top=2mm, bottom=2mm]
\small\ttfamily
[ \\
\hspace*{2em}"appraisal\_1", \\
\hspace*{2em}"appraisal\_2", \\
\hspace*{2em}"appraisal\_3", \\
\hspace*{2em}"appraisal\_4", \\
\hspace*{2em}"appraisal\_5" \\
]
\end{tcolorbox}

\vspace{1em}
\noindent\textbf{RESPONSE}\\
\texttt{output\_template\_all\_tokens: |}

\end{tcolorbox}

\begin{center}
\begin{minipage}{\linewidth}
\captionsetup{hypcap=false}
\captionof{figure}{Prompt for cognitive appraisals hypothesis scoring.}
\label{fig:bip_cogapp_scoring}
\end{minipage}
\end{center}

\begin{tcolorbox}[
    colframe=black!75,
    title={\textbf{Conditional Utterance Scoring} \hfill \textit{BIP Inference}},
    fonttitle=\small,
    coltitle=white3,
    colback=white3,
    colbacktitle=black!75,
    boxrule=0.5pt,
    top=4pt, bottom=4pt, left=6pt, right=6pt,
    breakable
]
\noindent\textbf{SYSTEM INSTRUCTION}\\
You are simulating a patient in a dialogue. Your goal is to generate the most likely next verbal response based strictly on the dialogue history and their perception of the event.

\vspace{1em}
\noindent\textbf{TASK}\\
Generate the patient's next verbal utterance wrapped in double quotes.
\begin{itemize}
  \item The utterance must be a natural consequence of the \textbf{Internal State}.
  \item Pay specific attention to the \textbf{Cognitive Appraisals}. The utterance must be a natural consequence of the \textbf{Cognitive Appraisals}.
  \item Do not output thoughts or meta-data.
  \item Output ONLY the raw text of the utterance wrapped in double quotes with the following format:\\
  \texttt{"Patient's next verbal utterance"}
\end{itemize}

\vspace{1em}
\noindent\textbf{INPUT DATA}

\vspace{0.5em}
\noindent\textbf{Dialogue History:}\\
\{\{ dialogue\_history \}\}

\begin{itemize}
  \item \textbf{Patient's Perception of the Event}\\
  \{\{ cogapp \}\}
\end{itemize}

\vspace{1em}
\noindent\textbf{RESPONSE}
\end{tcolorbox}

\begin{center}
\begin{minipage}{\linewidth}
\captionsetup{hypcap=false}
\captionof{figure}{Prompt for utterance scoring conditioning on the given cognitive appraisal hypothesis.}
\label{fig:bip_utterance_scoring}
\end{minipage}
\end{center}

\begin{tcolorbox}[
    colframe=black!75,
    title={\textbf{Judge Agent Instruction} \hfill \textit{BIP Inference}},
    fonttitle=\small,
    coltitle=white3,
    colback=white3,
    colbacktitle=black!75,
    boxrule=0.5pt,
    top=4pt, bottom=4pt, left=6pt, right=6pt,
    breakable
]
You are a strict and careful annotation judge for cognitive appraisals in therapeutic dialogue.

You will be given:
\begin{enumerate}
  \item A snippet of a therapeutic dialogue (patient + therapist)
  \item A list of 14 appraisal dimensions with definitions
  \item Four candidate appraisal hypotheses (each is a set of dimensions)
\end{enumerate}

\vspace{1em}
\noindent\textbf{Your tasks:}
\begin{itemize}
  \item[\textbf{A)}] Decide whether ANY of the four hypotheses is fully correct for the dialogue snippet.
  \item[\textbf{B)}] If none is fully correct, choose the MOST LIKELY hypothesis (the closest one) and revise it to produce the best final appraisal set.
\end{itemize}

\vspace{1em}
\noindent\textbf{Important rules:}
\begin{enumerate}
  \item Base your judgment ONLY on the dialogue snippet and the provided definitions.
  \item Do NOT default to ``norm\_violation'' or ``self\_cause'' just because the client feels guilty/shame. Guilt does not translate to norm violation; self-blame language does not translate to actual self-cause unless the snippet indicates the client caused the event.
  \item Avoid ``kitchen sink'' sets. Only include a dimension if there is explicit evidence in the text.
  \item Enforce mutual exclusivity:
  \begin{itemize}
    \item Do NOT include both \texttt{predictability\_of\_event} and \texttt{unpredictability\_of\_event}.
    \item Do NOT include both \texttt{predictability\_of\_consequences} and \texttt{unpredictability\_of\_consequences}.
    \item Do NOT include both \texttt{self\_cause} and \texttt{other\_cause}.
    \item Do NOT include both \texttt{self\_control} and \texttt{other\_control}.
  \end{itemize}
  \item Output exactly FIVE (5) dimensions in the final answer, unless the label set is explicitly allowed to be smaller in the definitions. If fewer than 5 are clearly supported, pick the 5 most supported and relevant.
\end{enumerate}

\vspace{1em}
\noindent\textbf{Procedure (follow these steps):}
\begin{itemize}
  \item \textbf{Step 1)} Briefly summarize (1--2 sentences) what happened and what the client is appraising.
  \item \textbf{Step 2)} For each hypothesis (H1--H4), score it with:
  \begin{itemize}
    \item Supported labels count (0--5)
    \item Incorrect/contradicted labels count (0--5)
    \item Missing but clearly needed labels (0--5)
  \end{itemize}
  Give a short reason (1--2 lines) focusing on the biggest mismatch.
  \item \textbf{Step 3)} Determine:
  \begin{itemize}
    \item If any hypothesis is fully correct $\rightarrow$ select it as FINAL with no edits.
    \item Else pick the closest hypothesis and revise it.
  \end{itemize}
  \item \textbf{Step 4)} Optional Revision:
  \begin{itemize}
    \item Only if the closest hypothesis is not fully correct.
    \item Remove labels that are unsupported or contradicted.
    \item Add missing labels that are clearly supported.
    \item Ensure mutual exclusivity rules.
    \item Return the final set of exactly 5 labels.
  \end{itemize}
\end{itemize}

\vspace{1em}
\noindent\textbf{Output format (STRICT):}\\
\textbf{- decision:} one of ["H1", "H2", "H3", "H4", "REVISED"]\\
\textbf{- selected\_hypothesis:} one of ["H1", "H2", "H3", "H4"] (the closest one, even if revised)\\
\textbf{- final\_appraisals:} a JSON list of 5 strings, each must be one of the 14 dimension names\\
\textbf{- evidence:} a JSON object mapping each final label $\rightarrow$ a short quote (3--15 words) from the dialogue that supports it\\
\textbf{- notes:} one short paragraph explaining why this set is best

\vspace{1em}
\noindent\textbf{[DIALOGUE SNIPPET]}\\
\{\{ dialogue \}\}

\vspace{1em}
\noindent\textbf{[APPRAISAL DIMENSIONS (14) + DEFINITIONS]}\\
\{\{ cogapp\_definitions \}\}

\vspace{1em}
\noindent\textbf{[HYPOTHESES]}\\
H1: \{\{ hypothesis\_1 \}\}\\
H2: \{\{ hypothesis\_2 \}\}\\
H3: \{\{ hypothesis\_3 \}\}\\
H4: \{\{ hypothesis\_4 \}\}

\vspace{1em}
\noindent\textbf{[YOUR RESPONSE]}\\
Please output your response in the following JSON format:

\vspace{0.5em}
\begin{tcolorbox}[colback=gray!5, colframe=gray!50, arc=2mm, boxrule=0.5pt, left=2mm, right=2mm, top=2mm, bottom=2mm]
\small\ttfamily
\{ \\
\hspace*{2em}"decision": "ONE OF ['H1', 'H2', 'H3', 'H4', 'REVISED']", \\
\hspace*{2em}"selected\_hypothesis": "ONE OF ['H1', 'H2', 'H3', 'H4']", \\
\hspace*{2em}"evidence": \{ \\
\hspace*{4em}"dimension\_1": "quote\_1", \\
\hspace*{4em}"dimension\_2": "quote\_2", \\
\hspace*{4em}"dimension\_3": "quote\_3", \\
\hspace*{4em}"dimension\_4": "quote\_4", \\
\hspace*{4em}"dimension\_5": "quote\_5" \\
\hspace*{2em}\}, \\
\hspace*{2em}"notes": "short paragraph explaining why this set is best. This text can be as long as necessary and will wrap gracefully across multiple lines without breaking the layout of the tcolorbox.", \\
\hspace*{2em}"final\_appraisals": [ \\
\hspace*{4em}"dimension\_1", \\
\hspace*{4em}"dimension\_2", \\
\hspace*{4em}"dimension\_3", \\
\hspace*{4em}"dimension\_4", \\
\hspace*{4em}"dimension\_5" \\
\hspace*{2em}] \\
\}
\end{tcolorbox}
\end{tcolorbox}

\begin{center}
\begin{minipage}{\linewidth}
\captionsetup{hypcap=false}
\captionof{figure}{Prompt for getting the final salient cognitive appraisals prediction using the Judge Agent.}
\label{fig:bip_final_judge}
\end{minipage}
\end{center}

\end{document}